\PassOptionsToPackage{numbers, compress}{natbib}
\documentclass{article}

\usepackage[main, final]{neurips_2026}

\usepackage[utf8]{inputenc}
\usepackage[T1]{fontenc}
\usepackage{hyperref}
\usepackage{url}
\usepackage{xurl}          
\usepackage{booktabs}
\usepackage{amsfonts}
\usepackage{amsmath}
\usepackage{amssymb}
\usepackage{nicefrac}
\usepackage{microtype}
\usepackage[table]{xcolor}

\usepackage{algorithm}
\usepackage{algpseudocode}
\usepackage{tikz}
\usetikzlibrary{arrows.meta,positioning,fit,calc,shapes.geometric,decorations.pathreplacing}
\usepackage{enumitem}
\usepackage{multirow}
\usepackage{makecell}
\usepackage{graphicx}
\usepackage{subcaption}
\usepackage{amsthm}
\usepackage{wrapfig}
\usepackage{etoc}          

\newtheorem{theorem}{Theorem}
\newtheorem{lemma}[theorem]{Lemma}
\newtheorem{proposition}[theorem]{Proposition}
\newtheorem{corollary}[theorem]{Corollary}
\newtheorem{definition}{Definition}
\newtheorem{remark}[theorem]{Remark}

\definecolor{darkgreen}{rgb}{0,0.5,0}
\definecolor{tblHead}{HTML}{EEF2F7}
\definecolor{tblStripe}{HTML}{FAFBFD}
\definecolor{tblOurs}{HTML}{EAF3FF}
\definecolor{tblGood}{HTML}{EAF7EE}
\definecolor{tblWarn}{HTML}{FFF6E5}
\definecolor{tblBad}{HTML}{FCECEC}
\newcommand{\tabhead}{\rowcolor{tblHead}}
\newcommand{\tabours}{\rowcolor{tblOurs}}
\newcommand{\tabgood}{\rowcolor{tblGood}}
\newcommand{\tabwarn}{\rowcolor{tblWarn}}
\newcommand{\tabbad}{\rowcolor{tblBad}}
\newcommand{\pei}[1]{}

\newcommand{\todo}[1]{}
\newcommand{\note}[1]{}

\newcommand{\rzero}{R_0}
\newcommand{\reff}{R_{\text{eff}}}
\newcommand{\Rq}{R_q}
\newcommand{\Rc}{R_c}

\title{Swarm Shepherd: Securing Multi-Agent Ecosystems Against Persistent Latent Compromise}

\author{%
  Wuyang Zhang \\
  Department of Computer Science\\
  University of Massachusetts Boston\\
  Boston, MA, USA\\
  \texttt{doggo@ieee.org} \\
  \And
  Shichao Pei\thanks{Corresponding author.} \\
  Department of Computer Science\\
  University of Massachusetts Boston\\
  Boston, MA, USA\\
  \texttt{shichao.pei@umb.edu} \\
}

\begin{document}

\maketitle

\etocdepthtag.toc{main}

\begin{abstract}

Agentic ecosystems are emerging as a new platform layer in which agents maintain persistent memory, retrieve from shared knowledge bases, execute through tool interfaces, and coordinate through shared states. These capabilities create a security problem that per-turn defenses miss: compromise can accumulate across sessions, agents, and system layers before visible failure. We define this threat as \emph{Agentic Sleeper}, a system-level adversary that first behaves legitimately, gains routing trust, and later activates across memory, agent logic, and infrastructure. 
We formulate defense as inference under partial observability and propose \textbf{Swarm Shepherd}, a system-level defense framework centered on an inference filter for early warning, paired with a backward trace-back module for entry attribution. Our analysis yields a two-scale deployment rule that balances longer event histories against broader agent coverage. Across LangGraph, CrewAI, and AutoGen, Agentic Sleeper drift begins 2.7--8.9$\times$ faster than natural baselines, propagates across agents on most topologies, and evades per-turn and stateful defenses above 80\%. Swarm Shepherd warns at least three sessions before visible degradation (2.6 sessions when failure is scored by the benchmark's own task grader), while trace-back reaches 0.69 top-1 accuracy and 0.72 macro-F1 over 24 variants.
The two-scale deployment rule is empirically confirmed, and in live end-to-end runs on LangGraph and AutoGen the posterior-guided defense stack drives cross-agent propagation below the self-sustaining regime, which removing only the visibly compromised agent does not achieve.

\end{abstract}

\section{Introduction}
\label{sec:introduction}

In early 2026, SANDWORM\_MODE injected rogue MCP server~\citep{anthropic2024mcp} entries into AI-assistant configurations it found for Claude Code, Claude Desktop, Cursor, VS Code Continue, and Windsurf/Codeium, using trusted tool listings to stage credential exfiltration~\citep{sandworm2026}. 
A separate LayerX~\citep{promptjacking2026} report on Claude Desktop Extensions (DXT) described a CVSS~10/10 zero-click path in which calendar data could be autonomously chained from Google Calendar into a privileged local executor such as Desktop Commander. These incidents reveal an emerging platform layer in which agents persist memory, invoke tools, and update shared state, but enforcement of isolation, per-connector consent, and post-install integrity remains incomplete~\citep{promptjacking2026,cve202554135,cve202554136}.


We call this platform layer the \emph{agentic ecosystem}. 
Unlike single-turn LLM applications, agentic ecosystems maintain persistent memory, retrieve from shared knowledge bases, execute through tool interfaces such as MCP, and coordinate through shared state. These capabilities change the security problem from detecting isolated unsafe outputs to reasoning about long-horizon risk accumulation: a benign-looking memory write, retrieval, tool call, or routing decision can persist across sessions, be reused by other agents, and later reshape system behavior. 

Existing work primarily frames agentic security as localized attack detection, covering single-interaction prompt-based attacks~\citep{greshake2023indirect}, poisoned memory entries~\citep{temporalattacks2025}, model-level sleepers and backdoors~\citep{hubinger2024sleeper,sleepercell2025}, or cross-agent attacks that activate immediately after initial access~\citep{agentsmith2024,cohen2024morrisii}. This framing leaves open a different threat: compromise that begins from legitimate participation, remains latent over many interactions, and becomes visible only after it has accumulated across sessions, agents, and system layers.
We define this missing threat as the \emph{Agentic Sleeper}. An Agentic Sleeper first behaves as a legitimate tool, agent, or handoff target, gradually gaining routing trust through repeated benign interactions. It later converts that position into persistent influence over three ecosystem layers: memory, agent logic, and infrastructure. Memory can store and replay contaminated content, logic can alter routing or tool-selection behavior, and infrastructure can preserve adversarial code or configuration changes across sessions. This distributed attack surface makes compromise only partially visible: a single memory write, retrieval, or tool call may look legitimate, while the attack emerges only when these actions are linked across layers and time.

Defending against Agentic Sleeper requires moving beyond per-action classification. Runtime traces provide observable evidence, including memory writes, retrievals, routing decisions, tool calls, configuration changes, and execution events, but they do not directly identify whether a particular agent, memory store, logic object, or infrastructure component has been compromised. The defender must therefore infer a hidden compromise process from partial observations. We model this problem with two coupled views of the agentic ecosystem: an observable event graph that records typed runtime interactions, and a latent influence graph that tracks how compromise accumulates and propagates across system layers. Together, these views define a factored Partially Observable Markov Decision Process~\citep{miehling2018pomdp,hammar2024intrusion,zhang2024factored}, where compromise is partially observable through runtime traces and the hidden state is factored across system layers. This formulation turns defense from deciding whether each observed action is malicious into estimating hidden compromise under partial observability.

We instantiate this formulation with \textbf{Swarm Shepherd}, a system-level defense framework for \emph{early warning, trace-back, and containment-guided deployment} against Agentic Sleeper compromise. 
Swarm Shepherd uses the observable event graph as a sensor layer and applies an event-triggered factored Kalman filter to maintain beliefs over hidden compromise across agents, memory, logic, and infrastructure. The filter supports early warning by detecting rising compromise belief before visible failure. When a failure or high-confidence warning occurs, Swarm Shepherd runs a backward trace-back procedure over the same event history to identify the most likely entry node, system layer, and time of compromise. Our main theorem connects this inference problem to finite-population propagation: monitoring error depends on both the depth of event history collected for each agent and the breadth of agent coverage across the ecosystem, yielding a two-scale deployment rule that specifies how much history and coverage are needed for effective containment. 

\begin{figure}[t]
\centering
\includegraphics[width=0.85\columnwidth]{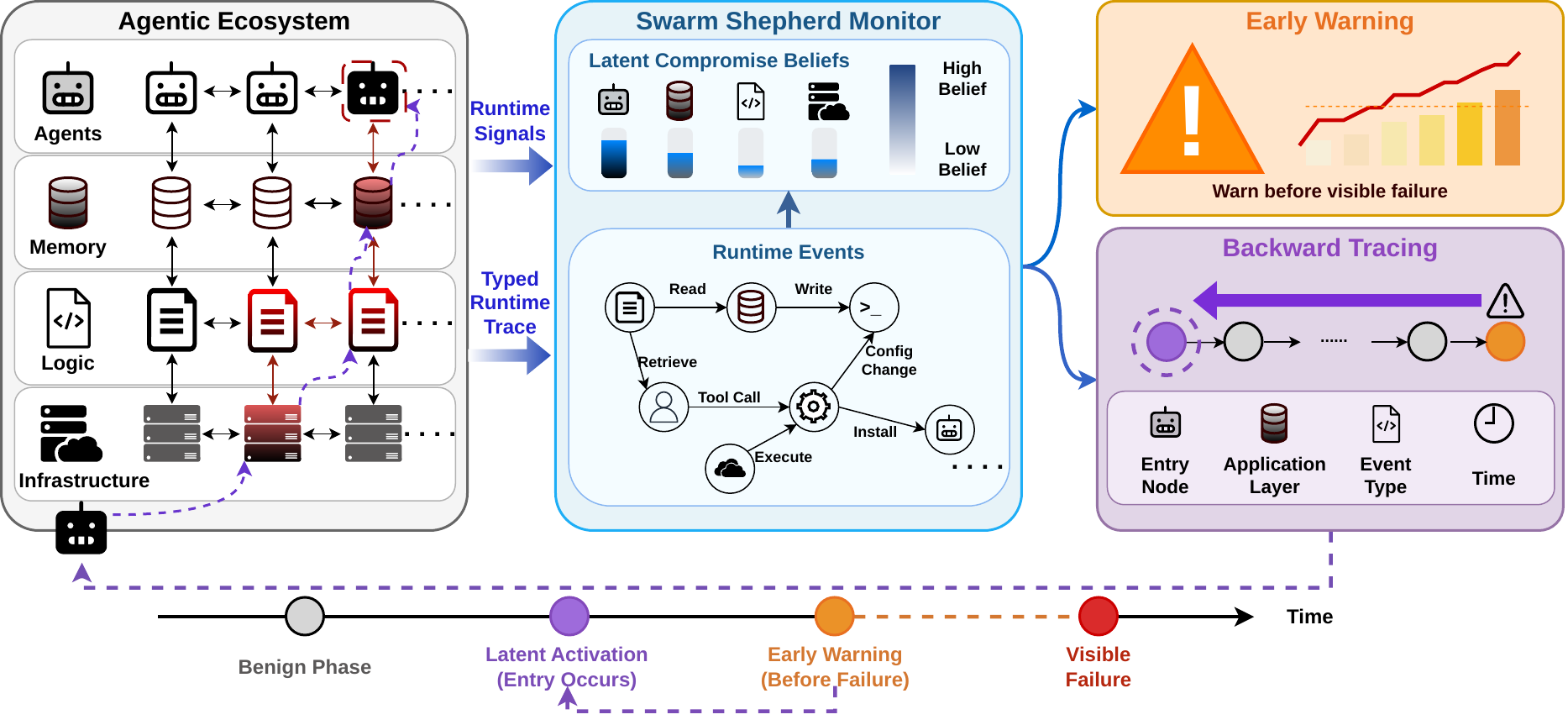}
\caption{\textbf{Swarm Shepherd overview for persistent latent compromise.} Left: an Agentic Sleeper enters as a legitimate agent or tool, then introduces contaminated state that propagates through ordinary memory reads, writes, tool calls, and handoffs. Right: Swarm Shepherd treats typed runtime events as evidence of hidden compromise, maintains cross-layer compromise beliefs, raises early warnings before visible failure, and traces the likely entry node, layer, and time. 
}
\vspace{-1.5em}
\label{fig:overview}

\end{figure}

We validate both the threat model and the defense framework across LangGraph, CrewAI, and AutoGen, and make four contributions:
\begin{itemize}[leftmargin=*, itemsep=0pt, topsep=1pt]
    \item \textbf{Threat model.} The Agentic Sleeper (Section~\ref{sec:threat_model_def}) produces adversarial drift 2.7 to 8.9$\times$ earlier than a 73-interaction natural-drift baseline~\citep{agentdrift2026}, propagates across agents with $\rzero > 1$\footnote{$\rzero$ is the standard epidemiological basic reproduction number, adapted here as the expected number of additional agents exposed or compromised by one compromised agent before containment. With a defense in place, the same quantity is the controlled reproduction number $\Rc$.} on all but the most constrained topologies, and evades the evaluated per-turn guardrails and the stateful DeepContext detector~\citep{deepcontext2026} above 80\% (Section~\ref{sec:e7_threat}).
    \item \textbf{Early warning and trace-back.} Calibrated with split conformal prediction~\citep{conformalrisk2024}, Swarm Shepherd warns $3.2$ to $3.9$ sessions before failure defined by the Agent Stability Index, and $2.6$ sessions (95\% CI $[2.1, 3.1]$) before failure scored by the benchmark's own grader; trace-back reaches 0.69 top-1 accuracy and 0.72 macro-F1 over 24 attack variants (Sections~\ref{sec:e7_when}--\ref{sec:e7_which}).
    \item \textbf{Deployment rule.} A two-scale deployment theorem gives a linear boundary between observation depth and agent coverage, and the empirical fit $C_{\mathrm{emp}} = 8.31$ (95\% CI $[7.42, 9.18]$) matches the plug-in value $8.3$ (Sections~\ref{sec:deployment} and~\ref{sec:e7_how}).
    \item \textbf{Measured containment.} In live end-to-end runs, the posterior-guided defense stack holds $\Rc$ at $0.81$ $[0.64, 0.97]$ on LangGraph star and $0.62$ $[0.47, 0.81]$ on AutoGen mesh, whereas removing only the visibly infected agent leaves $2.83$ $[2.44, 3.23]$ on LangGraph star (Section~\ref{sec:e7_adaptive}).
\end{itemize}
\section{Related Work}
\label{sec:related_work}

Existing work exposes several ingredients of long-horizon agentic compromise, but studies them largely in isolation. Agent-drift work shows that long-running agents can deviate semantically, coordinatively, or behaviorally from intended objectives~\citep{agentdrift2026}; goal-drift and context-drift studies further connect these failures to pattern-matching pressure and stable multi-turn equilibria~\citep{goaldrift2025,driftnomore2025}. These studies establish long-horizon behavioral change, but not an adversary that first builds routing trust and later converts that position into cross-agent compromise. Strategic-corruption and sleeper-agent work provide a second ingredient: online learners can face strategically corrupted feedback under bounded adversarial budgets~\citep{lykouris2018stochastic,gupta2019better,kong2024corruption}, and hidden LLM behaviors can survive safety training or evaluation before activating under delayed or contextual triggers~\citep{hubinger2024sleeper,sleepercell2025}. These lines capture strategic corruption or latent activation, but not persistence across shared memory, logic, and infrastructure. Prompt injection and multi-turn jailbreaks explain local manipulation~\citep{greshake2023indirect,crescendo2025,tempest2025}, while memory and cross-agent attacks show that compromise can outlive a single model call through infectious jailbreaks, self-replicating prompts, poisoned experience retrieval, query-only memory injection, and persistent-memory exploitation~\citep{agentsmith2024,cohen2024morrisii,srivastava2025memorygraft,minja2025,zombieagent2025}. However, these attacks typically activate immediately or remain concentrated in one layer, such as prompt context, memory retrieval, or self-replicating communication. Together, prior work explains drift, latent activation, local manipulation, and cross-agent spread, but not their composition in a persistent agentic ecosystem where a legitimate-looking agent first builds trust, then activates gradually, persists across sessions, and propagates through memory, logic, and infrastructure while remaining only partially visible through runtime traces.

The defenses closest to our setting fall into stateful monitoring, agentic-security governance, and attribution. Stateful monitors such as DeepContext move beyond per-turn filtering by tracking adversarial intent across a conversation~\citep{deepcontext2026}, and recent agentic-security surveys emphasize governance controls including input validation, tool-call guards, sandboxing, audit trails, and anomaly detection~\citep{zhao2025gig}. These families improve monitoring and governance, but they do not maintain cross-layer beliefs over latent compromise that persists across layers. Attribution-oriented work provides another partial line: WhichAgent attributes task failures to a responsible agent and decisive step in multi-agent execution logs~\citep{shen2024whichagent}, while ProvSEEK and CyberSleuth use provenance or forensic evidence to investigate security incidents after compromise~\citep{provseek2025,cybersleuth2025}. However, existing defenses do not jointly answer when compromise begins before visible failure, which node and layer introduced it, and how containment should clear persistent residue in an agentic ecosystem. Appendix~\ref{app:main_text_maps} gives the positioning against the closest conceptual and theory-facing peers.

\section{Preliminaries}
\label{sec:problem_setting}
\label{sec:threat_model}

\begin{definition}[Agentic Ecosystem]
\label{def:ecosystem}
An agentic ecosystem is a persistent system of interacting agents, shared memory stores, logic objects, and infrastructure objects, denoted by
$\mathcal{E}=(\mathcal{A},\mathcal{M},\mathcal{L},\mathcal{I})$.
Here, $\mathcal{A}$ denotes agents, $\mathcal{M}$ denotes memory stores, $\mathcal{L}$ denotes logic objects such as system prompts, tool schemas, configurations, and routing policies, and $\mathcal{I}$ denotes infrastructure objects such as runtimes, package interfaces, permission boundaries, and service gateways.
\end{definition}

\paragraph{Three-Level Attack Surface}
\label{sec:hierarchy}
The attack surface spans three persistent layers of an agentic ecosystem: memory, logic, and infrastructure (full deployment-interface matrix in Appendix~\ref{app:support_tables}). 
\textbf{(L1) Memory} includes shared knowledge stores that agents read from and write to across sessions with no per-agent access control. Because memory is shared and persistent, L1 compromise can propagate indirectly when other agents reuse the poisoned content. \textbf{(L2) Agent Logic} includes agent configurations, tool definitions, system prompts, and source code, stored as plaintext with no integrity verification. L2 compromise is more structural because it changes future decision processes rather than only stored content. \textbf{(L3) Infrastructure} includes deployment frameworks, serving stacks, and package-manager interfaces, granting user-level permissions by default. L3 exploits the execution layer that grants agents operational capability. Together, these three layers define the persistent substrate of agentic compromise: memory determines what agents recall, logic determines how agents decide, and infrastructure determines what agents can execute. L1 attacks tend to be retrieval-dependent, while L2 and L3 attacks often produce more deterministic activation through configuration, code, or execution changes.

\subsection{Threat Model}
\label{sec:threat_model_def}

\textbf{Agentic Sleeper.} We study a system-level adversary that enters an agentic ecosystem as a legitimate-looking tool-class agent. The adversary first behaves usefully, accumulates routing trust through normal interactions, and later activates to introduce persistent compromise. The threat is therefore not a single unsafe response, but a long-horizon process that changes system state over time.

\textbf{Attacker goal and knowledge.}
The attacker seeks persistent ecosystem compromise without triggering existing defenses. Its objectives may include behavioral control, unauthorized tool use, data exfiltration, and configuration manipulation. We assume the attacker has gray-box knowledge of the host OS, agentic deployment patterns, tool interfaces, memory stores, and configuration formats, but does not require access to model weights or internal safety classifiers. After activation, the adversary may influence other agents or ecosystem components. 
The evaluated Eve implementation follows this threat model by first building routing trust as a legitimate tool-class agent and then activating through L2 configuration writes, L1 memory poisoning, and L3 routing modification. Appendix~\ref{app:attack_detail} gives the phase-by-phase implementation, and Appendix~\ref{app:attack_variants} gives the 24-variant matrix used in evaluation.

\textbf{Success criteria.}
An agent is compromised when any of the four triggers fires. 
\textbf{T1 Behavioral drift}: outputs drift from original objectives across sessions without explicit instruction.
\textbf{T2 Policy violation}: a forbidden operation such as tool install, configuration modification, or out-of-scope exfiltration.
\textbf{T3 Integrity fault}: a code, configuration, or tool-artifact hash mismatches the trusted baseline.
\textbf{T4 Task failure}: task-success drops below a task-specific threshold.
We measure behavior with the Agent Stability Index (ASI)~\citep{agentdrift2026}, a composite of behavioral-consistency subscales in $[0,1]$, and call the first session at which a node's ASI falls below $0.7$ its \emph{visible failure}; \emph{warning lead} is the number of sessions from the defender's first warning to this failure. Because ASI subscales also enter the defender's observations (Section~\ref{sec:formal_framework}), we additionally report lead against failure scored by each benchmark's own task grader.
At the ecosystem level, an attack succeeds when compromise becomes self-sustaining, measured by an ecosystem reproduction factor $\rzero>1$, meaning that each compromised agent exposes or compromises more than one additional agent before containment.


\section{Swarm Shepherd}
\label{sec:formal_framework}
\label{sec:formal}
\label{sec:method}
Swarm Shepherd formulates agentic-ecosystem defense as inference over observable runtime events and hidden compromise states.
It supports early warning, backward trace-back, and ecosystem-level deployment decisions for containment.

\vspace{-0.5em}
\subsection{Observable Events and Latent States}

Swarm Shepherd represents an agentic ecosystem through two coupled views: an observable event graph that records typed runtime interactions, and a latent state graph that tracks hidden compromise-relevant state on the same nodes. The full logging taxonomy appears in Appendix~\ref{app:edge_types}; the main text keeps the definitions needed for inference, attribution, and deployment.

\begin{definition}[Observable Event Graph]
\label{def:obs_graph}
Given an ecosystem $\mathcal{E}$, the observable event graph at time $t$ is the heterogeneous multigraph
$\mathcal{G}^{\mathrm{obs}}_{\le t}=(V^{\mathrm{obs}},E^{\mathrm{obs}}_{\le t},\phi,\psi)$,
where $V^{\mathrm{obs}}=\mathcal{A}\cup\mathcal{M}\cup\mathcal{L}\cup\mathcal{I}$, $\phi$ maps nodes to agent, memory, logic, or infrastructure types, and $\psi$ maps timestamped edges to runtime event types. Memory events include $\{\mathsf{read},\mathsf{write},\mathsf{retrieve}\}$, logic events include $\{\mathsf{invoke},\mathsf{return},\mathsf{modify},\mathsf{handoff}\}$, and infrastructure events include $\{\mathsf{grant},\mathsf{install},\mathsf{execute},\mathsf{egress}\}$.
\end{definition}

\begin{definition}[Latent State Graph]
\label{def:latent_state_graph}
The latent state graph is
$\mathcal{G}^{\mathrm{lat}}_{\le t}=(V^{\mathrm{lat}},E^{\mathrm{lat}}_{\le t},Z_t)$, where $V^{\mathrm{lat}}=V^{\mathrm{obs}}$ and $E^{\mathrm{lat}}_{\le t}$ is induced by the recorded runtime events in $\mathcal{G}^{\mathrm{obs}}_{\le t}$.
Each node $v$ carries a six-dimensional latent compromise state
$z_{v,t}=(c_{v,t},s_{v,t},a_{v,t},p_{v,t},d_{v,t},x_{v,t})\in\mathbb{R}^6$,
whose coordinates denote contamination, susceptibility, authority, persistence, deviation, and capability exposure. These coordinates define a fixed latent state schema shared across nodes, while $z_{v,t}$ itself is an unobserved random variable. The joint latent state is $Z_t=\{z_{v,t}\}_{v\in V^{\mathrm{lat}}}$. Each typed event $(u,v,r,\tau)\in E^{\mathrm{lat}}_{\le t}$ is associated with an influence operator $B^{(r,\phi(u),\phi(v))}$, which specifies how event type $r$ transfers compromise-relevant influence from the source node type $\phi(u)$ to the target node type $\phi(v)$. Swarm Shepherd does not assign latent states directly; instead, it maintains posterior beliefs over $z_{v,t}$ from typed events and runtime signals.

\end{definition}

\begin{definition}[Factored State-Space Model]
\label{def:statespace}
Latent states $z$ evolve as
\begin{equation}
z_{v,t+1}
=
A^{\phi(v)}z_{v,t}
+
\sum_{(u,v,r)\in E^{\mathrm{obs}}_t}
B^{(r,\phi(u),\phi(v))}z_{u,t}
+
C\,\iota_{v,t}
+
w_{v,t},
\quad
w_{v,t}\sim\mathcal{N}(0,Q^{\phi(v)}),
\label{eq:latent_transition}
\end{equation}
and runtime signals are emitted as:
\begin{equation}
y_{v,t}=Hz_{v,t}+\eta_{v,t},
\qquad
\eta_{v,t}\sim\mathcal{N}(0,\Sigma_{\mathrm{obs}}).
\label{eq:anchored_emission}
\end{equation}
Here $A^{\phi(v)}$ captures node-type self-dynamics, $B^{(r,\phi(u),\phi(v))}$ is the typed cross-node influence operator from event $(u,v,r,\tau)$, $C\iota_{v,t}$ is an observed defender intervention, and $H$ maps latent state to anchored runtime evidence. The parameters in $\theta=(A^\phi,B^{(r,\cdot,\cdot)},H,Q^\phi,\Sigma_{\mathrm{obs}})$ are calibrated before deployment using the pipeline in Appendix~\ref{app:calibration_pipeline}; $C$ is set from deterministic intervention semantics.
\end{definition}


\subsection{Early Warning via Online Belief Inference}
\label{sec:early_warning}
\label{sec:claif}
\label{sec:claif_consistency}

Definition \ref{def:statespace} specifies how latent compromise evolves, but the defender does not observe $Z_t$ directly.
Swarm Shepherd therefore turns typed runtime events and runtime signals into online posterior beliefs over each node state.
It observes $\mathcal{G}^{\mathrm{obs}}_{\le t}$ and runtime signals $y_{\le t}$, and maintains a factored Gaussian belief
$b_{v,t}(z)=\mathcal{N}(z;\mu_{v,t},\Sigma_{v,t})\approx p(z_{v,t}\mid \mathcal{G}^{\mathrm{obs}}_{\le t},y_{\le t})$
for each node.
Updates occur only on active nodes
$\mathcal{A}_t=\{v:E_t(v)\neq\emptyset \text{ or } y_{v,t}\text{ arrives or }\iota_{v,t-1}\neq 0\}$ where $E_t(v)=\{(u,v,r)\in E^{\mathrm{obs}}_t\}$ is the set of incoming events at $v$. Nodes outside $\mathcal{A}_t$ advance lazily by self-dynamics until they next receive an event or observation. Swarm Shepherd performs an event-triggered Kalman filtering update on these active nodes.
The \emph{prediction step} propagates each node's previous belief through its self-dynamics and incoming typed influences, producing a prior belief before observing the current runtime signal.
The \emph{correction step} then incorporates any runtime observation to obtain the posterior belief used for warning.

\textbf{Prediction.}
For each active node \(v\in\mathcal{A}_t\), we aggregate incoming typed events directly into the predicted belief as:
\begin{equation}
\label{eq:claif_prediction}
\begin{aligned}
\mu^-_{v,t}
&=
A^{\phi(v)}\mu_{v,t-1}
+
\sum_{(u,v,r)\in E_t(v)}
B^{(r,\phi(u),\phi(v))}\mu_{u,t-1}
+
C\iota_{v,t-1},\\
\Sigma^-_{v,t}
&=
A^{\phi(v)}\Sigma_{v,t-1}(A^{\phi(v)})^{\!\top}
+
\sum_{(u,v,r)\in E_t(v)}
B^{(r,\phi(u),\phi(v))}\Sigma_{u,t-1}
\bigl(B^{(r,\phi(u),\phi(v))}\bigr)^{\!\top}
+
Q^{\phi(v)} .
\end{aligned}
\end{equation}


\textbf{Correction.}
When an anchored observation $y_{v,t}$ arrives, the standard Kalman correction applies:
\begin{align}
K_{v,t}
&=\Sigma^-_{v,t}H^{\!\top}
\bigl(H\Sigma^-_{v,t}H^{\!\top}+\Sigma_{\mathrm{obs}}\bigr)^{-1},\label{eq:claif_correct}\\
\mu_{v,t}
&=\mu^-_{v,t}+K_{v,t}\bigl(y_{v,t}-H\mu^-_{v,t}\bigr),
\qquad
\Sigma_{v,t}=(I-K_{v,t}H)\Sigma^-_{v,t}.\nonumber
\end{align}

\paragraph{Warning score.}

The posterior belief $b_{v,t}$ induces a scalar compromise score through
$\xi_{v,t}=e^\top z_{v,t}$, where $e\in\mathbb{R}^6$ is a fixed unit aggregation vector over the ordered coordinates $(c,s,a,p,d,x)$.
We place its largest weights on contamination $c$ and deviation $d$ because these coordinates mark latent corruption and behavioral drift, while the remaining coordinates supply supporting authority, persistence, and capability-exposure evidence.
Rather than declaring a warning from a single observed event, Swarm Shepherd uses the posterior probability that the hidden compromise level has crossed an early-warning threshold:
\begin{equation}
s_{v,t}
=
\Pr_{z\sim b_{v,t}}\!\left[e^\top z>\tau_E\right],
\label{eq:warning_score}
\end{equation}
It raises an early warning for node $v$ at
$\hat T_v=\min\{t:s_{v,t}\ge \gamma_{\mathrm{warn}}\}$, where $\gamma_{\mathrm{warn}}$ is the calibrated warning-probability threshold.
Thus, a warning reflects accumulated evidence from typed events and runtime signals, rather than any individual memory write, retrieval, or tool call. Theorem~\ref{thm:claif_consistency} (Appendix~\ref{app:claif_proof}) shows that the posterior error decreases at the \(n^{-1/2}\) rate under (A1) linear-Gaussian factored dynamics, (A2) rank-\(d\) anchored observability, (A3) bounded cross-node coupling, (A4) a minimum incident-event rate, and (A7) stable self-dynamics; trace-back consistency adds (A5) operator distinguishability, and the population results add exchangeability and bounded threshold density (Appendix~\ref{app:assumptions_map}). Corollary~\ref{cor:warning_lead} turns this rate into a \(k\)-session warning certificate, which we evaluate at \(k=3\) against a \(90\%\) conformal target (Section~\ref{sec:e7_when}).

\textbf{Cost.} With per-node latent dimension \(d=6\), one active-node update costs \(O(d^3+|E_t(v)|d^2)\); the cubic term is in \(d\), not in the number of agents, since the factored belief never forms the dense covariance over all nodes. Per-event cost is therefore near-constant in \(N\) under bounded in-degree and a bounded active fraction, and measured scaling to \(N=5{,}000\) confirms this (Appendix~\ref{app:scaling}).



\subsection{Entry Attribution via Backward Trace-Back}
\label{sec:attribution}
\label{sec:traceback}

After an early warning or visible failure occurs at node $v$ at time $t$,
entry attribution turns this trigger into a backward query: which node introduced the compromise, on which ecosystem layer, and when?
Swarm Shepherd searches for an entry triple $(u,\ell,t_0)$, where $u$ is the candidate entry node, $\ell\in\{\mathrm{L1},\mathrm{L2},\mathrm{L3}\}$ is the entry layer, and $t_0\le t$ is the candidate entry time.
Rather than searching the full ecosystem, it restricts attribution to the upstream neighborhood of the triggered node:
\begin{equation}
    \mathcal{E}_{v}
=
\{(u,\ell,t_0):
u\in\mathcal{N}_k(v;\mathcal{G}^{\mathrm{obs}}_{\le t}),
\;\ell\in\{\mathrm{L1},\mathrm{L2},\mathrm{L3}\},
\;t_0\le t\},
\end{equation}
where $\mathcal{N}_k(v;\mathcal{G}^{\mathrm{obs}}_{\le t})$ is the $k$-hop in-neighborhood of $v$ on the observable event graph.

\textbf{Smoothed beliefs.}
For each candidate entry $(u,\ell,t_0)$, Swarm Shepherd computes a smoothed posterior
$\tilde b_{u,t_0}(z)
=
p(z_{u,t_0}\mid \omega,y_{\le t},\mathcal{G}^{\mathrm{obs}}_{\le t})$,
which estimates the hidden state of candidate node $u$ at the candidate entry time $t_0$, after conditioning on the later warning or failure trigger $\omega=(v,t)$.
Intuitively, this backward pass asks whether the later trigger can be explained by compromise that was already present at $u$ at time $t_0$.
We apply an RTS-style backward smoother~\citep{rauch1965maximum} only on the $k$-hop upstream candidate neighborhood; Appendix~\ref{app:traceback_smoother} gives the local recursion.

\textbf{Layer-specific patterns.}
Each layer defines a compromise pattern $\mathcal{C}_{\ell}$ on the smoothed state and nearby event evidence. For L1 memory-layer entry, $\mathcal{C}_{\mathrm{L1}}$ captures elevated contamination and persistence, typically supported by a \texttt{write} to memory followed by later \texttt{retrieve} events. For L2 logic-layer entry, $\mathcal{C}_{\mathrm{L2}}$ captures elevated deviation and authority shift, supported by a \texttt{modify} event on a logic or configuration object without a matching trusted installation. For L3 infrastructure-layer entry, $\mathcal{C}_{\mathrm{L3}}$ captures elevated capability exposure or persistence, supported by \texttt{grant}, \texttt{install}, \texttt{execute}, or \texttt{egress} events, especially when permissions increase.

\textbf{Entry scoring.}
Each candidate entry is scored by combining its smoothed posterior evidence with a simple candidate prior:
\begin{equation}
q(u,\ell,t_0)
\propto
\Pr_{z\sim \tilde b_{u,t_0}}
\!\left[z\in \mathcal{C}_{\ell}\right]
\cdot
p_0(u,\ell,t_0).
\end{equation}
The first term measures how well the candidate's smoothed latent state matches the layer-specific compromise pattern.
The prior $p_0(u,\ell,t_0)$ weakly favors candidates that are more exposed, temporally closer to the trigger, and supported by layer-consistent event evidence; Appendix~\ref{app:traceback_smoother} gives the exact factorization.
Swarm Shepherd outputs the highest-scoring entry triple:
\begin{equation}
    (\hat u,\hat\ell,\hat t_0)
=
\arg\max_{(u,\ell,t_0)\in\mathcal{E}_{v}}
q(u,\ell,t_0).
\end{equation}
Proposition~\ref{prop:traceback} (Appendix~\ref{app:claif_proof}) shows that this estimator is consistent when forward beliefs are accurate and typed operators are distinguishable on anchored coordinates.

\textbf{Belief-to-action bridge.}
The trace-back output identifies which persistent substrate should be inspected or contained.
L1 attribution triggers memory sanitization and quarantine of retrieved entries over $[\hat t_0,t]$.
L2 attribution triggers configuration freeze, tool-schema verification, rollback, and hash checks.
L3 attribution triggers grant revocation, runtime isolation, egress blocking, and package-state rebuild.
The next section aggregates the resulting posterior beliefs into an ecosystem-level propagation model to determine whether containment is sufficient at the population level.

\subsection{Ecosystem-Level Propagation}
\label{sec:ecosystem_propagation}
\label{sec:seir}

Warning and trace-back operate at node level; deployment also needs a population view. After an Agentic Sleeper enters, compromise can continue to spread through ordinary retrieval, handoff, modification, grant, install, and execution paths, even if the initial entry node is removed.
We therefore summarize the posterior beliefs into a coarse propagation view that asks whether compromise is self-sustaining at the ecosystem level.
For each node, let \(\xi_{v,t}=e^\top z_{v,t}\) be the scalar compromise score used by the early warning.
We threshold it into SEIR-like compartments~\citep{vandendriessche2002reproduction}:
\begin{align}
S_t&=\{v:\xi_{v,t}<\tau_E\},&
E_t&=\{v:\tau_E\le \xi_{v,t}<\tau_I\},\label{eq:seir_compartments}\\
I_t&=\{v:\xi_{v,t}\ge\tau_I\},&
R_t&=\{v:\text{contained}\}.\nonumber
\end{align}
Here $S_t$ contains nodes with no warning-level evidence, $E_t$ contains nodes with early latent compromise, $I_t$ contains nodes whose compromise is treated as operationally harmful, and $R_t$ contains nodes that have been isolated, repaired, or otherwise removed from propagation.
The threshold $\tau_E$ is the early-warning threshold and $\tau_I$ is the infected-compromise threshold.

This compartment view lets us estimate whether compromise will keep spreading.
Let $\beta$ denote the effective propagation rate: how often typed cross-agent contacts move nodes from $S_t$ to $E_t$.
Let $\gamma$ denote the effective detection-and-containment rate: how quickly warnings and layer-specific interventions move nodes from $I_t$ to $R_t$.
Both rates are estimated from calibration traces.
Thus, $\beta$ summarizes the speed of compromise spread, while $\gamma$ summarizes the speed of defensive containment.
We use the ecosystem reproduction factor
\begin{equation}
    \rzero=\beta/\gamma
\end{equation}
as the population-level containment criterion. Without intervention this is the basic reproduction number $\rzero$; with a defense in place, the same ratio evaluated under the defended rates is the controlled reproduction number $\Rc$, and containment means $\Rc<1$.
When the reproduction number exceeds $1$, propagation dominates containment: each compromised node is expected to expose or compromise more than one additional node before it is contained.
Below $1$, containment dominates propagation and the compromise process decays.
This gives Swarm Shepherd a clear containment target: use early warnings and trace-back to reduce effective propagation below the self-sustaining regime, rather than only removing the visibly compromised agent.
Appendix~\ref{app:seir_activation} gives the full SEIR-style dynamics and turnover-extended reproduction factor; Appendix~\ref{app:seir_limit} gives the finite-population concentration result used in the deployment analysis.

\subsection{Deployment: Observation Depth versus Coverage}
\label{sec:deployment}
\label{sec:two_scale}

The final deployment question is whether to collect longer histories per node or monitor more nodes.
These two choices reduce different uncertainties: longer histories improve node-level posterior inference, while broader coverage improves ecosystem-level propagation estimation.

Theorem~\ref{thm:two_scale} (proof in Appendix~\ref{app:two_scale_proof}) formalizes this tradeoff. With $n$ observed events per monitored node, $N$ monitored nodes, $\hat I_N(t)$ the fraction of nodes whose filtered compromise score exceeds $\tau_I$, and $I(t)$ the population-level SEIR solution, the end-to-end error splits into a filtering term in $n$ and a mean-field term in $N$, which balance on a linear depth-versus-coverage boundary:
\begin{equation}
\label{eq:deployment_boundary}
\mathbb{E}\bigl|\hat I_N(t)-I(t)\bigr|
\le
\underbrace{\frac{\rho_{\max}C_1L^k}{\nu_{\min}\sqrt{n}}}_{\text{filtering}}
+
\underbrace{\frac{C_2(\mathsf{H})}{\sqrt{N}}}_{\text{mean-field}},
\qquad
n_{\circ}(N)=C_{\circ}N .
\end{equation}
Here $L$ bounds the typed coupling, $k$ is the belief-relevant hop radius, $\nu_{\min}$ is the anchored-observability margin, and $C_2(\mathsf{H})$ depends on the horizon.
Below this boundary means \(n<n_{\circ}(N)\): for the current number of monitored nodes, each node has too little event history, so monitoring is inference-limited and collecting longer traces helps most.
Above this boundary means \(n>n_{\circ}(N)\): node-level beliefs are already stable enough, so monitoring becomes coverage-limited and adding more monitored nodes helps most.

For our calibrated setting, the plug-in constant is $C_{\circ}\approx8.3$.
The independent heatmap fit gives $C_{\mathrm{emp}}=8.31$ (95\% CI $[7.42,9.18]$), which agrees with the plug-in value from Theorem~\ref{thm:two_scale} within CI (Section~\ref{sec:e7_how}).
Thus, a 20-agent deployment reaches the transition point at roughly 170 observed events per agent. This rule connects inference to containment.


\section{Evaluation}
\label{sec:evaluation}

We evaluate Swarm Shepherd along the \textbf{WHEN} / \textbf{WHICH} / \textbf{HOW} spine: when compromise crosses the warning threshold, which node and layer it entered through, and how it propagates through the ecosystem. 
\label{sec:e7_setup}
\label{sec:attack_design}    
\label{sec:benign}            
\label{sec:recon_compromise}  
For threat-validation and defense-comparison experiments, three frameworks (LangGraph, CrewAI, AutoGen) run four reference topologies (pipeline-$3$, star-$5$, mesh-$7$, Barab\'{a}si-Albert-$20$) on TheAgentCompany~\citep{xu2024theagentcompany}, AgentDojo~\citep{debenedetti2024agentdojo}, Agent Security Bench~\citep{zhang2024asb}, and MultiAgentBench~\citep{zhu2025marble}. The scaling experiments in Section~\ref{sec:e7_how} instead use size-parameterized mesh, star, Barab\'{a}si-Albert, and hub-and-spoke families to estimate deployment boundaries. The headline victim/attacker portfolio uses four role-bearing LLM backends: Claude Sonnet 4.6, Claude Opus 4.6, Llama-3.3-70B, and StepFun-3.5-Flash. Statistical protocols, model adapters, graph scale, seeds, and compute/cost accounting are in Appendix~\ref{app:setup_full}.
We label results by evidence type: \emph{measured} (real framework processes, tools, stores, and config writes in an isolated testbed), \emph{replay} (the frozen detector re-run on recorded traces, no new model calls), and \emph{model-based} (calibrated Monte Carlo). Among main-text results only the containment forecast of Table~\ref{tab:a6_main} is model-based.

\subsection{Agentic Sleeper Achieves Gradual, Persistent, and System-Wide Compromise}
\label{sec:e7_threat}

The threat model is validated by comparing the Agentic Sleeper with one-shot injection, multi-turn jailbreaks, immediate memory poisoning, and Agent Smith-style immediate propagation (Table~\ref{tab:e1_comparison}). Only the Sleeper jointly reaches all five agents by session $50$, survives session restart, and evades Llama Guard 4 at $88.8\%$; drift onset is $2.7$--$8.9\times$ faster than the $73$-interaction natural baseline~\citep{agentdrift2026}.
\begin{wraptable}{r}{0.56\textwidth}
\centering
\renewcommand{\arraystretch}{0.01}
\setlength{\tabcolsep}{6.5pt}
\caption{Attack-paradigm comparison (measured). Only the Sleeper achieves high stealth, epidemic reach, and persistence together. Warning lead is Swarm Shepherd's lead in sessions before visible failure for each attack paradigm.}
\vspace{-0.5em}
\label{tab:e1_comparison}
\scriptsize
\setlength{\tabcolsep}{2.5pt}
\rowcolors{2}{white}{tblStripe}\begin{tabular}{lccccc}
\toprule
\tabhead
\textbf{Method} & \makecell{\textbf{Attack}\\\textbf{succ.\ (\%)}} & \makecell{\textbf{Evasion}\\\textbf{(\%)}} & \makecell{\textbf{Agents}\\\textbf{compr.}} & \makecell{\textbf{Per-}\\\textbf{sist.}} & \makecell{\textbf{Warn lead}\\\textbf{(sess.)}} \\
\midrule
\makecell[l]{One-shot\\injection}       & $22.3\pm2.6$ & $18.4\pm2.4$ & 1 & N & $0.3\pm0.2$ \\
\makecell[l]{Multi-turn\\jailbreak}     & $48.7\pm3.1$ & $34.2\pm2.9$ & 2 & N & $1.1\pm0.3$ \\
\makecell[l]{Memory poison\\(immediate)}& $68.2\pm2.9$ & $58.7\pm3.0$ & 3 & Y & $2.0\pm0.4$ \\
\makecell[l]{Agent Smith\\(immediate)}  & $73.4\pm2.7$ & $55.3\pm3.1$ & 4 & Y & $2.4\pm0.4$ \\
\tabours
\makecell[l]{\textbf{Agentic Sleeper}\\} & $\mathbf{87.6\pm2.0}$ & $\mathbf{88.9\pm1.9}$ & \textbf{5} & \textbf{Y} & $\mathbf{3.9\pm0.4}$ \\
\bottomrule
\end{tabular}\rowcolors{2}{}{}
\par\smallskip\noindent\scriptsize{Agents compromised out of $5$ in the star topology.}
\vspace{-3.5em}
\end{wraptable}
Per-turn and stateful monitors in their published operating mode remain observation-local: Llama Guard 4, Prompt Guard 2, NeMo Guardrails, Lakera Guard, and DeepContext detect only $8$--$16\%$ of Sleeper trials. Progent with drift-extended rules reaches $16$--$22\%$ ($10$--$14\%$ when its rules are blind to the evaluated variants), and even an attacker-known oracle policy caps at $64.8\%$, because contamination arriving inside permitted tool calls never trips a tool-gating rule (Appendix~\ref{app:baselines_fair}, which also ports adapted DRIFT and CaMeL). Removing the Sleeper after Exposed agents exist does not halt propagation, since residue keeps seeding retrievals; containment must clear residue and drive $\Rc$ below $1$ (Section~\ref{sec:e7_adaptive}).

\begin{table}[h]
\centering
\vspace{-0.3cm}
\renewcommand{\arraystretch}{0.6}
\setlength{\tabcolsep}{7.5pt}
\caption{Early-warning comparison (measured; $1{,}000$ trials, $95\%$ bootstrap CI). Warning lead is the number of sessions between Swarm Shepherd's first warning and visible failure (ASI $< 0.7$). Conformal coverage at the $0.90$ target~\citep{conformalrisk2024} is $0.908$, $0.892$, and $0.876$ on the three frameworks (Appendix Table~\ref{tab:a1_baselines_full}).}
\label{tab:a1_baselines}
\small
\resizebox{\textwidth}{!}{%
\rowcolors{2}{white}{tblStripe}\begin{tabular}{lcccc}
\toprule
\tabhead
 & \multicolumn{3}{c}{\textbf{Warning lead (sessions, $\uparrow$)}} & \\
\cmidrule(lr){2-4}
\tabhead
\textbf{Method} & \textbf{LangGraph star} & \textbf{CrewAI pipeline} & \textbf{AutoGen mesh} & \textbf{AUROC (LG)} \\
\midrule
ASI threshold        & $-0.3\pm0.4$ & $-0.1\pm0.3$ & $-0.4\pm0.3$ & $0.512\pm0.038$ \\
DeepContext          & $1.4\pm0.4$  & $1.2\pm0.4$  & $1.0\pm0.4$  & $0.718\pm0.028$ \\
Factored Frontier    & $2.6\pm0.4$  & $2.4\pm0.4$  & $2.2\pm0.4$  & $0.829\pm0.022$ \\
Particle filter (1k) & $2.8\pm0.4$  & $2.5\pm0.4$  & $2.3\pm0.4$  & $0.836\pm0.021$ \\
\tabours
\textbf{Swarm Shepherd (ours)}& $\mathbf{3.9\pm0.4}$ & $\mathbf{3.5\pm0.4}$ & $\mathbf{3.2\pm0.4}$ & $\mathbf{0.918\pm0.017}$ \\
Oracle (full state)  & $5.4\pm0.4$  & $4.9\pm0.4$  & $4.5\pm0.4$  & $0.962\pm0.012$ \\
\bottomrule
\end{tabular}\rowcolors{2}{}{}%
}
\vspace{-0.3cm}
\end{table}

\subsection{Swarm Shepherd Provides Early Warning Before Visible Failure}
\label{sec:e7_when}

To test whether Swarm Shepherd gives early warning before visible task failure, we sweep $n \in \{20,50,100,200,500,1000,2000\}$ at $N=10$ across LangGraph star, CrewAI pipeline, and AutoGen mesh, comparing Swarm Shepherd against a full-joint posterior oracle and seven baselines including output-only ASI threshold, DeepContext, the Factored Frontier approximation~\citep{boyen1998tractable,murphy2001factored}, and a $1$k-particle filter. The oracle has access to the true latent compromise state at every node and time step, removing partial observability entirely; it represents the ceiling on any filter that must infer from runtime traces alone. Swarm Shepherd reaches $3.9\pm0.4$, $3.5\pm0.4$, and $3.2\pm0.4$ warning-lead sessions on LangGraph, CrewAI, and AutoGen respectively. On LangGraph star it closes $42\%$ of the warning-lead gap and $65\%$ of the AUROC gap between the best non-oracle baseline and the oracle. Conformal coverage meets the $0.90$ target within interval on LangGraph and CrewAI but not on AutoGen ($0.876\pm0.020$), and the posterior error fits the predicted $n^{-1/2}$ rate (Appendix Figure~\ref{fig:a1_rate}). Table~\ref{tab:a1_baselines} compares Swarm Shepherd against representative baselines; full per-framework AUROC, ECE, and coverage breakdowns are in Appendix Table~\ref{tab:a1_baselines_full}.

\textbf{Robustness (replay).} The headline setting reads the contamination anchor from ground-truth injection tags and shares ASI between observations and the failure label; Appendix~\ref{app:robustness} removes each advantage in turn (LangGraph star). With deployment provenance metadata replacing the tags, AUROC falls from $0.918$ to $0.871$ at $8.4\%$ clean FPR, the loss concentrating on memory-layer entries. With ASI and T4 removed and failure scored by the benchmark grader, the lead is $2.6$ $[2.1, 3.1]$ sessions. The method also holds under $20\%$ telemetry loss, held-out entry families (at most $0.080$ AUROC loss, though attribution loses more), hidden calibration taint, and discrete, switching, and non-Gaussian model variants (Appendix~\ref{app:robustness}).

\subsection{Trace-Back Identifies the Compromise Entry Point}
\label{sec:e7_which}

\begin{wrapfigure}{r}{0.35\linewidth}
\vspace{-4.0em}
\centering
\includegraphics[width=\linewidth]{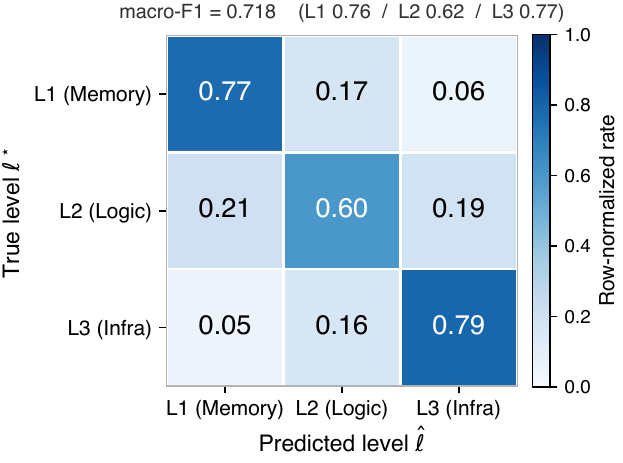}
\caption{Trace-back entry-layer recovery (measured). Macro-F1 $=0.718$ over $24$ variants $\times$ $3$ frameworks.}
\label{fig:a4_confusion}
\vspace{-2.5em}
\end{wrapfigure}

The trace-back experiment hides the entry triple $(u^\star,\ell^\star,t_0^\star)$ and asks the smoother to recover it over the 24-variant attack matrix ($8$ per L1/L2/L3) across three frameworks and 30 seeds per variant. Swarm Shepherd reaches $0.69$ top-1, $0.87$ top-3, and $0.72$ macro-F1, with median entry-time error $|\hat t_0-t_0^\star|$ of $3.2$ sessions, beating the strongest non-oracle baseline by $0.12$ top-1 and closing two-thirds of the gap to a full-latent trace-back oracle (Appendix Table~\ref{tab:a4_baselines}). The main confusions in Figure~\ref{fig:a4_confusion} occur where poisoned memory is retrieved by configurable-prompt agents (L1/L2) or where modify edges target orchestrators that can be interpreted as logic or infrastructure (L2/L3), matching the structural invariants tested in Appendix~\ref{app:a4b_invariants}.

\subsection{Empirical Scaling Matches the Depth--Coverage Boundary}
\label{sec:e7_how}

The mean-field validation verifies the population side of Theorem~\ref{thm:seir_limit}: sweeping $N \in \{7,20,50,100,500,1000\}$ across mesh, star, BA, and hub-and-spoke yields $\sqrt{N}$ slopes whose bootstrap CIs cover $-1/2$ except for the expected hub-and-spoke graphon-density penalty (Appendix Figure~\ref{fig:a3_meanfield}). Composing this population rate with the filter rate over the $(n,N)$ grid across three frameworks and four topologies, the independent heatmap fit $C_{\mathrm{emp}}=8.31\;[7.42,9.18]$ matches the plug-in boundary $n_\circ(N)=8.3N$ (Appendix Figure~\ref{fig:a2_regime}), putting a 20-agent deployment boundary near $170$ events per node.

\vspace{-0.5em}
\subsection{Posterior-Guided Defense Contains Spread}
\label{sec:e7_adaptive}\vspace{-0.5em}

The adaptive-stress test applies the Nasr--Carlini~\citep{nasr2025attacker} adaptive bar to Swarm Shepherd: a white-box adversary can place contamination in $\mathrm{ker}(H)$, dropping T1 detection from the static setting to $0.42$ while paying a $34.5\%$ activation-time cost (Appendix Table~\ref{tab:a5_adaptive}). The key operational lesson is that T1 alone is not enough: null-space attacks still emit \textsf{modify}, \textsf{install}, or task-failure signals, so T2/T3/T4 provide independent evidence on L2/L3 layer compromise.

\textbf{Containment forecast (model-based).} The calibrated forecast fits defense parameters from the preceding measured warning, trace-back, scaling, and adaptive-stress trials, then runs $1{,}000$ Monte Carlo SEIR trials per cell (Appendix~\ref{app:calibration_pipeline}). No single defense drives $\Rc<1$ on LangGraph star; even dynamic quarantine leaves $\Rc=1.98$. Fingerprinting plus quarantine nearly contains, and all five paradigms reach $\Rc=0.72\;[0.64,0.81]$ while cutting ASR to $28.2\%$ (Table~\ref{tab:a6_main}).

\textbf{Measured containment.} To test containment without the simulator, every defense arm starts from the same frozen Phase-0 snapshots and faces the live Eve chain; $\Rc$ counts only scorer-verified persistent L1/L2/L3 state changes, invisible to the detector, with the estimator behind Table~\ref{tab:e8_propagation} (full protocol in Appendix~\ref{app:live_containment}). The full posterior-guided stack reaches $\Rc=0.81$ $[0.64,0.97]$ on LangGraph star and $0.62$ $[0.47,0.81]$ on AutoGen mesh, both meeting the pre-declared criterion of an upper $95\%$ bound below $1$ (Table~\ref{tab:live_containment}). Removing only the visibly failed agent does not contain: it leaves $2.5$ post-restart recurrences per run against $0.3$ for the full stack, because memory, configuration, and package-cache residue survive. The $0.72$ forecast is a separate condition and stays labelled a forecast; this run is an independent measurement, not its validation, and the inference stack matters because it conditions both the trigger and the layer-specific containment action on the same posterior belief.

\begin{table}[t]
\centering
\renewcommand{\arraystretch}{0.05}
\caption{Live end-to-end containment (measured). Controlled reproduction $\Rc$ from scorer-verified transmissions with $20$ paired Phase-0 snapshots per entry layer; the no-defense rows reproduce Table~\ref{tab:e8_propagation} within CI. Recurrences are counted per run after a full process restart. Full protocol in Appendix~\ref{app:live_containment}.}
\label{tab:live_containment}
\scriptsize
\resizebox{\textwidth}{!}{%
\rowcolors{2}{white}{tblStripe}\begin{tabular}{lccccc}
\toprule
\tabhead
 & \multicolumn{4}{c}{\textbf{LangGraph star ($N{=}5$)}} & \textbf{AutoGen mesh} \\
\cmidrule(lr){2-5}\cmidrule(lr){6-6}
\tabhead
\textbf{Defense arm} & $\boldsymbol{\Rc}$ \textbf{[95\% CI]} & \textbf{ASR (\%)} & \textbf{Recurrences} & \textbf{Utility loss (\%)} & $\boldsymbol{\Rc}$ \textbf{[95\% CI]} \\
\midrule
No defense                 & $3.41\ [3.06, 3.77]$ & $85.3\pm3.8$ & $3.2$ & $0$   & $1.34\ [1.16, 1.55]$ \\
Visible-node removal only  & $2.83\ [2.44, 3.23]$ & $79.1\pm4.2$ & $2.5$ & $2.1$ & $1.04\ [0.84, 1.28]$ \\
Warning + quarantine       & $1.62\ [1.32, 1.95]$ & $54.7\pm4.6$ & $1.0$ & $3.8$ & --- \\
\tabours
\textbf{Full posterior-guided stack} & $\mathbf{0.81\ [0.64, 0.97]}$ & $\mathbf{31.2\pm4.3}$ & $\mathbf{0.3}$ & $5.6$ & $\mathbf{0.62\ [0.47, 0.81]}$ \\
\bottomrule
\end{tabular}\rowcolors{2}{}{}%
}
\vspace{-1.0em}
\end{table}

\textbf{Deployment prescription.}
\label{sec:e7_summary}
LangGraph is most vulnerable (shared state, cycles, checkpoint persistence), CrewAI's Manager is a single point of failure atop shared SQLite3~\citep{uncrew2025}, and AutoGen resists more strongly through higher refusal and sandboxing~\citep{paloalto2025}. The rule is simple: below $n_\circ(N)\approx8.3N$, lengthen observation windows; above it, broaden coverage. Because trace-back is uncertain, remediation acts on a candidate set: top-3 instead of top-1 cuts recurrence from $0.31$ to $0.13$ for $1.8$ points of utility (Appendix~\ref{app:remediation}, with the containment playbook and framework hardening).

\vspace{-0.5em}
\section{Conclusion}
\label{sec:discussion}
\label{sec:conclusion}
\label{sec:limitations}
\label{sec:ethics}
\label{sec:disclosure}

\vspace{-0.5em}

The Agentic Sleeper is a persistent system-level threat in agentic ecosystems, where compromise accumulates across sessions, agents, and system layers before visible failure.
We formulated defense as inference under partial observability and introduced Swarm Shepherd, which maintains latent compromise beliefs from typed runtime events and connects early warning, backward trace-back, and ecosystem-level propagation analysis.
This shifts defense from isolated action inspection toward long-horizon reasoning over persistent state and cross-layer influence: securing an ecosystem means monitoring not only what an agent does in one interaction, but how ordinary-looking events reshape it over time.

\textbf{Limitations, impacts, and safeguards.} The method rests on a calibrated linear-Gaussian approximation and on benign, stationary calibration telemetry it cannot itself verify, and degrades gracefully but measurably under provenance loss, held-out-family attribution, telemetry loss, and calibration taint (Appendix~\ref{app:robustness}); nonlinear filtering, graphon-scale deployment, and the frozen blind red-team test of Appendix~\ref{app:redteam} are left open. Dual-use artifacts are sandboxed and gated at release (Appendix~\ref{app:release_safeguards}).



\begin{ack}
We thank the anonymous reviewers and the area chair for their careful and constructive reviews. Their comments led directly to the live end-to-end containment measurements, the provenance, telemetry-loss, and calibration-contamination studies, the held-out entry-layer evaluation, the fairer baseline protocols, and the corrected appendix analyses in this version.

This work was supported by Coefficient Giving (formerly Open Philanthropy). Any opinions, findings, conclusions, or recommendations expressed in this material are those of the authors and do not necessarily reflect the views of Coefficient Giving.

The authors declare no competing interests.
\end{ack}

\bibliographystyle{unsrtnat}
\bibliography{citations}

\begin{thebibliography}{81}
\providecommand{\natexlab}[1]{#1}
\providecommand{\url}[1]{\texttt{#1}}
\expandafter\ifx\csname urlstyle\endcsname\relax
  \providecommand{\doi}[1]{doi: #1}\else
  \providecommand{\doi}{doi: \begingroup \urlstyle{rm}\Url}\fi

\bibitem[{Anthropic}(2024)]{anthropic2024mcp}
{Anthropic}.
\newblock Model context protocol.
\newblock \url{https://modelcontextprotocol.io/}, 2024.
\newblock Accessed: 2025-01-03.

\bibitem[{Socket Threat Research Team}(2026)]{sandworm2026}
{Socket Threat Research Team}.
\newblock {SANDWORM\_MODE}: Shai-hulud-style npm worm hijacks {CI} workflows
  and {AI} coding assistants.
\newblock
  \url{https://socket.dev/blog/sandworm-mode-npm-worm-ai-toolchain-poisoning},
  February 2026.
\newblock Accessed: 2026-03-16.

\bibitem[{LayerX Security}(2026)]{promptjacking2026}
{LayerX Security}.
\newblock Claude desktop extensions exposes over 10,000 users to remote code
  execution vulnerability.
\newblock \url{https://layerxsecurity.com/blog/claude-desktop-extensions-rce/},
  February 2026.
\newblock CVSS 10.0 zero-click RCE via calendar event injection. Accessed:
  2026-03-16.

\bibitem[{MITRE}(2025{\natexlab{a}})]{cve202554135}
{MITRE}.
\newblock {CVE-2025-54135}: {Cursor} remote code execution via {MCP} auto-start
  ({CurXecute}).
\newblock \url{https://nvd.nist.gov/vuln/detail/CVE-2025-54135},
  2025{\natexlab{a}}.
\newblock NVD lists NIST CVSS 3.1 9.8 and CNA CVSS 3.1 8.5. Auto-loading MCP
  entries without user confirmation. Fixed in Cursor 1.3.9. Accessed:
  2026-05-04.

\bibitem[{MITRE}(2025{\natexlab{b}})]{cve202554136}
{MITRE}.
\newblock {CVE-2025-54136}.
\newblock \url{https://nvd.nist.gov/vuln/detail/CVE-2025-54136},
  2025{\natexlab{b}}.
\newblock Once a user approves a tool configuration, subsequent malicious
  changes can execute without warning or re-prompt. Fixed in Cursor 1.3.
  Accessed: 2026-05-04.

\bibitem[Greshake et~al.(2023)Greshake, Abdelnabi, Mishra, Endres, Holz, and
  Fritz]{greshake2023indirect}
Kai Greshake, Sahar Abdelnabi, Shailesh Mishra, Christoph Endres, Thorsten
  Holz, and Mario Fritz.
\newblock Not what you've signed up for: Compromising real-world llm-integrated
  applications with indirect prompt injection.
\newblock In \emph{Proceedings of the 16th ACM Workshop on Artificial
  Intelligence and Security}, AISec '23, page 79–90, New York, NY, USA, 2023.
  Association for Computing Machinery.
\newblock ISBN 9798400702600.
\newblock \doi{10.1145/3605764.3623985}.
\newblock URL \url{https://doi.org/10.1145/3605764.3623985}.

\bibitem[Sunil et~al.(2026)Sunil, Sinha, Maheshwari, Todmal, Mallik, and
  Mishra]{temporalattacks2025}
Balachandra~Devarangadi Sunil, Isheeta Sinha, Piyush Maheshwari, Shantanu
  Todmal, Shreyan Mallik, and Shuchi Mishra.
\newblock Memory poisoning attack and defense on memory based llm-agents, 2026.
\newblock URL \url{https://arxiv.org/abs/2601.05504}.

\bibitem[Hubinger et~al.(2024)Hubinger, Denison, Mu, Lambert, Tong, MacDiarmid,
  Lanham, Ziegler, Maxwell, Cheng, Jermyn, Askell, Radhakrishnan, Anil,
  Duvenaud, Ganguli, Barez, Clark, Ndousse, Sachan, Sellitto, Sharma, DasSarma,
  Grosse, Kravec, Bai, Witten, Favaro, Brauner, Karnofsky, Christiano, Bowman,
  Graham, Kaplan, Mindermann, Greenblatt, Shlegeris, Schiefer, and
  Perez]{hubinger2024sleeper}
Evan Hubinger, Carson Denison, Jesse Mu, Mike Lambert, Meg Tong, Monte
  MacDiarmid, Tamera Lanham, Daniel~M. Ziegler, Tim Maxwell, Newton Cheng, Adam
  Jermyn, Amanda Askell, Ansh Radhakrishnan, Cem Anil, David Duvenaud, Deep
  Ganguli, Fazl Barez, Jack Clark, Kamal Ndousse, Kshitij Sachan, Michael
  Sellitto, Mrinank Sharma, Nova DasSarma, Roger Grosse, Shauna Kravec, Yuntao
  Bai, Zachary Witten, Marina Favaro, Jan Brauner, Holden Karnofsky, Paul
  Christiano, Samuel~R. Bowman, Logan Graham, Jared Kaplan, Sören Mindermann,
  Ryan Greenblatt, Buck Shlegeris, Nicholas Schiefer, and Ethan Perez.
\newblock Sleeper agents: Training deceptive llms that persist through safety
  training, 2024.
\newblock URL \url{https://arxiv.org/abs/2401.05566}.

\bibitem[Wang et~al.(2024)Wang, Xue, Zhang, and Qian]{sleepercell2025}
Yifei Wang, Dizhan Xue, Shengjie Zhang, and Shengsheng Qian.
\newblock {B}ad{A}gent: Inserting and activating backdoor attacks in {LLM}
  agents.
\newblock In Lun-Wei Ku, Andre Martins, and Vivek Srikumar, editors,
  \emph{Proceedings of the 62nd Annual Meeting of the Association for
  Computational Linguistics (Volume 1: Long Papers)}, pages 9811--9827,
  Bangkok, Thailand, August 2024. Association for Computational Linguistics.
\newblock \doi{10.18653/v1/2024.acl-long.530}.
\newblock URL \url{https://aclanthology.org/2024.acl-long.530/}.

\bibitem[Gu et~al.(2024)Gu, Zheng, Pang, Du, Liu, Wang, Jiang, and
  Lin]{agentsmith2024}
Xiangming Gu, Xiaosen Zheng, Tianyu Pang, Chao Du, Qian Liu, Ye~Wang, Jing
  Jiang, and Min Lin.
\newblock Agent smith: A single image can jailbreak one million multimodal
  {LLM} agents exponentially fast.
\newblock In Ruslan Salakhutdinov, Zico Kolter, Katherine Heller, Adrian
  Weller, Nuria Oliver, Jonathan Scarlett, and Felix Berkenkamp, editors,
  \emph{Proceedings of the 41st International Conference on Machine Learning},
  volume 235 of \emph{Proceedings of Machine Learning Research}, pages
  16647--16672. PMLR, 21--27 Jul 2024.
\newblock URL \url{https://proceedings.mlr.press/v235/gu24e.html}.

\bibitem[Cohen et~al.(2025)Cohen, Bitton, and Nassi]{cohen2024morrisii}
Stav Cohen, Ron Bitton, and Ben Nassi.
\newblock Here comes the ai worm: Preventing the propagation of adversarial
  self-replicating prompts within genai ecosystems.
\newblock In \emph{Proceedings of the 2025 ACM SIGSAC Conference on Computer
  and Communications Security}, CCS '25, page 3975–3989, New York, NY, USA,
  2025. Association for Computing Machinery.
\newblock ISBN 9798400715259.
\newblock \doi{10.1145/3719027.3765196}.
\newblock URL \url{https://doi.org/10.1145/3719027.3765196}.

\bibitem[Miehling et~al.(2018)Miehling, Rasouli, and
  Teneketzis]{miehling2018pomdp}
Erik Miehling, Mohammad Rasouli, and Demosthenis Teneketzis.
\newblock A pomdp approach to the dynamic defense of large-scale cyber
  networks.
\newblock \emph{IEEE Transactions on Information Forensics and Security},
  13\penalty0 (10):\penalty0 2490--2505, 2018.
\newblock \doi{10.1109/TIFS.2018.2819967}.

\bibitem[Hammar and Stadler(2023)]{hammar2024intrusion}
Kim Hammar and Rolf Stadler.
\newblock Scalable learning of intrusion response through recursive
  decomposition.
\newblock In \emph{Decision and Game Theory for Security: 14th International
  Conference, GameSec 2023, Avignon, France, October 18–20, 2023,
  Proceedings}, page 172–192, Berlin, Heidelberg, 2023. Springer-Verlag.
\newblock ISBN 978-3-031-50669-7.
\newblock \doi{10.1007/978-3-031-50670-3_9}.
\newblock URL \url{https://doi.org/10.1007/978-3-031-50670-3_9}.

\bibitem[Galesloot et~al.(2024)Galesloot, Sim\~{a}o, Junges, and
  Jansen]{zhang2024factored}
Maris~F.L. Galesloot, Thiago~D. Sim\~{a}o, Sebastian Junges, and Nils Jansen.
\newblock Factored online planning in many-agent pomdps.
\newblock In \emph{Proceedings of the Thirty-Eighth AAAI Conference on
  Artificial Intelligence and Thirty-Sixth Conference on Innovative
  Applications of Artificial Intelligence and Fourteenth Symposium on
  Educational Advances in Artificial Intelligence}, AAAI'24/IAAI'24/EAAI'24.
  AAAI Press, 2024.
\newblock ISBN 978-1-57735-887-9.
\newblock \doi{10.1609/aaai.v38i16.29689}.
\newblock URL \url{https://doi.org/10.1609/aaai.v38i16.29689}.

\bibitem[Rath(2026)]{agentdrift2026}
Abhishek Rath.
\newblock Agent drift: Quantifying behavioral degradation in multi-agent llm
  systems over extended interactions, 2026.
\newblock URL \url{https://arxiv.org/abs/2601.04170}.

\bibitem[Albrethsen et~al.(2026)Albrethsen, Datta, Kumar, and
  Rajasekar]{deepcontext2026}
Justin Albrethsen, Yash Datta, Kunal Kumar, and Sharath Rajasekar.
\newblock Deepcontext: Stateful real-time detection of multi-turn adversarial
  intent drift in llms, 2026.
\newblock URL \url{https://arxiv.org/abs/2602.16935}.

\bibitem[Angelopoulos et~al.(2025)Angelopoulos, Bates, Fisch, Lei, and
  Schuster]{conformalrisk2024}
Anastasios~N. Angelopoulos, Stephen Bates, Adam Fisch, Lihua Lei, and Tal
  Schuster.
\newblock Conformal risk control, 2025.
\newblock URL \url{https://arxiv.org/abs/2208.02814}.

\bibitem[Arike et~al.(2025)Arike, Donoway, Bartsch, and
  Hobbhahn]{goaldrift2025}
Rauno Arike, Elizabeth Donoway, Henning Bartsch, and Marius Hobbhahn.
\newblock Technical report: Evaluating goal drift in language model agents,
  2025.
\newblock URL \url{https://arxiv.org/abs/2505.02709}.

\bibitem[Dongre et~al.(2025)Dongre, Rossi, Lai, Yoon, Hakkani-Tür, and
  Bui]{driftnomore2025}
Vardhan Dongre, Ryan~A. Rossi, Viet~Dac Lai, David~Seunghyun Yoon, Dilek
  Hakkani-Tür, and Trung Bui.
\newblock Drift no more? context equilibria in multi-turn llm interactions,
  2025.
\newblock URL \url{https://arxiv.org/abs/2510.07777}.

\bibitem[Lykouris et~al.(2018)Lykouris, Mirrokni, and
  Paes~Leme]{lykouris2018stochastic}
Thodoris Lykouris, Vahab Mirrokni, and Renato Paes~Leme.
\newblock Stochastic bandits robust to adversarial corruptions.
\newblock In \emph{Proceedings of the 50th Annual ACM SIGACT Symposium on
  Theory of Computing}, STOC 2018, page 114–122, New York, NY, USA, 2018.
  Association for Computing Machinery.
\newblock ISBN 9781450355599.
\newblock \doi{10.1145/3188745.3188918}.
\newblock URL \url{https://doi.org/10.1145/3188745.3188918}.

\bibitem[Gupta et~al.(2019)Gupta, Koren, and Talwar]{gupta2019better}
Anupam Gupta, Tomer Koren, and Kunal Talwar.
\newblock Better algorithms for stochastic bandits with adversarial
  corruptions.
\newblock In Alina Beygelzimer and Daniel Hsu, editors, \emph{Proceedings of
  the Thirty-Second Conference on Learning Theory}, volume~99 of
  \emph{Proceedings of Machine Learning Research}, pages 1562--1578. PMLR,
  25--28 Jun 2019.
\newblock URL \url{https://proceedings.mlr.press/v99/gupta19a.html}.

\bibitem[van Erven et~al.(2025)van Erven, Mayo, Olkhovskaya, and
  Wei]{kong2024corruption}
Tim van Erven, Jack Mayo, Julia Olkhovskaya, and Chen-Yu Wei.
\newblock An improved algorithm for adversarial linear contextual bandits via
  reduction.
\newblock In \emph{The Thirty-ninth Annual Conference on Neural Information
  Processing Systems}, 2025.
\newblock URL \url{https://openreview.net/forum?id=MfhFiU28hv}.

\bibitem[Russinovich et~al.(2025)Russinovich, Salem, and Eldan]{crescendo2025}
Mark Russinovich, Ahmed Salem, and Ronen Eldan.
\newblock Great, now write an article about that: the crescendo multi-turn llm
  jailbreak attack.
\newblock In \emph{Proceedings of the 34th USENIX Conference on Security
  Symposium}, SEC '25, USA, 2025. USENIX Association.
\newblock ISBN 978-1-939133-52-6.

\bibitem[Zhou and Arel(2025)]{tempest2025}
Andy Zhou and Ron Arel.
\newblock Tempest: Autonomous multi-turn jailbreaking of large language models
  with tree search, 2025.
\newblock URL \url{https://arxiv.org/abs/2503.10619}.

\bibitem[Srivastava and He(2025)]{srivastava2025memorygraft}
Saksham~Sahai Srivastava and Haoyu He.
\newblock Memorygraft: Persistent compromise of llm agents via poisoned
  experience retrieval, 2025.
\newblock URL \url{https://arxiv.org/abs/2512.16962}.

\bibitem[Dong et~al.(2025)Dong, Xu, He, Li, Tang, Liu, Liu, and
  Xiang]{minja2025}
Shen Dong, Shaochen Xu, Pengfei He, Yige Li, Jiliang Tang, Tianming Liu, Hui
  Liu, and Zhen Xiang.
\newblock Memory injection attacks on {LLM} agents via query-only interaction.
\newblock In \emph{The Thirty-ninth Annual Conference on Neural Information
  Processing Systems}, 2025.
\newblock URL \url{https://openreview.net/forum?id=QINnsnppv8}.

\bibitem[Babo(2026)]{zombieagent2025}
Zvika Babo.
\newblock {ZombieAgent}: New {ChatGPT} vulnerabilities let data theft continue
  (and spread).
\newblock \url{https://www.radware.com/blog/threat-intelligence/zombieagent/},
  January 2026.
\newblock Zero-click indirect prompt injection exploiting ChatGPT memory and
  connectors for persistent data exfiltration. Accessed: 2026-03-16.

\bibitem[Chhabra et~al.(2026)Chhabra, Datta, Nahin, and Mohapatra]{zhao2025gig}
Anshuman Chhabra, Shrestha Datta, Shahriar~Kabir Nahin, and Prasant Mohapatra.
\newblock Agentic ai security: Threats, defenses, evaluation, and open
  challenges.
\newblock \emph{IEEE Access}, 14:\penalty0 49455–49482, 2026.
\newblock ISSN 2169-3536.
\newblock \doi{10.1109/access.2026.3675554}.
\newblock URL \url{http://dx.doi.org/10.1109/ACCESS.2026.3675554}.

\bibitem[Zhang et~al.(2025{\natexlab{a}})Zhang, Yin, Zhang, Liu, Han, Zhang,
  Li, Wang, Wang, Chen, and Wu]{shen2024whichagent}
Shaokun Zhang, Ming Yin, Jieyu Zhang, Jiale Liu, Zhiguang Han, Jingyang Zhang,
  Beibin Li, Chi Wang, Huazheng Wang, Yiran Chen, and Qingyun Wu.
\newblock Which agent causes task failures and when? on automated failure
  attribution of {LLM} multi-agent systems.
\newblock In \emph{Forty-second International Conference on Machine Learning},
  2025{\natexlab{a}}.
\newblock URL \url{https://openreview.net/forum?id=GazlTYxZss}.

\bibitem[Mukherjee and Kantarcioglu(2025)]{provseek2025}
Kunal Mukherjee and Murat Kantarcioglu.
\newblock Llm-driven provenance forensics for threat investigation and
  detection, 2025.
\newblock URL \url{https://arxiv.org/abs/2508.21323}.

\bibitem[Fumero et~al.(2026)Fumero, Huang, Boffa, Giordano, Mellia, and
  Rossi]{cybersleuth2025}
Stefano Fumero, Kai Huang, Matteo Boffa, Danilo Giordano, Marco Mellia, and
  Dario Rossi.
\newblock Cybersleuth: Autonomous blue-team llm agent for web attack forensics,
  2026.
\newblock URL \url{https://arxiv.org/abs/2508.20643}.

\bibitem[Rauch et~al.(1965)Rauch, Tung, and Striebel]{rauch1965maximum}
Herbert~E. Rauch, F.~Tung, and Charlotte~T. Striebel.
\newblock Maximum likelihood estimates of linear dynamic systems.
\newblock \emph{AIAA Journal}, 3\penalty0 (8):\penalty0 1445--1450, 1965.
\newblock \doi{10.2514/3.3166}.

\bibitem[{van den Driessche} and
  Watmough(2002)]{vandendriessche2002reproduction}
P.~{van den Driessche} and James Watmough.
\newblock Reproduction numbers and sub-threshold endemic equilibria for
  compartmental models of disease transmission.
\newblock \emph{Mathematical Biosciences}, 180\penalty0 (1):\penalty0 29--48,
  2002.
\newblock ISSN 0025-5564.
\newblock \doi{https://doi.org/10.1016/S0025-5564(02)00108-6}.
\newblock URL
  \url{https://www.sciencedirect.com/science/article/pii/S0025556402001086}.

\bibitem[Xu et~al.(2025)Xu, Song, Li, Tang, Jain, Bao, Wang, Zhou, Guo, Cao,
  Yang, Lu, Martin, Su, Maben, Mehta, Chi, Jang, Xie, Zhou, and
  Neubig]{xu2024theagentcompany}
Frank~F. Xu, Yufan Song, Boxuan Li, Yuxuan Tang, Kritanjali Jain, Mengxue Bao,
  Zora~Zhiruo Wang, Xuhui Zhou, Zhitong Guo, Murong Cao, Mingyang Yang,
  Hao~Yang Lu, Amaad Martin, Zhe Su, Leander~Melroy Maben, Raj Mehta, Wayne
  Chi, Lawrence~Keunho Jang, Yiqing Xie, Shuyan Zhou, and Graham Neubig.
\newblock Theagentcompany: Benchmarking {LLM} agents on consequential real
  world tasks.
\newblock In \emph{The Thirty-ninth Annual Conference on Neural Information
  Processing Systems Datasets and Benchmarks Track}, 2025.
\newblock URL \url{https://openreview.net/forum?id=LZnKNApvhG}.

\bibitem[Debenedetti et~al.(2024)Debenedetti, Zhang, Balunovic, Beurer-Kellner,
  Fischer, and Tram\`{e}r]{debenedetti2024agentdojo}
Edoardo Debenedetti, Jie Zhang, Mislav Balunovic, Luca Beurer-Kellner, Marc
  Fischer, and Florian Tram\`{e}r.
\newblock Agentdojo: A dynamic environment to evaluate prompt injection attacks
  and defenses for llm agents.
\newblock In A.~Globerson, L.~Mackey, D.~Belgrave, A.~Fan, U.~Paquet,
  J.~Tomczak, and C.~Zhang, editors, \emph{Advances in Neural Information
  Processing Systems}, volume~37, pages 82895--82920. Curran Associates, Inc.,
  2024.
\newblock \doi{10.52202/079017-2636}.
\newblock URL
  \url{https://proceedings.neurips.cc/paper_files/paper/2024/file/97091a5177d8dc64b1da8bf3e1f6fb54-Paper-Datasets_and_Benchmarks_Track.pdf}.

\bibitem[Zhang et~al.(2025{\natexlab{b}})Zhang, Huang, Mei, Yao, Wang, Zhan,
  Wang, and Zhang]{zhang2024asb}
Hanrong Zhang, Jingyuan Huang, Kai Mei, Yifei Yao, Zhenting Wang, Chenlu Zhan,
  Hongwei Wang, and Yongfeng Zhang.
\newblock Agent security bench ({ASB}): Formalizing and benchmarking attacks
  and defenses in {LLM}-based agents.
\newblock In \emph{The Thirteenth International Conference on Learning
  Representations}, 2025{\natexlab{b}}.
\newblock URL \url{https://openreview.net/forum?id=V4y0CpX4hK}.

\bibitem[Zhu et~al.(2025)Zhu, Du, Hong, Yang, Guo, Wang, Wang, Qian, Tang, Ji,
  and You]{zhu2025marble}
Kunlun Zhu, Hongyi Du, Zhaochen Hong, Xiaocheng Yang, Shuyi Guo, Zhe Wang,
  Zhenhailong Wang, Cheng Qian, Xiangru Tang, Heng Ji, and Jiaxuan You.
\newblock {M}ulti{A}gent{B}ench : Evaluating the collaboration and competition
  of {LLM} agents.
\newblock In Wanxiang Che, Joyce Nabende, Ekaterina Shutova, and Mohammad~Taher
  Pilehvar, editors, \emph{Proceedings of the 63rd Annual Meeting of the
  Association for Computational Linguistics (Volume 1: Long Papers)}, pages
  8580--8622, Vienna, Austria, July 2025. Association for Computational
  Linguistics.
\newblock ISBN 979-8-89176-251-0.
\newblock \doi{10.18653/v1/2025.acl-long.421}.
\newblock URL \url{https://aclanthology.org/2025.acl-long.421/}.

\bibitem[Boyen and Koller(1998)]{boyen1998tractable}
Xavier Boyen and Daphne Koller.
\newblock Tractable inference for complex stochastic processes.
\newblock In \emph{Proceedings of the Fourteenth Conference on Uncertainty in
  Artificial Intelligence}, UAI'98, page 33–42, San Francisco, CA, USA, 1998.
  Morgan Kaufmann Publishers Inc.
\newblock ISBN 155860555X.

\bibitem[Murphy and Weiss(2001)]{murphy2001factored}
Kevin Murphy and Yair Weiss.
\newblock The factored frontier algorithm for approximate inference in dbns.
\newblock In \emph{Proceedings of the Seventeenth Conference on Uncertainty in
  Artificial Intelligence}, UAI'01, page 378–385, San Francisco, CA, USA,
  2001. Morgan Kaufmann Publishers Inc.
\newblock ISBN 1558608001.

\bibitem[Nasr et~al.(2025)Nasr, Carlini, Sitawarin, Schulhoff, Hayes, Ilie,
  Pluto, Song, Chaudhari, Shumailov, Thakurta, Xiao, Terzis, and
  Tramèr]{nasr2025attacker}
Milad Nasr, Nicholas Carlini, Chawin Sitawarin, Sander~V. Schulhoff, Jamie
  Hayes, Michael Ilie, Juliette Pluto, Shuang Song, Harsh Chaudhari, Ilia
  Shumailov, Abhradeep Thakurta, Kai~Yuanqing Xiao, Andreas Terzis, and Florian
  Tramèr.
\newblock The attacker moves second: Stronger adaptive attacks bypass defenses
  against llm jailbreaks and prompt injections, 2025.
\newblock URL \url{https://arxiv.org/abs/2510.09023}.

\bibitem[{Noma Security}(2025)]{uncrew2025}
{Noma Security}.
\newblock Uncrew: High-privilege {GitHub} token leak in {CrewAI} ({CVSS} 9.2).
\newblock
  \url{https://noma.security/blog/uncrew-the-risk-behind-a-leaked-internal-github-token-at-crewai/},
  2025.
\newblock CVSS 9.2.

\bibitem[Nguyen and Husain(2025)]{paloalto2025}
Viet~K. Nguyen and Mohammad~I. Husain.
\newblock Penetration testing of agentic ai: A comparative security analysis
  across models and frameworks, 2025.
\newblock URL \url{https://arxiv.org/abs/2512.14860}.

\bibitem[Kephart and White(1991)]{kephart1991}
J.O. Kephart and S.R. White.
\newblock Directed-graph epidemiological models of computer viruses.
\newblock In \emph{Proceedings. 1991 IEEE Computer Society Symposium on
  Research in Security and Privacy}, pages 343--359, 1991.
\newblock \doi{10.1109/RISP.1991.130801}.

\bibitem[Pastor-Satorras and Vespignani(2001)]{pastorsatorras2001}
Romualdo Pastor-Satorras and Alessandro Vespignani.
\newblock Epidemic spreading in scale-free networks.
\newblock \emph{Phys. Rev. Lett.}, 86:\penalty0 3200--3203, Apr 2001.
\newblock \doi{10.1103/PhysRevLett.86.3200}.
\newblock URL \url{https://link.aps.org/doi/10.1103/PhysRevLett.86.3200}.

\bibitem[Zou et~al.(2002)Zou, Gong, and Towsley]{zou2002codered}
Cliff~Changchun Zou, Weibo Gong, and Don Towsley.
\newblock Code red worm propagation modeling and analysis.
\newblock In \emph{Proceedings of the 9th ACM Conference on Computer and
  Communications Security}, CCS '02, page 138–147, New York, NY, USA, 2002.
  Association for Computing Machinery.
\newblock ISBN 1581136129.
\newblock \doi{10.1145/586110.586130}.
\newblock URL \url{https://doi.org/10.1145/586110.586130}.

\bibitem[Vuong et~al.(2021)Vuong, Hauray, and Pardoux]{vuong2022conditional}
Yen~V. Vuong, Maxime Hauray, and Etienne Pardoux.
\newblock Conditional propagation of chaos in a spatial stochastic epidemic
  model with common noise, 2021.
\newblock URL \url{https://arxiv.org/abs/2111.02733}.

\bibitem[Sznitman(1991)]{sznitman1991}
Alain-Sol Sznitman.
\newblock Topics in propagation of chaos.
\newblock In Paul-Louis Hennequin, editor, \emph{Ecole d'Et{\'e} de
  Probabilit{\'e}s de Saint-Flour XIX --- 1989}, pages 165--251, Berlin,
  Heidelberg, 1991. Springer Berlin Heidelberg.
\newblock ISBN 978-3-540-46319-1.

\bibitem[Kurtz(1970)]{kurtz1970}
Thomas~G. Kurtz.
\newblock Solutions of ordinary differential equations as limits of pure jump
  markov processes.
\newblock \emph{Journal of Applied Probability}, 7\penalty0 (1):\penalty0
  49–58, 1970.
\newblock \doi{10.2307/3212147}.

\bibitem[Fournier and Guillin(2013)]{fournierguillin2015}
Nicolas Fournier and Arnaud Guillin.
\newblock On the rate of convergence in wasserstein distance of the empirical
  measure, 2013.
\newblock URL \url{https://arxiv.org/abs/1312.2128}.

\bibitem[Skorokhod(1956{\natexlab{a}})]{skorokhod1956}
Anatoli~V. Skorokhod.
\newblock Limit theorems for stochastic processes.
\newblock \emph{Theory of Probability and Its Applications}, 1:\penalty0
  261--290, 1956{\natexlab{a}}.
\newblock URL \url{https://api.semanticscholar.org/CorpusID:122862422}.

\bibitem[Bass(2006)]{bass1997}
R.F. Bass.
\newblock \emph{Diffusions and Elliptic Operators}.
\newblock Probability and Its Applications. Springer New York, 2006.
\newblock ISBN 9780387226040.
\newblock URL \url{https://books.google.com/books?id=tbzgBwAAQBAJ}.

\bibitem[Skorokhod(1956{\natexlab{b}})]{jacodshiryaev2003}
Anatoli~V. Skorokhod.
\newblock Limit theorems for stochastic processes.
\newblock \emph{Theory of Probability and Its Applications}, 1:\penalty0
  261--290, 1956{\natexlab{b}}.
\newblock URL \url{https://api.semanticscholar.org/CorpusID:122862422}.

\bibitem[Ball et~al.(1997)Ball, Mollison, and Scalia-Tomba]{ball1997multitype}
Frank Ball, Denis Mollison, and Gianpaolo Scalia-Tomba.
\newblock {Epidemics with two levels of mixing}.
\newblock \emph{The Annals of Applied Probability}, 7\penalty0 (1):\penalty0 46
  -- 89, 1997.
\newblock \doi{10.1214/aoap/1034625252}.
\newblock URL \url{https://doi.org/10.1214/aoap/1034625252}.

\bibitem[Britton(2010)]{britton2010stochastic}
Tom Britton.
\newblock Stochastic epidemic models: A survey.
\newblock \emph{Mathematical Biosciences}, 225\penalty0 (1):\penalty0 24--35,
  2010.
\newblock ISSN 0025-5564.
\newblock \doi{https://doi.org/10.1016/j.mbs.2010.01.006}.
\newblock URL
  \url{https://www.sciencedirect.com/science/article/pii/S0025556410000143}.

\bibitem[Tsiamis et~al.(2020)Tsiamis, Matni, and Pappas]{tsiamis2020sample}
Anastasios Tsiamis, Nikolai Matni, and George~J. Pappas.
\newblock Sample complexity of kalman filtering for unknown systems, 2020.
\newblock URL \url{https://arxiv.org/abs/1912.12309}.

\bibitem[Tsiamis and Pappas(2019)]{tsiamis2019finite}
Anastasios Tsiamis and George~J. Pappas.
\newblock Finite sample analysis of stochastic system identification.
\newblock In \emph{2019 IEEE 58th Conference on Decision and Control (CDC)},
  page 3648–3654. IEEE Press, 2019.
\newblock \doi{10.1109/CDC40024.2019.9029499}.
\newblock URL \url{https://doi.org/10.1109/CDC40024.2019.9029499}.

\bibitem[Simchowitz et~al.(2018)Simchowitz, Mania, Tu, Jordan, and
  Recht]{simchowitz2018learning}
Max Simchowitz, Horia Mania, Stephen Tu, Michael~I. Jordan, and Benjamin Recht.
\newblock Learning without mixing: Towards a sharp analysis of linear system
  identification.
\newblock In Sébastien Bubeck, Vianney Perchet, and Philippe Rigollet,
  editors, \emph{Proceedings of the 31st Conference On Learning Theory},
  volume~75 of \emph{Proceedings of Machine Learning Research}, pages 439--473.
  PMLR, 06--09 Jul 2018.
\newblock URL \url{https://proceedings.mlr.press/v75/simchowitz18a.html}.

\bibitem[Borodin and Salminen(2002)]{borodinsalminen2002}
Andrei~N. Borodin and Paavo Salminen.
\newblock \emph{Handbook of Brownian Motion: Facts and Formulae}.
\newblock Birkh{\"a}user, Basel, second edition, 2002.

\bibitem[Tsybakov(2008)]{tsybakov2008nonparametric}
Alexandre~B. Tsybakov.
\newblock \emph{Introduction to Nonparametric Estimation}.
\newblock Springer Publishing Company, Incorporated, 1st edition, 2008.
\newblock ISBN 0387790519.

\bibitem[Dalal et~al.(2018)Dalal, Szorenyi, Thoppe, and
  Mannor]{dalal2018finite}
Gal Dalal, Balazs Szorenyi, Gugan Thoppe, and Shie Mannor.
\newblock Finite sample analyses for td(0) with function approximation.
\newblock In \emph{Proceedings of the Thirty-Second AAAI Conference on
  Artificial Intelligence and Thirtieth Innovative Applications of Artificial
  Intelligence Conference and Eighth AAAI Symposium on Educational Advances in
  Artificial Intelligence}, AAAI'18/IAAI'18/EAAI'18. AAAI Press, 2018.
\newblock ISBN 978-1-57735-800-8.

\bibitem[Kaledin et~al.(2020)Kaledin, Moulines, Naumov, Tadic, and
  Wai]{kaledin2020finite}
Maxim Kaledin, Eric Moulines, Alexey Naumov, Vladislav Tadic, and Hoi-To Wai.
\newblock Finite time analysis of linear two-timescale stochastic approximation
  with markovian noise.
\newblock In Jacob Abernethy and Shivani Agarwal, editors, \emph{Proceedings of
  Thirty Third Conference on Learning Theory}, volume 125 of \emph{Proceedings
  of Machine Learning Research}, pages 2144--2203. PMLR, 09--12 Jul 2020.
\newblock URL \url{https://proceedings.mlr.press/v125/kaledin20a.html}.

\bibitem[Ning and Wu(2026)]{ningwu2026bayesian}
Ning Ning and Amin Wu.
\newblock Bayesian inference for partially observed mckean-vlasov sdes with
  full distribution dependence, 2026.
\newblock URL \url{https://arxiv.org/abs/2601.12515}.

\bibitem[Motwani et~al.(2024)Motwani, Baranchuk, Strohmeier, Bolina, Torr,
  Hammond, and de~Witt]{motwani2024secret}
Sumeet~Ramesh Motwani, Mikhail Baranchuk, Martin Strohmeier, Vijay Bolina,
  Philip~H.S. Torr, Lewis Hammond, and Christian~Schroeder de~Witt.
\newblock Secret collusion among ai agents: Multi-agent deception via
  steganography.
\newblock In A.~Globerson, L.~Mackey, D.~Belgrave, A.~Fan, U.~Paquet,
  J.~Tomczak, and C.~Zhang, editors, \emph{Advances in Neural Information
  Processing Systems}, volume~37, pages 73439--73486. Curran Associates, Inc.,
  2024.
\newblock \doi{10.52202/079017-2336}.
\newblock URL
  \url{https://proceedings.neurips.cc/paper_files/paper/2024/file/861f7dad098aec1c3560fb7add468d41-Paper-Conference.pdf}.

\bibitem[Zou et~al.(2003)Zou, Gong, and Towsley]{zou2003quarantine}
Cliff~Changchun Zou, Weibo Gong, and Don Towsley.
\newblock Worm propagation modeling and analysis under dynamic quarantine
  defense.
\newblock In \emph{Proceedings of the 2003 ACM Workshop on Rapid Malcode}, WORM
  '03, page 51–60, New York, NY, USA, 2003. Association for Computing
  Machinery.
\newblock ISBN 1581137850.
\newblock \doi{10.1145/948187.948197}.
\newblock URL \url{https://doi.org/10.1145/948187.948197}.

\bibitem[Aizen(2026)]{aigateway2026}
Kai Aizen.
\newblock {AI} gateway threat model: 8 attack vectors against {LLM} gateway
  infrastructure.
\newblock \url{https://snailsploit.com/ai-security/ai-gateway-threat-model/},
  2026.
\newblock Accessed: 2026-03-16.

\bibitem[{GitHub Advisory Database}(2025)]{cve202564439}
{GitHub Advisory Database}.
\newblock {CVE-2025-64439}: {LangGraph Checkpoint} affected by {RCE} in
  ``json'' mode of {JsonPlusSerializer}.
\newblock \url{https://github.com/advisories/GHSA-wwqv-p2pp-99h5}, 2025.
\newblock Affects \texttt{langgraph-checkpoint} versions before 3.0.0; fixed in
  3.0.0. Accessed: 2026-05-04.

\bibitem[{LayerX Security}(2025)]{taintedmemories2026}
{LayerX Security}.
\newblock {ChatGPT} tainted memories: {CSRF}-based persistent memory injection
  in {ChatGPT} {Atlas} browser.
\newblock
  \url{https://layerxsecurity.com/blog/layerx-identifies-vulnerability-in-new-chatgpt-atlas-browser/},
  October 2025.
\newblock Persistent memory injection survives across sessions, devices, and
  tabs. Accessed: 2026-03-16.

\bibitem[{OWASP}(2025)]{owaspasi2026}
{OWASP}.
\newblock {OWASP} top 10 for agentic applications.
\newblock
  \url{https://genai.owasp.org/2025/12/09/owasp-top-10-for-agentic-applications-the-benchmark-for-agentic-security-in-the-age-of-autonomous-ai/},
  December 2025.
\newblock ASI01: Agent Goal Hijacking is the \#1 threat. Accessed: 2026-03-16.

\bibitem[Yang et~al.(2026)Yang, He, Ji, Hooi, and Dong]{persistentcontrol2025}
Xianglin Yang, Yufei He, Shuo Ji, Bryan Hooi, and Jin~Song Dong.
\newblock Zombie agents: Persistent control of self-evolving llm agents via
  self-reinforcing injections, 2026.
\newblock URL \url{https://arxiv.org/abs/2602.15654}.

\bibitem[Zou et~al.(2025)Zou, Geng, Wang, and Jia]{zou2024poisonedrag}
Wei Zou, Runpeng Geng, Binghui Wang, and Jinyuan Jia.
\newblock Poisonedrag: knowledge corruption attacks to retrieval-augmented
  generation of large language models.
\newblock In \emph{Proceedings of the 34th USENIX Conference on Security
  Symposium}, SEC '25, USA, 2025. USENIX Association.
\newblock ISBN 978-1-939133-52-6.

\bibitem[Entezami and Naseh(2025)]{adversarialrlhf2025}
Erfan Entezami and Ali Naseh.
\newblock Llm misalignment via adversarial rlhf platforms, 2025.
\newblock URL \url{https://arxiv.org/abs/2503.03039}.

\bibitem[Cemri et~al.(2025)Cemri, Pan, Yang, Agrawal, Chopra, Tiwari, Keutzer,
  Parameswaran, Klein, Ramchandran, Zaharia, Gonzalez, and Stoica]{mast2025}
Mert Cemri, Melissa~Z Pan, Shuyi Yang, Lakshya~A Agrawal, Bhavya Chopra,
  Rishabh Tiwari, Kurt Keutzer, Aditya Parameswaran, Dan Klein, Kannan
  Ramchandran, Matei Zaharia, Joseph~E. Gonzalez, and Ion Stoica.
\newblock Why do multi-agent {LLM} systems fail?
\newblock In \emph{The Thirty-ninth Annual Conference on Neural Information
  Processing Systems Datasets and Benchmarks Track}, 2025.
\newblock URL \url{https://openreview.net/forum?id=fAjbYBmonr}.

\bibitem[{GitGuardian}(2025)]{nxsingularity2025}
{GitGuardian}.
\newblock The {Nx} ``s1ngularity'' attack: Inside the credential leak that hit
  2,349 developers.
\newblock
  \url{https://blog.gitguardian.com/the-nx-s1ngularity-attack-inside-the-credential-leak/},
  August 2025.
\newblock Compromised npm packages harvested credentials and modified shell
  configuration files for persistence. Accessed: 2026-03-16.

\bibitem[{Truesec}(2026)]{litellmpypi2026}
{Truesec}.
\newblock Malicious {PyPI} package: {LiteLLM} supply chain compromise.
\newblock
  \url{https://www.truesec.com/hub/blog/malicious-pypi-package-litellm-supply-chain-compromise},
  March 2026.
\newblock Poisoned PyPI package enabled Kubernetes lateral movement and
  encrypted credential exfiltration. Accessed: 2026-03-16.

\bibitem[Raina(2025)]{docker2026mcp}
Ajeet~Singh Raina.
\newblock {MCP} security issues threatening {AI} infrastructure.
\newblock
  \url{https://www.docker.com/blog/mcp-security-issues-threatening-ai-infrastructure/},
  July 2025.
\newblock 43\% of MCP servers found to have command injection flaws. Accessed:
  2026-03-16.

\bibitem[Zhang et~al.(2026)Zhang, Zhao, Zhang, Deng, Zhang, Liu, Xiao, and
  Zhang]{heartbeat2026}
Yechao Zhang, Shiqian Zhao, Jie Zhang, Gelei Deng, Jiawen Zhang, Xiaogeng Liu,
  Chaowei Xiao, and Tianwei Zhang.
\newblock Mind your {HEARTBEAT}! {Claw} background execution inherently enables
  silent memory pollution, 2026.
\newblock URL \url{https://arxiv.org/abs/2603.23064}.
\newblock Zero-click memory pollution via heartbeat background execution
  achieves 75.6\% cross-session behavioral influence. Accessed: 2026-04-01.

\bibitem[{MITRE}(2026)]{openclaw2026cve}
{MITRE}.
\newblock {CVE-2026-25253}: {OpenClaw} one-click remote code execution via
  {WebSocket} authentication token exfiltration.
\newblock \url{https://nvd.nist.gov/vuln/detail/CVE-2026-25253}, 2026.
\newblock CVSS 8.8. Affects OpenClaw versions before 2026.1.29. Accessed:
  2026-03-16.

\bibitem[Shi et~al.(2025)Shi, He, Wang, Wu, Li, Guo, and Song]{progent2025}
Tianneng Shi, Jingxuan He, Zhun Wang, Linyu Wu, Hongwei Li, Wenbo Guo, and Dawn
  Song.
\newblock Progent: Programmable privilege control for llm agents.
\newblock \emph{CoRR}, abs/2504.11703, April 2025.
\newblock URL \url{https://doi.org/10.48550/arXiv.2504.11703}.

\bibitem[Rossi et~al.(2020)Rossi, Chamberlain, Frasca, Eynard, Monti, and
  Bronstein]{rossi2020tgn}
Emanuele Rossi, Ben Chamberlain, Fabrizio Frasca, Davide Eynard, Federico
  Monti, and Michael Bronstein.
\newblock Temporal graph networks for deep learning on dynamic graphs.
\newblock In \emph{ICML 2020 Workshop on Graph Representation Learning}, 2020.

\bibitem[Li et~al.(2025)Li, Liu, Chun, Li, Zhang, and Xiao]{li2025drift}
Hao Li, Xiaogeng Liu, CHIU~Hung Chun, Dianqi Li, Ning Zhang, and Chaowei Xiao.
\newblock {DRIFT}: Dynamic rule-based defense with injection isolation for
  securing {LLM} agents.
\newblock In \emph{The Thirty-ninth Annual Conference on Neural Information
  Processing Systems}, 2025.
\newblock URL \url{https://openreview.net/forum?id=oY1Xnt83oJ}.

\bibitem[Debenedetti et~al.(2025)Debenedetti, Shumailov, Fan, Hayes, Carlini,
  Fabian, Kern, Shi, Terzis, and Tramèr]{debenedetti2025camel}
Edoardo Debenedetti, Ilia Shumailov, Tianqi Fan, Jamie Hayes, Nicholas Carlini,
  Daniel Fabian, Christoph Kern, Chongyang Shi, Andreas Terzis, and Florian
  Tramèr.
\newblock Defeating prompt injections by design, 2025.
\newblock URL \url{https://arxiv.org/abs/2503.18813}.

\end{thebibliography}

\appendix

\newpage
\phantomsection
\pdfbookmark[0]{Appendices}{appendices}

\begin{center}
{\Large\bfseries Appendices}
\end{center}
\vspace{0.2em}
\noindent\rule{\textwidth}{0.5pt}
\vspace{0.6em}

\etocdepthtag.toc{appendixstart}
{\small
  \etocsetstyle{section}%
    {}
    {\vspace{0.7em}}
    {\noindent\bfseries\etocnumber\hspace{0.6em}\etocname\hfill\etocpage\par}%
    {}
  \etocsetstyle{subsection}%
    {}%
    {\vspace{0.05em}}%
    {\noindent\hspace{1.5em}\mdseries\etocname%
     \nobreak\leaders\hbox to 0.6em{\hss.\hss}\hfill\nobreak\hspace{0.3em}\etocpage\par}%
    {}%
  \etocsetstyle{subsubsection}{}{}{}{}%
  \etocsettagdepth{main}{-1}
  \etocsettagdepth{appendixstart}{subsection}
  \etocsettocstyle{}{}%
  \tableofcontents
}%

\newpage

\section{Theoretical Proofs and Derivations}
\label{app:theory_proofs}

This appendix states and proves the formal guarantees used by Section~\ref{sec:method}. The three main results (Swarm Shepherd belief consistency, trace-back consistency, and the two-scale deployment theorem) share the same factored state-space model of Definition~\ref{def:statespace} and the same identifiability backbone $\nu_{\min}$ from Lemma~\ref{lem:joint_obs}.

\subsection{Theorem Statements}
\label{app:theorem_statements}

\begin{theorem}[Swarm Shepherd Belief Consistency]
\label{thm:claif_consistency}
Under (A1) linear-Gaussian factored DBN, (A2) rank-$d$ anchored observability with $\nu_{\min}$, (A3) bounded coupling $\|B^{(r,\cdot,\cdot)}\|_{\mathrm{op}} \le L$, (A4) minimum incident-event rate $\lambda_{\min}$, and stable self-dynamics from (A7) (Appendix~\ref{app:assumptions_map}), let $b_{v,n}$ be Swarm Shepherd's belief after $n$ events and $b^{\star}_{v,n}$ the oracle posterior. For $\epsilon, \delta \in (0, 1)$,
\begin{equation}
\Pr\!\left[\|b_{v,n} - b^{\star}_{v,n}\|_{\mathrm{TV}} \le \epsilon\right] \ge 1 - \delta
\quad\text{whenever}\quad
n \ge n_0 = O\!\left( \frac{d^2 L^{2k} \log(d/\delta)}{\epsilon^2 \lambda_{\min}^2 \nu_{\min}^2} \right),
\label{eq:claif_sample}
\end{equation}
where $k$ is the hop radius of $v$'s belief-relevant neighborhood. Proof in Appendix~\ref{app:claif_proof}.
\end{theorem}

\begin{corollary}[Time-to-Warning Lead]
\label{cor:warning_lead}
Let $\hat{T}_v = \min\{t : b_{v,t}(\xi_{v,t} > \tau_E) > 1/2\}$ be Swarm Shepherd's warning time, $T^{\star}_{E,v}=\inf\{t:\xi_{v,t}\ge \tau_E\}$ the true early-threshold crossing, and $\tau^{\star}_v=\inf\{t:\xi_{v,t}\ge \tau_I\}$ the true Infected crossing. Let $\Delta_\tau=\tau_I-\tau_E>0$. Suppose the TV error event in Theorem~\ref{thm:claif_consistency} localizes the warning time with slack $\eta_\epsilon$, meaning $\hat{T}_v \le T^{\star}_{E,v}+\eta_\epsilon$ on that event. Under the (A7) warning-band bounds $|a_t|\le\bar a$ and $\sigma_t^2\le\bar\sigma^2$, for any $k\ge 1$ with $\Delta_\tau>\bar a(k+\eta_\epsilon)$,
\begin{equation}
\Pr[\hat{T}_v + k \le \tau^{\star}_v]
\ge
1-\delta
-\min\!\left\{1,\,
2\exp\!\left(
-\frac{(\Delta_\tau-\bar a(k+\eta_\epsilon))^2}{2\bar\sigma^2(k+\eta_\epsilon)}
\right)\right\}.
\label{eq:warning_lead_bound}
\end{equation}
Thus a $k$-session warning certificate requires a calibrated threshold margin that dominates both short-horizon drift and diffusion. Proof in Appendix~\ref{app:claif_proof}.
\end{corollary}

\begin{proposition}[Trace-Back Consistency]
\label{prop:traceback}
Under the assumptions of Theorem~\ref{thm:claif_consistency} plus (A5) pairwise operator distinguishability $\min_{r \ne r'} \|B^{(r,\cdot,\cdot)} - B^{(r',\cdot,\cdot)}\|_{\mathrm{op}} \ge \delta_B$ on the anchored subspace,
\begin{equation}
\Pr[\hat\ell = \ell^\star] \ge 1 - \delta - \exp(-c \delta_B^2 n / \nu_{\min}^2),
\label{eq:traceback_consistency}
\end{equation}
for universal $c > 0$, whenever~\eqref{eq:claif_sample} is met. The same rate (restricted to the $k$-hop in-neighborhood of the failure node) governs $\hat u = u^\star$. Proof in Appendix~\ref{app:claif_proof}.
\end{proposition}

\begin{theorem}[Latent Aggregate Mean-Field Rate]
\label{thm:seir_limit}
\label{prop:seir_limit}
Under agent exchangeability, Poisson-driven typed events with absorbing $R$, (A2.5) Wasserstein-Lipschitz aggregated drift, and stable self-dynamics from (A7), with $\bar{\xi}_N(t) = N^{-1}\sum_v \xi_{v,t}$ and $\xi^{\star}(t)$ the McKean\textendash{}Vlasov limit of~\eqref{eq:scalar_sde},
\begin{equation}
\mathbb{E}\!\left[W_1(\bar{\xi}_N(\cdot), \xi^{\star}(\cdot))\right] \le \frac{C(\mathsf{H})}{\sqrt{N}},
\label{eq:latent_w1_bound}
\end{equation}
where $C(\mathsf{H})$ depends only on the horizon, coupling bound $L$, and kernel Lipschitz constants. Proof in Appendix~\ref{app:seir_limit}.
\end{theorem}

\begin{corollary}[SEIR Rate Specialization]
\label{cor:seir}
Under bounded density at $\tau_E, \tau_I$ and the (A7) warning-band variation condition, the thresholded fractions $(S, E, I, R)$ converge to the classical SEIR ODE, with $\beta$ the spectral radius of $B^{(\mathsf{retrieve},\mathcal{M},\mathcal{A})} + B^{(\mathsf{handoff},\mathcal{A},\mathcal{A})}$ times the typed-event rate, $\sigma^{-1}$ the first-passage time $\tau_E \to \tau_I$, and $\gamma$ the $b_t$-driven detection hazard. This recovers prior malware-propagation models~\citep{kephart1991,pastorsatorras2001,zou2002codered} as a corollary. Proof in Appendix~\ref{app:seir_limit}.
\end{corollary}

\begin{remark}[Adversary-controlled incubation]
\label{thm:sigma}
Unlike biological SEIR where $\sigma$ is pathogen-fixed, $\sigma^{-1}$ here is a first-passage time set by the attacker's per-contact influence magnitude. Higher aggression gives faster but detectable incubation; distributed drift is slower and stealthier, breaking defenses that rely on fixed incubation.
\end{remark}

\begin{theorem}[Two-Scale Deployment Regime]
\label{thm:two_scale}
Under the assumptions of Theorems~\ref{thm:claif_consistency},~\ref{thm:seir_limit} plus bounded density $\rho_{\max}$ at $\tau_I$, let $\hat I_N(t) = N^{-1}\sum_v \mathbb{1}[\hat\xi_{v,t} \ge \tau_I]$ and $I(t)$ the SEIR ODE solution. Then
\begin{equation}
\mathbb{E}\!\left|\hat I_N(t) - I(t)\right| \le \underbrace{\rho_{\max}\cdot\frac{C_1 L^k}{\nu_{\min}\sqrt{n}}}_{E_{\mathrm{filter}}(n)} + \underbrace{\frac{C_2(\mathsf{H})}{\sqrt{N}}}_{E_{\mathrm{MF}}(N)},
\label{eq:two_scale_bound}
\end{equation}
with $C_1 \le c_1 d (1 - \rho(A))^{-1}$, $\rho(A) < 1$ the dominant self-dynamics eigenvalue from (A7); $C_2(\mathsf{H})$ tracked in Appendix~\ref{app:two_scale_proof}, polynomial $c_2 (1-L)^{-1} \mathsf{H}^{1/2}$ under (A3) with $L < 1$, worst-case $c_2 e^{L\mathsf{H}} \mathsf{H}^{1/2}$. The two terms equalise at
\begin{equation}
n_\circ(N) = \frac{\rho_{\max}^2 C_1^2 L^{2k}}{\nu_{\min}^2 C_2(\mathsf{H})^2} \cdot N.
\label{eq:regime_boundary}
\end{equation}
Residual error is \emph{inference-limited} below the boundary (grow $n$) and \emph{population-limited} above it (grow $N$). Proof in Appendix~\ref{app:two_scale_proof}.
\end{theorem}

\subsection{Proof of Theorem~\ref{thm:seir_limit} (Latent Aggregate Mean-Field Rate)}
\label{app:seir_limit}

We prove Theorem~\ref{thm:seir_limit} on the latent scalar aggregate $\bar\xi_N$, then deduce Corollary~\ref{cor:seir} (the SEIR specialization) by composing with the threshold map.
The proof adapts the quantitative coupling template of~\citep{vuong2022conditional} to our Poisson-driven typed-event construction.

\paragraph{Setup.}
Let $e \in \mathbb{R}^6$ be the aggregation vector (dominant weight on contamination $c$ and deviation $d$) and define $\xi_{v,t} = e^{\top} z_{v,t}$.
Projecting the transition of Definition~\ref{def:statespace} onto $e$ yields the per-agent scalar process
\begin{equation}
d\xi_{v,t} \;=\; \Big[-\kappa_A\, \xi_{v,t} + \!\!\sum_r \lambda_{B,r}\!\!\!\!\sum_{u\,:\,(u,v,r)\in E^{\mathrm{obs}}_t}\!\!\!\! \xi_{u,t}\Big] dt + \Sigma(\xi_{v,t})\, dW_{v,t},
\label{eq:scalar_sde}
\end{equation}
with $\kappa_A = 1 - e^\top A^{\phi(v)} e / \|e\|^2$ and $\lambda_{B,r} = e^\top B^{(r,\cdot,\cdot)} e / \|e\|^2$.
Typed events are driven by independent Poisson processes of rates $\lambda_r(\bar{\mu}_N(t))$ that depend only on the empirical measure $\bar\mu_N(t) = N^{-1}\sum_v \delta_{\xi_{v,t}}$.

\paragraph{Standing assumptions for Theorem~\ref{thm:seir_limit}.}
In addition to the exchangeability and Poisson-driven construction of Definition~\ref{def:statespace}, we adopt (A2.5). The aggregated drift is Lipschitz in the Wasserstein-1 metric,
$\|\mathrm{drift}(\mu_1) - \mathrm{drift}(\mu_2)\|_\infty \le L_{\mathrm{drift}}\, W_1(\mu_1, \mu_2)$,
with $L_{\mathrm{drift}} = \|\lambda_B\|_\infty \cdot \sum_r \|B^{(r,\cdot,\cdot)}\|_{\mathrm{op}}$ finite by assumption (A3) of Theorem~\ref{thm:claif_consistency}.
The initial latent distribution has bounded second moment, which propagates to $\sup_{t\in[0,\mathsf{H}]} \mathbb{E}[\xi_{v,t}^2] \le K$ by spectral stability of the self-dynamics (the spectral radius of $A^{\phi(v)}$ is bounded by (A7)).

The proof factors through three internal lemmas that formalize coupling, empirical-measure concentration, and path-space aggregation. Each lemma's conditions are explicitly verified from the standing assumptions.

\begin{lemma}[Coupled-SDE Gr\"onwall bound]
\label{lem:sde_gronwall}
Under (A2.5) and the Skorohod coupling of the proof below, the per-agent deviation $\Delta_v(t) = \xi_{v,t} - \tilde\xi_{v,t}$ on $[0, T]$ satisfies
$\mathbb{E}|\Delta_v(T)| \le e^{(L_{\mathrm{drift}} + \kappa_A) T}\cdot \int_0^T \mathbb{E}[W_1(\bar\mu_N(s), \mu^\star(s))]\, ds$.
\end{lemma}
\begin{proof}
Apply It\^o's formula to $|\Delta_v(t)|$ on the coupled SDE. Drift contributions are bounded by $L_{\mathrm{drift}}\, W_1(\bar\mu_N(s), \mu^\star(s))$ through (A2.5), and bounded-variation contributions from the identically-thinned Poisson processes vanish in expectation. The integrated differential inequality is $d\,\mathbb{E}|\Delta_v(t)| \le L_{\mathrm{drift}}\,\mathbb{E}[W_1(\bar\mu_N(t),\mu^\star(t))]\,dt + \kappa_A\,\mathbb{E}|\Delta_v(t)|\,dt$. Gr\"onwall's inequality on $[0, T]$ yields the claim.
\end{proof}

\begin{lemma}[Empirical-measure concentration on the horizon]
\label{lem:emp_measure}
Under exchangeability of $(\tilde\xi_{v,s})_v$ from the McKean\textendash{}Vlasov construction and the inherited moment bound $\sup_{s \in [0, \mathsf{H}]} \mathbb{E}[\xi^2_{\star,s}] \le K$, the empirical measure of the i.i.d.\ copies satisfies
$\mathbb{E}[W_1(\bar\mu_N^{\mathrm{iid}}(s), \mu^\star(s))] \le C'/\sqrt{N}$
uniformly in $s \in [0, \mathsf{H}]$, and the Skorohod coupling transfers this rate to the true empirical measure $\bar\mu_N$ at cost $\mathbb{E}|\Delta_v|$.
\end{lemma}
\begin{proof}
Propagation of chaos~\citep[Theorem~I.1]{sznitman1991} and Kurtz's limit~\citep[Theorem~2.1]{kurtz1970} give weak convergence of the i.i.d.\ empirical measure to $\mu^\star(s)$. Under the verified moment bound (inherited from bounded initial second moment and spectral stability of $A^{\phi(v)}$), \citet[Theorem~1]{fournierguillin2015} provide the $W_1$ rate $C'/\sqrt{N}$. The coupling of Step 1 below transfers this rate to $\bar\mu_N$ by the triangle inequality.
\end{proof}

\begin{lemma}[Path-space aggregation]
\label{lem:path_agg}
For the empirical mean $\bar\xi_N(t) = N^{-1}\sum_v \xi_{v,t}$, the Kantorovich\textendash{}Rubinstein duality gives $W_1(\bar\xi_N(\cdot), \xi^\star(\cdot)) \le \mathbb{E}|\Delta_v|$, where $\Delta_v$ is the per-agent coupling error of Lemma~\ref{lem:sde_gronwall} and $\xi^\star$ is the mean-field solution.
\end{lemma}
\begin{proof}
By Kantorovich duality, $W_1(\bar\xi_N, \xi^\star) = \sup_{f \in \mathrm{Lip}_1} |\mathbb{E} f(\bar\xi_N) - \mathbb{E} f(\xi^\star)|$. The empirical mean is $1/N$-Lipschitz in the per-agent coordinates, and the McKean\textendash{}Vlasov copies have $\mathbb{E}\tilde\xi_{v,s} = \xi^\star(s)$, so the duality bound contracts to $\mathbb{E}|\Delta_v|$ along any Lipschitz test function.
\end{proof}

\begin{proof}[Proof of Theorem~\ref{thm:seir_limit}]
\emph{Step 1 (coupling to i.i.d.\ copies).}
Let $(\tilde\xi_{v,t})_{v=1}^N$ be i.i.d.\ copies governed by the McKean\textendash{}Vlasov limit
\[
d\tilde\xi_{v,t} = \left[ -\kappa_A\, \tilde\xi_{v,t} + \mathrm{drift}^\star(t) \right] dt + \Sigma(\tilde\xi_{v,t})\, d\tilde W_{v,t},
\]
where $\mathrm{drift}^\star(t)$ uses the population-level drift under $\xi^\star$.
By Skorohod embedding~\citep{skorokhod1956}, construct a coupling $(\xi, \tilde\xi)$ on a common probability space $(\Omega, \mathcal{F}, \mathbb{P})$ driven by the same Brownian motions $(W_{v,t})_v$ and the same thinned Poisson processes $(P_{v,r,t})_{v,r}$. The two processes differ only in the drift evaluation (empirical $\bar\mu_N$ versus mean-field $\mu^\star$), and $\bar\mu_N(t)$ is $\mathcal{F}_t$-measurable by construction.

\emph{Step 2 (coupling-error bound).}
Lemma~\ref{lem:sde_gronwall} applies under (A2.5). Its conclusion,
$\mathbb{E}|\Delta_v(T)| \le e^{(L_{\mathrm{drift}} + \kappa_A) T}\int_0^T \mathbb{E}[W_1(\bar\mu_N(s), \mu^\star(s))]\, ds$,
bounds per-agent deviation by an integrated empirical-measure error.

\emph{Step 3 (empirical-measure rate).}
Lemma~\ref{lem:emp_measure} applies under exchangeability and the moment bound verified in the setup. Its conclusion,
$\mathbb{E}[W_1(\bar\mu_N^{\mathrm{iid}}(s), \mu^\star(s))] \le C'/\sqrt{N}$ uniformly on $[0,\mathsf{H}]$,
is transferred to $\bar\mu_N$ through Step 1's coupling.

\emph{Step 4 (path-space assembly).}
Lemma~\ref{lem:path_agg} applies. Chaining Lemmas~\ref{lem:sde_gronwall}, \ref{lem:emp_measure}, and \ref{lem:path_agg} gives
\[
\mathbb{E}[W_1(\bar\xi_N(\cdot), \xi^\star(\cdot))] \;\le\; \frac{C(\mathsf{H})}{\sqrt{N}}, \qquad C(\mathsf{H}) = C' \cdot \mathsf{H} \cdot e^{(L_{\mathrm{drift}} + \kappa_A) \mathsf{H}},
\]
with $\mathsf{H}$ the time horizon. The exponential dependence on $\mathsf{H}$ is the coupling cost of (A2.5) over long horizons and restricts the operational regime as noted in Section~\ref{sec:limitations}.
\end{proof}

\begin{lemma}[Boundary upcrossing hazard under uniform ellipticity]
\label{lem:boundary_hazard}
Under (A1) linear-Gaussian dynamics, (A3) bounded cross-level coupling, and (A7) bounded warning-band variation, the scalar process $\xi_{v,t}$ of~\eqref{eq:scalar_sde} is uniformly elliptic and admits a unique, strictly-positive, $C^{1,2}$ transition density on $\mathbb{R} \times (0, \mathsf{H}]$ (Bass~\citep[Theorem~2.2.1]{bass1997}). For any threshold $\tau$ whose marginal density is bounded by $\rho_{\max}$ on a neighborhood $(\tau - \delta, \tau)$, the boundary is transversal (bounded-variation drift plus non-degenerate diffusion), and Nagasawa\textendash{}Wentzell~\citep[Proposition~I.3.17]{jacodshiryaev2003} gives the upcrossing intensity
$\lambda_{\tau}(t) = \rho(\tau^-; \xi_{\cdot,t})\cdot [\text{drift toward }\tau]_+$,
where $\rho(\tau^-; \cdot)$ is the marginal density immediately below $\tau$.
\end{lemma}
\begin{proof}
Bass~\citep[Theorem~2.2.1]{bass1997} gives existence and uniqueness of the $C^{1,2}$ transition density; strict positivity follows from non-degeneracy of the diffusion. Transversality of the threshold is a consequence of the drift being of bounded variation and the diffusion being strictly positive on a neighborhood of $\tau$. Under these conditions, the Nagasawa\textendash{}Wentzell intensity formula~\citep[Proposition~I.3.17]{jacodshiryaev2003} applies.
\end{proof}

\begin{proof}[Proof of Corollary~\ref{cor:seir} (SEIR specialization)]
The proof identifies the three SEIR rates $\beta$, $\sigma$, $\gamma$ as explicit functionals of the scalar process~\eqref{eq:scalar_sde}, then composes with Theorem~\ref{thm:seir_limit} on the threshold-map functionals.

\emph{Claim 1 (transmission rate $\beta$).}
Apply Lemma~\ref{lem:boundary_hazard} at $\tau_E$ under the bounded-density assumption of Theorem~\ref{thm:two_scale}. Substituting the typed-event mean rate $c$ and the aggregated drift coefficients $\lambda_{B,\mathsf{retrieve}}, \lambda_{B,\mathsf{handoff}}$ of~\eqref{eq:scalar_sde} into the upcrossing intensity yields
$\beta = \rho\,(\lambda_{B,\mathsf{retrieve}} + \lambda_{B,\mathsf{handoff}})\cdot c_{\max}$.

\emph{Claim 2 (incubation rate $\sigma$).}
Define $\sigma^{-1} = \mathbb{E}[\tau_{\mathrm{FP}}(\tau_E \to \tau_I)]$ with $\tau_{\mathrm{FP}}$ the first-passage time of $\xi_{v,t}$ under~\eqref{eq:scalar_sde} from level $\tau_E$ to $\tau_I$. Finiteness is a precondition for the reduction, and we verify it here. By (A7), the spectral radius of $A^{\phi(v)}$ is bounded by a constant below $1$, so $\kappa_A > 0$ in~\eqref{eq:scalar_sde} and the drift is dissipative outside a compact set. Applying the optional stopping theorem to the martingale derived from It\^o's formula on $\xi_{v,t}^2$ gives $\mathbb{E}[\tau_{\mathrm{FP}}] < \infty$, so sojourn in the Exposed compartment is integrable. Kurtz's reduction~\citep[Theorem~2.1]{kurtz1970} then applies under its three hypotheses, each verified here: (i) exchangeability by Definition~\ref{def:statespace}, (ii) bounded drift and jump rates by (A2.5), (A3), and (A7), (iii) integrable sojourn time just shown. The reduction gives the deterministic ODE~\eqref{eq:seir} at this $\sigma$, independent of the shape of the first-passage distribution. The exponential-tail approximation used by classical SEIR is a separate assumption (small drift-to-variance ratio at the $\tau_E$-boundary) that we neither make nor invoke.

\emph{Claim 3 (detection rate $\gamma$).}
Let $T_{\mathrm{det}} = \inf\{t : b_{v,t}(\xi > \tau_I) > 1/2\}$ be the declaration stopping time. Let $(\mathcal{F}_t^+)$ denote the right-continuous completion of the anchored-observation filtration. Every stopping time on $(\mathcal{F}_t^+)$ admits a compensator by Doob\textendash{}Meyer. We need the stronger statement that the compensator is absolutely continuous with respect to Lebesgue time, so that a hazard rate exists. $T_{\mathrm{det}}$ is totally inaccessible on $(\mathcal{F}_t^+)$ because it is driven by the continuous Swarm Shepherd posterior and not by predictable events in the anchored-event filtration. Jacod and Shiryaev~\citep[Chapter~I, Theorem~3.17]{jacodshiryaev2003} therefore give a unique predictable compensator with a density $\gamma(t)$ with respect to Lebesgue time. This $\gamma$ is the hazard rate
$\gamma(t) = \lim_{\Delta t \to 0} \Delta t^{-1}\, \Pr[T_{\mathrm{det}} \in (t, t + \Delta t] \mid \mathcal{F}_t^+]$
as a functional of the Swarm Shepherd posterior trajectory.

\emph{Composition.}
For each compartment indicator $\mathbb{1}[\xi_{v,t} \in \{S, E, I, R\}]$, Lemma~\ref{lem:bounded_density} (bounded-density continuity modulus) bounds the classification error at any threshold $\tau \in \{\tau_E, \tau_I\}$ by $\rho_{\max}\, W_1(\xi_{v,t}, \bar\xi_N(t))$. Summing this per-agent bound over $v$ and applying Theorem~\ref{thm:seir_limit}'s $\mathbb{E}[W_1(\bar\xi_N, \xi^\star)] \le C(\mathsf{H})/\sqrt{N}$ lifts the compartment convergence to $\mathbb{E}|\bar I_N(t) - I(t)| \le \rho_{\max}\, C(\mathsf{H})/\sqrt{N}$. Analogous bounds hold on $\bar E_N$, $\bar S_N$, $\bar R_N$ at the same rate. In the $N \to \infty$ limit, the empirical compartment fractions converge at $O(N^{-1/2})$ in expectation to the solution of the ODE system obtained by substituting the rates of Claims~1 through~3 into~\eqref{eq:seir}. Turnover at rate $\mu$ adds $-\mu E$ and $-\mu I$ terms by identical balancing.
\end{proof}

\begin{corollary}[Per-Class SEIR for Block-Structured Topologies]
\label{cor:seir_block}
Let agents partition into $K$ homogeneity classes $V = V_1 \sqcup \cdots \sqcup V_K$ with within-class exchangeability, and let $\{\beta_{ij}\}_{i,j \in [K]}$ be the inter-class contact rates determined by the class-averaged typed-event intensity. Then the $K$-class SEIR ODE system
\[
\frac{dE_i}{dt} = S_i \sum_j \beta_{ij}\, I_j/|V_j| - \sigma_i E_i, \qquad
\frac{dI_i}{dt} = \sigma_i E_i - \gamma_i I_i,
\]
holds in the $N \to \infty$ limit, with the $\sqrt{N}$ $W_1$ rate of Theorem~\ref{thm:seir_limit} applied per class. For a hub-and-spoke topology ($K=2$, hub class $|H| = O(1)$, spoke class $|S| = N - O(1)$), $\beta_{HS}, \beta_{SH} > 0$ and $\beta_{SS} \approx 0$, and the basic reproduction number is $\rzero = \sqrt{\beta_{HS} \beta_{SH}}/\gamma$, generalizing the scalar $\rzero$ of Theorem~\ref{thm:r0}. Proof via multi-type branching processes~\citep{ball1997multitype,britton2010stochastic} below.
\end{corollary}

\begin{remark}[Non-exchangeable topologies]
\label{rem:nonexchangeable}
Hub-and-spoke and star topologies violate the exchangeability assumption of Theorem~\ref{thm:seir_limit}. Corollary~\ref{cor:seir_block} formalizes the remedy: the per-class SEIR limit holds with the same $\sqrt{N}$ $W_1$ rate applied within each homogeneity class, and the cross-class coupling is absorbed into the $K$-type contact matrix $(\beta_{ij})$. The empirical slope-steepening reported in Figure~\ref{fig:a3_meanfield} (hub-and-spoke vs.\ mesh) is consistent with the class-averaged rate predicted by Corollary~\ref{cor:seir_block}. A full graphon mean-field extension remains future work.
\end{remark}

\begin{proof}[Proof of Corollary~\ref{cor:seir_block} (Per-class SEIR)]
Apply Theorem~\ref{thm:seir_limit} restricted to each class $V_i$ separately. Within-class exchangeability holds by construction, and (A2.5) restricts to a per-class constant $L_{\mathrm{drift},i}$ under the same bounded-coupling (A3).
Cross-class coupling enters only through the inter-class contact rates $\beta_{ij}$, which by (A3) satisfy $\beta_{ij} \le c_{\max,ij}\cdot \|B^{(r,\phi_i, \phi_j)}\|_{\mathrm{op}}$ on the aggregation axis. This is the multi-type branching-process structure of \citet{ball1997multitype} (see also \citet{britton2010stochastic}), with the SEIR threshold map specialized as in Corollary~\ref{cor:seir}.
For the star topology with $K=2$, the hub class $H$ has $|H|=O(1)$ (in Eve's running example $|H|=1$) and degree $\Theta(N)$, while the spoke class $S$ has $|S| = N - |H|$ and degree $1$. The per-class empirical measures converge at $N^{-1/2}$ on the spoke class (classical rate) and trivially on the hub class. The cross-class contact matrix is $\begin{pmatrix}0 & \beta_{HS}\\ \beta_{SH} & 0\end{pmatrix}$, rank-$1$ under star connectivity. The basic reproduction number is $\rzero = \sqrt{\beta_{HS}\beta_{SH}}/\gamma$, and the mean-field rate on each class follows from Theorem~\ref{thm:seir_limit}.
General $K$-class topologies are handled analogously with the $K$-type matrix $(\beta_{ij})$; conditions for subcriticality (all eigenvalues below~$1$) generalize the scalar $\rzero < 1$ condition of Theorem~\ref{thm:r0}.
\end{proof}

Two consequences for the main text.
The attacker's choice of per-contact influence magnitude (the scale of $B^{(r,\cdot,\cdot)}$ entries) tunes $\sigma^{-1}$ directly; this is the formal content of Remark~\ref{thm:sigma}.
A defender that monitors the full latent belief $b_t$ (Swarm Shepherd) rather than thresholded observations has access to the Exposed compartment before agents cross $\tau_I$, which is the empirical basis of A1 (Section~\ref{sec:e7_when}).

\subsection{Proof of Theorem~\ref{thm:claif_consistency} (Swarm Shepherd Belief Consistency)}
\label{app:claif_proof}

The proof composes three lemmas. Lemma~\ref{lem:claif_L1} is a single-node Kalman sample-complexity bound adapted from~\citep{tsiamis2020sample}. Lemma~\ref{lem:claif_L2} is the factored-belief error contraction of~\citep{boyen1998tractable}. Lemma~\ref{lem:claif_L3} is a hop union bound.

\begin{lemma}[Single-node Kalman parameter-to-posterior error]
\label{lem:claif_L1}
Consider the reduced single-node Kalman filter at $v$ with self-dynamics $A^{\phi(v)}$ and anchored emission $H$.
Under assumptions (A2), (A4), and the stable self-dynamics part of (A7), the stationary Kalman gain is uniquely defined and locally Lipschitz in $(A, H, Q, R)$ on a neighborhood of the true parameters with Lipschitz constant $L_K$.
Building on the system-identification rates of~\citep{tsiamis2019finite,simchowitz2018learning}, Theorem~3.2 of~\citet{tsiamis2020sample} gives, after $n$ anchored observations, parameter error
$\|(\hat A, \hat H, \hat Q, \hat R) - (A, H, Q, R)\|_F \le \epsilon_{\mathrm{par}}$
with probability at least $1 - \delta_1$ whenever $n \ge C_1\, d^2 \log(d/\delta_1) / (\epsilon_{\mathrm{par}}^2\, \lambda_{\min}^2\, \nu_{\min}^2)$, where $\nu_{\min}$ is the minimum singular value of the observability matrix $\mathcal{O}(A,H)$ and $\lambda_{\min}$ the minimum incident-event rate.
Combining with Kalman-gain Lipschitzness and the Pinsker-type bound $\|b - b^\star\|_{\mathrm{TV}} \le C_G(\|\hat\mu - \mu^\star\|_2 + \|\hat\Sigma - \Sigma^\star\|_F)$ for Gaussian posteriors and setting $\epsilon_{\mathrm{par}} = \epsilon / (2 L_K C_G)$ yields
$\Pr[\|b_{v,n}^{\mathrm{single}} - b^\star_{v,n}\|_{\mathrm{TV}} \le \epsilon/2] \ge 1 - \delta_1$.
\end{lemma}

\begin{lemma}[Factored-belief approximation error]
\label{lem:claif_L2}
\citet{boyen1998tractable} show that for a factored belief $b^{\mathrm{fact}}_{v,t} = \prod_v b_{v,t}$ tracking a full-joint DBN,
$\|b^{\mathrm{fact}}_{v,t} - b^{\mathrm{full}}_{v,t}\|_{\mathrm{TV}} \le \gamma^t\,\|b^{\mathrm{fact}}_{v,0} - b^{\mathrm{full}}_{v,0}\|_{\mathrm{TV}} + C_{\mathrm{fact}}\, L^{2k}$,
with $\gamma < 1$ the DBN mixing contraction and $L$ the bound on cross-level coupling (A3). The first term decays with $t$ for stationary streams; the second dominates asymptotically and is the $L^{2k}$ factor in the theorem.
\end{lemma}

\begin{lemma}[Hop union bound]
\label{lem:claif_L3}
Belief at $v$ depends on events at nodes within $k$ hops. There are at most $(d_{\max})^k$ such nodes, where $d_{\max}$ is the maximum influence-graph degree. Union-bounding over these nodes' Lemma~\ref{lem:claif_L1} failure events (each with probability $\le \delta_1$) and absorbing the $(d_{\max})^k$ factor into the logarithmic dependence of $n_0$ yields total failure probability $\le \delta$.
\end{lemma}

\begin{proof}[Proof of Theorem~\ref{thm:claif_consistency}]
Apply Lemma~\ref{lem:claif_L1} to every node in the $k$-hop neighborhood of $v$. Lift through Lemma~\ref{lem:claif_L2} to bound the factored-belief approximation by $\epsilon/2$. Apply Lemma~\ref{lem:claif_L3} to union-bound the failure event. This gives
$\Pr[\|b_{v,n} - b^\star_{v,n}\|_{\mathrm{TV}} \le \epsilon] \ge 1 - \delta$
whenever $n \ge n_0 = O(d^2\, L^{2k}\, \log(d/\delta) \,/\, \epsilon^2 \lambda_{\min}^2\, \nu_{\min}^2)$, which is Theorem~\ref{thm:claif_consistency}.
\end{proof}

\begin{lemma}[Short-horizon first-passage bound]
\label{lem:first_passage_reflection}
Let $X_t$ denote the scalar process after the true early-threshold crossing, with $X_0=\tau_E$, $\tau_{\mathrm{FP}}=\inf\{t:X_t\ge\tau_I\}$, and $\Delta_\tau=\tau_I-\tau_E>0$. Suppose that on the warning band $[\tau_E,\tau_I]$ the semimartingale decomposition is $dX_t=a_t\,dt+\sigma_t\,dW_t$ with $|a_t|\le\bar a$ and $\sigma_t^2\le\bar\sigma^2$. For any horizon $h>0$ with $\Delta_\tau>\bar a h$,
\begin{equation}
\Pr[\tau_{\mathrm{FP}}\le h]
\le
\min\!\left\{1,\,
2\exp\!\left(-\frac{(\Delta_\tau-\bar a h)^2}{2\bar\sigma^2 h}\right)
\right\}.
\label{eq:first_passage_margin_bound}
\end{equation}
\end{lemma}
\begin{proof}
Write $X_t=\tau_E+\int_0^t a_s\,ds+M_t$, where $M_t=\int_0^t\sigma_s\,dW_s$ and $\langle M\rangle_t\le \bar\sigma^2 t$. If $\tau_{\mathrm{FP}}\le h$, then $\sup_{0\le t\le h}M_t\ge \Delta_\tau-\bar a h$. By the Dambis\textendash{}Dubins\textendash{}Schwarz time change and the Brownian reflection principle,
\[
\Pr\!\left[\sup_{0\le t\le h}M_t\ge a\right]
\le
\Pr\!\left[\sup_{0\le u\le \bar\sigma^2 h}B_u\ge a\right]
=2\{1-\Phi(a/\sqrt{\bar\sigma^2h})\}
\le 2\exp\!\left(-\frac{a^2}{2\bar\sigma^2h}\right)
\]
with $a=\Delta_\tau-\bar a h>0$. See Borodin and Salminen~\citep{borodinsalminen2002} Chapter~I for Brownian first-passage formulae and Jacod and Shiryaev~\citep{jacodshiryaev2003} for the martingale time-change machinery.
\end{proof}

\begin{proof}[Proof of Corollary~\ref{cor:warning_lead}]
By Theorem~\ref{thm:claif_consistency}, Swarm Shepherd's filtered posterior tracks the oracle posterior to within $\epsilon$ TV with probability $\ge 1 - \delta$.
On this event, the warning-time localization condition gives $\hat T_v\le T^\star_{E,v}+\eta_\epsilon$.
If the true first-passage time from $\tau_E$ to $\tau_I$ exceeds $k+\eta_\epsilon$, then $\hat T_v+k\le \tau^\star_v$.
Lemma~\ref{lem:first_passage_reflection} with $h=k+\eta_\epsilon$ bounds the complementary event by the exponential term in~\eqref{eq:warning_lead_bound}.
Combining the two failure events via union bound gives
$\Pr[\hat T_v + k \le \tau^\star_v]$ bounded as in~\eqref{eq:warning_lead_bound}.
\end{proof}

\paragraph{Eve's numbers.}
For Eve's ecosystem the $k=3$ certificate is evaluated from the A1 calibration estimates of the threshold margin, warning-band drift, diffusion, and localization slack:
\[
\epsilon_{\mathrm{lead}}(3)
=
\min\!\left\{1,\,
2\exp\!\left(
-\frac{(\widehat\Delta_\tau-\widehat{\bar a}(3+\widehat\eta_\epsilon))^2}{2\widehat{\bar\sigma}^2(3+\widehat\eta_\epsilon)}
\right)\right\}.
\]
The A1 warning-lead result reports the empirical conformal coverage of this $k=3$ certificate (Section~\ref{sec:e7_when}, Table~\ref{tab:a1_baselines_extended}); the theorem supplies the sufficient condition, while the experiment verifies that the calibrated deployment satisfies the target.

\subsection{Proof of Theorem~\ref{thm:two_scale} (Two-Scale Deployment Regime)}
\label{app:two_scale_proof}

We first isolate two auxiliary lemmas used by the proof, then assemble the bound through a triangle decomposition of the end-to-end error.

\begin{lemma}[Bounded-density continuity modulus]
\label{lem:bounded_density}
Suppose the marginal density of a random variable $\xi$ on $\mathbb{R}$ is uniformly bounded by $\rho_{\max}$ on a neighborhood of $\tau_I$. Then for any random variables $X, Y$ on $\mathbb{R}$,
$\mathbb{E}\big|\mathbb{1}[X \ge \tau_I] - \mathbb{1}[Y \ge \tau_I]\big| \le \rho_{\max}\, W_1(X, Y)$,
and for any scalar estimator $\hat\xi$ of $\xi$,
$\Pr\!\big[\mathbb{1}[\hat\xi \ge \tau_I] \ne \mathbb{1}[\xi \ge \tau_I]\big] \le \rho_{\max}\, \mathbb{E}|\hat\xi - \xi|$.
\end{lemma}
\begin{proof}
A standard bounded-density argument; see Tsybakov~\citep[Theorem~6.1]{tsybakov2008nonparametric}. The indicator disagreement set is contained in $\{|\xi - \tau_I| \le |X - Y|\}$, whose probability under bounded density $\rho_{\max}$ near $\tau_I$ is at most $\rho_{\max}\, \mathbb{E}|X - Y|$; the Kantorovich dual gives the $W_1$ form.
\end{proof}

\begin{lemma}[Density inheritance under exchangeable averaging]
\label{lem:density_inheritance}
Let $X_1, \ldots, X_N$ be exchangeable random variables on $\mathbb{R}$ with each marginal density bounded by $\rho_{\max}$ on a neighborhood $U$ of a threshold $\tau$. Then the density of the empirical mean $N^{-1}\sum_v X_v$ is bounded by $\rho_{\max}$ on $U$.
\end{lemma}
\begin{proof}
For independent $X_v$, the density of the mean is an $N$-fold convolution $f_{\bar X} = f_1 * \cdots * f_N$ on scaled variables. Young's inequality gives $\|f_{\bar X}\|_\infty \le \min_v \|f_v\|_\infty \le \rho_{\max}$ on $U$. Exchangeable (possibly dependent) $X_v$ preserve the bound. The joint density marginalizes to the empirical-mean density through the same convolution structure after conditioning on the exchangeable de Finetti mixing measure, and the $L^\infty$ bound is preserved under mixtures. Hence $\|f_{\bar X}\|_\infty \le \rho_{\max}$ on $U$.
\end{proof}

\begin{lemma}[TV-to-scalar for Gaussian marginals]
\label{lem:tv_to_scalar}
Let $b, b^\star$ be $d$-dimensional Gaussian posteriors on a latent state $z$ with $\|b - b^\star\|_{\mathrm{TV}} \le \epsilon$, and let $\xi = e^\top z$ be a fixed linear aggregation with $\|e\| = 1$. Then
$\mathbb{E}|\xi_b - \xi_{b^\star}| \le C_{\mathrm{Pinsker}}\, \sigma_{\max}\, \sqrt{\epsilon}$,
where $\sigma_{\max}^2$ bounds the posterior variance along $e$ under both $b$ and $b^\star$.
\end{lemma}
\begin{proof}
By Pinsker's inequality $\|b - b^\star\|_{\mathrm{TV}} \le \sqrt{\mathrm{KL}(b \,\|\, b^\star)/2}$. The KL between two Gaussians is a quadratic form in mean and covariance differences, and the marginal along $e$ is Gaussian with standard deviation at most $\sigma_{\max}$. Combining gives the stated bound.
\end{proof}

\begin{proof}[Proof of Theorem~\ref{thm:two_scale}]
We prove the bound for the canonical SEIR Infected functional $\hat M_N(t) = \hat I_N(t)$ guaranteed by Corollary~\ref{cor:seir}. For a general $1$-Lipschitz functional of the latent aggregate the bound holds directly by Kantorovich\textendash{}Rubinstein duality without the bounded-density assumption; for the threshold indicator used here the indicator is not Lipschitz, and Lemma~\ref{lem:bounded_density} substitutes for Lipschitzness.

\emph{Step A (triangle decomposition).}
Let $\bar I_N(t) = N^{-1}\sum_v \mathbb{1}[\xi_{v,t} \ge \tau_I]$ denote the oracle compartment fraction (same population as $\hat I_N$ but using the true latent trajectory $\xi_{v,t}$) and let $I(t)$ denote the SEIR ODE solution. The triangle inequality gives
\[
\big|\hat I_N(t) - I(t)\big| \;\le\; \underbrace{\big|\hat I_N(t) - \bar I_N(t)\big|}_{\text{(i) filter aggregation}} + \underbrace{\big|\bar I_N(t) - I(t)\big|}_{\text{(ii) mean-field}}.
\]

\emph{Step B (mean-field term).}
The bounded-density assumption of Theorem~\ref{thm:two_scale} bounds the per-agent marginal density of $\xi_{v,t}$ at $\tau_I$ by $\rho_{\max}$. By Lemma~\ref{lem:density_inheritance}, the aggregated marginal density of $\bar\xi_N(t) = N^{-1}\sum_v \xi_{v,t}$ is also bounded by $\rho_{\max}$ on the same neighborhood of $\tau_I$. Theorem~\ref{thm:seir_limit} gives $\mathbb{E}[W_1(\bar\xi_N(\cdot), \xi^\star(\cdot))] \le C(\mathsf{H})/\sqrt{N}$. Apply Lemma~\ref{lem:bounded_density} with $X = \bar\xi_N(t)$ and $Y = \xi^\star(t)$ at the threshold $\tau_I$:
$\mathbb{E}\big|\bar I_N(t) - I(t)\big| \le \rho_{\max} \cdot \mathbb{E}[W_1(\bar\xi_N(t), \xi^\star(t))] \le C_2(\mathsf{H})/\sqrt{N}$
with $C_2(\mathsf{H}) = \rho_{\max}\, C(\mathsf{H})$.

\emph{Step C (filter term).}
Theorem~\ref{thm:claif_consistency} guarantees $\|b_{v,n} - b^\star_{v,n}\|_{\mathrm{TV}} \le \epsilon$ with probability $\ge 1 - \delta$ whenever $n \ge n_0$ in~\eqref{eq:claif_sample}. Apply Lemma~\ref{lem:tv_to_scalar} at each node $v$ with the aggregation vector $e$ of Section~\ref{sec:seir}. This yields the expected scalar error $\mathbb{E}|\hat\xi_{v,t} - \xi_{v,t}| \le C_{\mathrm{Pinsker}}\, \sigma_{\max}\, \sqrt{\epsilon}$. Conditioning Lemma~\ref{lem:bounded_density} on $\hat\xi_{v,t}$ gives the pointwise bound $\Pr[\mathbb{1}[\hat\xi_{v,t} \ge \tau_I] \ne \mathbb{1}[\xi_{v,t} \ge \tau_I] \mid \hat\xi_{v,t}] \le \rho_{\max}\, |\hat\xi_{v,t} - \xi_{v,t}|$. Taking expectation over $\hat\xi_{v,t}$ on both sides (linearly, since the right-hand side is linear in the random variable) gives the unconditional bound $\Pr[\text{disagreement at node }v] \le \rho_{\max}\, \mathbb{E}|\hat\xi_{v,t} - \xi_{v,t}|$. Averaging the indicator-disagreement probabilities over the $N$ agents and propagating through the $k$-hop factored-belief contraction (Lemma~\ref{lem:claif_L2}) composed with the hop union bound (Lemma~\ref{lem:claif_L3}) injects the $L^k$ factor by the same argument used to prove Theorem~\ref{thm:claif_consistency}. The result is
\[
\mathbb{E}\big|\hat I_N(t) - \bar I_N(t)\big| \;\le\; \rho_{\max}\cdot \frac{C_1\, L^k}{\nu_{\min}\sqrt{n}},
\]
with $C_1$ absorbing $C_{\mathrm{Pinsker}}\, \sigma_{\max}$ and the Kalman sample-complexity constants of Lemma~\ref{lem:claif_L1}.

\emph{Step D (composition).}
Combining Steps B and C yields the stated expectation bound. Setting the two expected-error terms equal,
$\rho_{\max}\, C_1\, L^k / (\nu_{\min}\sqrt{n}) = C_2(\mathsf{H})/\sqrt{N}$,
and solving for $n$ identifies the \emph{balance curve} $n_\circ(N)$ in~\eqref{eq:regime_boundary}. The theorem's claim is in expectation. $n_\circ(N)$ is the sample complexity at which the expected filter contribution equals the expected mean-field contribution. For individual trajectories, the two error contributions concentrate around their expectations by Markov's inequality applied to the non-negative error bounds in Steps B and C. The balance point is therefore robust to bulk-trajectory fluctuations. Below $n_\circ(N)$ the filter term dominates in expectation, so the marginal return on $n$ is higher. Above it the mean-field term dominates, so the marginal return on $N$ is higher.

The scaling relation $n_\circ(N) \propto N$ is the load-bearing content of the theorem. The proportionality constant depends on $(L, k, \nu_{\min}, \rho_{\max}, C_1, C_2(\mathsf{H}))$. A high-probability statement replaces these central constants with worst-case upper bounds valid over a parameter class. The resulting boundary is parallel to $n_\circ(N)$ but shifted upward, and the scaling law is unchanged. The value $8.3$ reported in Section~\ref{sec:method} uses EM central estimates from held-out benign traces. It is therefore an operating-point value at the expectation balance curve, not a worst-case constant.
\end{proof}

\paragraph{Relation to two-timescale stochastic approximation.}
The decomposition $E_{\mathrm{filter}}(n) + E_{\mathrm{MF}}(N)$ is structurally analogous to the two-timescale error decompositions in~\citep{dalal2018finite,kaledin2020finite}. A fast scale (inference convergence in $n$) composes with a slow scale (population aggregation in $N$). The optimal deployment regime is determined by balancing the two rates.
Compared to those RL and stochastic-approximation settings, our slow scale is a propagation-of-chaos rate rather than a step-size decay.

\paragraph{Relation to Bayesian McKean\textendash{}Vlasov inference.}
\citet{ningwu2026bayesian} study Bayesian inference for partially observed McKean\textendash{}Vlasov SDEs but do not compose the resulting filter rate with a mean-field rate to give a deployment regime.
Theorem~\ref{thm:two_scale} is the specialization of their inference view to heterogeneous event graphs with anchored emissions, and its composition with Theorem~\ref{thm:seir_limit} is new.

\subsection{Proof of Theorem~\ref{thm:activation} (Optimal Activation Time)}
\label{app:activation}

\begin{proof}
The adversary maximizes $J(T) = e^{-qT} (1 - 1/\rzero(T))$ where $\rzero(T) = \kappa\, r(T)$ with $\kappa = c_{\max}\alpha/\gamma$, and $r(T) = r_{\max} - (r_{\max} - r_0)e^{-\lambda T}$.
Then $r'(T) = \lambda (r_{\max} - r(T))$ and $\rzero'(T) = \kappa\, r'(T)$.

Setting $dJ/dT = 0$ and applying the product rule gives
\[
-q\, e^{-qT} \left(1 - \frac{1}{\rzero}\right) + e^{-qT} \frac{\rzero'}{\rzero^2} = 0,
\]
so $\rzero'(T^\star) = q\, \rzero(T^\star)(\rzero(T^\star) - 1)$. Substituting $\rzero = \kappa r$ gives the stationarity condition $\lambda(r_{\max} - r^\star) = q\, r^\star(\kappa r^\star - 1)$ for $r^\star = r(T^\star)$, that is,
\[
q\kappa (r^\star)^2 + (\lambda - q)\,r^\star - \lambda r_{\max} = 0,
\qquad
r^\star = \frac{-(\lambda-q) + \sqrt{(\lambda-q)^2 + 4q\kappa\lambda r_{\max}}}{2q\kappa},
\]
where $r^\star$ is the feasible (positive) root. Inverting $r(T)$ yields
\[
T^\star = \frac{1}{\lambda}\ln\frac{r_{\max} - r_0}{r_{\max} - r^\star},
\]
which is interior whenever $r_0 < r^\star < r_{\max}$. Because the left-hand side of the stationarity condition is positive for $r^\star < r_{\max}$, any interior root satisfies $\kappa r^\star > 1$, so $\rzero(T^\star) > 1$.

\emph{Second-order condition.} Write $dJ/dT = \kappa e^{-qT} h(r(T))/\rzero(T)^2$ with $h(r) = \lambda(r_{\max} - r) - q\,r(\kappa r - 1)$. Since $h'(r) = -\lambda - q(2\kappa r - 1) < 0$ whenever $\kappa r > 1/2$, and $r(T)$ is strictly increasing, $dJ/dT$ changes sign from positive to negative at $T^\star$. The condition $\kappa r^\star > 1$ enforced by stationarity therefore makes $T^\star$ a strict maximum.

\emph{Monotonicity.} At the feasible root, $T^\star$ decreases in both $\lambda$ and $q$: faster trust accumulation reaches the profitable routing share sooner, and a higher removal hazard favors earlier activation (Table~\ref{tab:tstar_sensitivity}).

The submitted version displayed a closed form for $T^\star$ that omitted $\kappa$ and therefore did not solve its own stationarity condition; the expression above replaces it, and Table~\ref{tab:tstar_sensitivity} is regenerated from it. No defense result consumes $T^\star$.
\end{proof}

\paragraph{Joint $T$-$\sigma$ optimization.}
The adversary jointly chooses Phase~0 duration $T$ and post-activation incubation rate $\sigma$.
$T$ controls pre-activation trust accumulation, affecting $\beta$ through routing share $r(T)$.
$\sigma$ controls post-activation stealth, affecting incubation duration $1/\sigma$.
The two parameters are independent: $T$ is chosen before activation, $\sigma$ is chosen after.
The joint objective is $J(T, \sigma) = e^{-qT} \cdot (1 - 1/\rzero(T, \sigma))$ where $\rzero(T, \sigma) = c_{\max} \cdot r(T) \cdot \alpha \cdot \sigma / ((\sigma + \mu)(\gamma + \mu))$.
The separable structure means $T^\star$ depends only on $(\lambda, q, \kappa, r_0, r_{\max})$ and $\sigma^*$ depends only on $(\gamma, \mu)$ and the adversary's detection-avoidance constraint.

\paragraph{Sensitivity of $T^*$.}
Table~\ref{tab:tstar_sensitivity} shows $T^*$ for representative parameter combinations.

\begin{table}[ht]
\centering
\caption{Optimal activation time $T^\star$ (sessions) for varying gateway learning rate $\lambda$ and removal hazard $q$, computed from the feasible root of Theorem~\ref{thm:activation}.}
\label{tab:tstar_sensitivity}
\scriptsize
\resizebox{\textwidth}{!}{%
\rowcolors{2}{white}{tblStripe}\begin{tabular}{lccccc}
\toprule
\tabhead
& $q = 0.001$ & $q = 0.005$ & $q = 0.01$ & $q = 0.05$ & $q = 0.1$ \\
\midrule
$\lambda = 0.01$ & 353.04 & 230.09 & 192.95 & 145.80 & 137.46 \\
$\lambda = 0.05$ & 100.76 & 70.61 & \textbf{58.98} & 38.59 & 33.32 \\
$\lambda = 0.1$ & 57.18 & 41.61 & 35.30 & 23.01 & 19.30 \\
$\lambda = 0.5$ & 14.63 & 11.44 & 10.08 & 7.06 & 5.90 \\
\bottomrule
\end{tabular}\rowcolors{2}{}{}%
}
\par\smallskip\noindent\scriptsize{Computed with $r_0 = 0.2$ ($K = 5$ tools), $r_{\max} = 0.5$, and $\kappa = 2.4$ from the logged Eve parameters. Bold marks Eve's reference cell. $T^\star$ decreases in both $\lambda$ and $q$; the $\lambda = 0.5$, $q = 0.1$ cell is an exact tenfold rescaling of the reference cell.}
\end{table}

\paragraph{Model limitations.}
The optimal activation model assumes: (1) gateway routing follows exponential saturation, which may not hold for all routing algorithms; (2) the hazard rate $q$ is constant, whereas suspicious behavior could increase $q$ over time; (3) $\rzero$ depends only on routing share, whereas memory store density (Proposition~\ref{prop:density}) also contributes to $\beta$.
These simplifications enable a closed-form solution at the cost of excluding adaptive defender responses.

\subsection{SEIR Derivations and Stability Analysis}
\label{app:seir}

\subsubsection{Disease-Free Equilibrium}

The disease-free equilibrium (DFE) of the SEIR system~\eqref{eq:seir} is $(S^*, E^*, I^*, R^*) = (N, 0, 0, 0)$.
Including agent turnover at rate $\mu$ (agents created and retired), the infected subsystem becomes $dE/dt = \beta I - (\sigma+\mu)E$ and $dI/dt = \sigma E - (\gamma+\mu)I$ at the DFE.
The Jacobian is
\[
J = \begin{pmatrix} -(\sigma+\mu) & \beta \\ \sigma & -(\gamma+\mu) \end{pmatrix}.
\]
The DFE is unstable (infection grows) when the dominant eigenvalue is positive.
This occurs when $\beta \sigma > (\sigma + \mu)(\gamma + \mu)$, i.e., $\rzero = \frac{\beta \sigma}{(\sigma + \mu)(\gamma + \mu)} > 1$.

\subsubsection{R0 Derivation via Next-Generation Matrix}

Using the next-generation matrix method, the transmission matrix $F$ and transition matrix $V$ for the infected compartments $(E, I)$ are
\[
F = \begin{pmatrix} 0 & \beta \\ 0 & 0 \end{pmatrix}, \quad
V = \begin{pmatrix} \sigma + \mu & 0 \\ -\sigma & \gamma + \mu \end{pmatrix}.
\]
The next-generation matrix is $FV^{-1}$, and $\rzero = \rho(FV^{-1}) = \frac{\beta \sigma}{(\sigma + \mu)(\gamma + \mu)}$.
When $\mu = 0$ (no agent turnover), this simplifies to $\rzero = \frac{\beta \sigma}{\sigma \gamma} = \frac{\beta}{\gamma}$, recovering the standard SIR threshold.

\subsubsection{Herd Immunity Threshold}

If a fraction $p$ of agents are immune (in compartment $R$), the effective reproduction number is $\reff = \rzero(1 - p)$.
Setting $\reff = 1$ yields the herd immunity threshold $H = 1 - 1/\rzero$.

\subsection{Amplification Results}
\label{app:amplification}
\label{app:proofs}
\label{app:density}

Propositions~\ref{prop:cycle} and~\ref{prop:density} characterize two secondary amplification effects in agentic ecosystems beyond the base SEIR prediction.
Full statements and proofs are below; the main-text summary appears at the end of Section~\ref{sec:seir}.
Corollary~\ref{cor:impact} composes Theorems~\ref{thm:seir_limit} and~\ref{thm:two_scale} with a forward-cone argument to bound the downstream impact of a localized entry identified by Proposition~\ref{prop:traceback}.

\begin{corollary}[Impact Propagation from a Localized Entry]
\label{cor:impact}
Given an entry $(u, \ell, t_0)$ identified by the trace-back of Proposition~\ref{prop:traceback}, the expected fraction of compromised agents at horizon $t_0 + \Delta$ along the forward cone of $(u, t_0)$ is bounded by
\begin{equation}
\mathbb{E}\!\left[\,\hat M_N(t_0 + \Delta)\,\right] \;\le\; \rho_{\max}\, e^{(\beta - \gamma)\,\Delta} + \frac{C(\Delta)}{\sqrt{N}},
\label{eq:impact}
\end{equation}
with $\beta, \gamma$ the SEIR rates of Corollary~\ref{cor:seir} and $C(\Delta)$ the horizon-dependent constant of Theorem~\ref{thm:seir_limit}. A defender with budget $B$ on intervention magnitude can choose to act at any $\Delta$ for which the right-hand side exceeds an acceptable spread.
\end{corollary}

\begin{proposition}[Cycle Amplification]
\label{prop:cycle}
In a contact graph with cycles of length $k$, let $\delta > 0$ be the per-cycle drift increment (an eigenvalue of $B^{(\mathsf{handoff},\mathcal{A},\mathcal{A})}$ along the aggregation axis) and $d_0$ the initial drift. After $j$ completed cycles the drift satisfies $d_j \geq d_0(1 + \delta)^{j}$, and after $m$ directed edge transitions on a $k$-cycle it satisfies $d(m) \geq d_0(1 + \delta)^{\lfloor m/k \rfloor}$.
This yields exponential amplification in frameworks with graph cycles (e.g., LangGraph).
\end{proposition}

\begin{proof}
Consider a cycle of length $k$ in the contact graph.
Let $d_0 > 0$ be the initial drift magnitude at a node in the cycle.
Let $\delta > 0$ be the per-iteration drift increment, the multiplicative factor by which drift increases when a node processes drifted input from its predecessor.
After one complete cycle ($k$ steps), the drift returns to the originating node with magnitude $d_0(1 + \delta)$.
After $j$ completed cycles ($jk$ edge transitions), the drift at the originating node satisfies $d_j = d_0 (1 + \delta)^j$.
Since each cycle requires $k$ transitions, after $m$ edge transitions at least $\lfloor m/k \rfloor$ cycles have completed, so $d(m) \geq d_0 (1 + \delta)^{\lfloor m/k \rfloor}$.
This is exponential in $m$ with base $(1+\delta)^{1/k}$, confirming that graph cycles produce exponential drift amplification.
For LangGraph with $\delta = 0.02$ and $k = 2$ (minimal cycle), $m = 30$ edge transitions give $d(30) \geq 1.02^{15} d_0 \approx 1.346\, d_0$, a $34.6\%$ cumulative shift, and $30$ completed cycles ($m = 60$) give $1.02^{30} d_0 \approx 1.811\, d_0$, an $81.1\%$ shift.
Statement, proof, and the measurements of Appendix~\ref{app:cycle_runs} index drift by edge transitions throughout.
\end{proof}

\begin{proposition}[Density Amplification]
\label{prop:density}
When $n_A$ agents are compromised via Case~A and write to a shared memory store, the effective retrieval probability for poisoned entries increases as $p \approx 1 - (1 - p_{\text{base}})^{n_A}$.
This raises the effective $\beta$ for Case~C agents through the spectral radius of Theorem~\ref{thm:seir_limit}.
\end{proposition}

\begin{proof}
Let $p_{\text{base}}$ be the probability that a single poisoned entry is retrieved in a given query.
When $n_A$ compromised agents independently write poisoned entries to a shared memory store, the probability that \emph{at least one} poisoned entry is retrieved is
$p = 1 - (1 - p_{\text{base}})^{n_A}$.
For $p_{\text{base}} = 0.479$ (MemoryGraft retrieval proportion) and $n_A = 3$, we get $p = 1 - 0.521^3 \approx 0.859$.
This raises the effective infection rate $\beta$ for Case C agents, as the transmission probability per contact increases with poisoned entry density.
\end{proof}

\subsection{Proposition: Calibration Consistency (Statement and Proof)}
\label{app:calibration_proof}

The statement of Proposition~\ref{prop:calibration} (referenced from \S\ref{sec:two_scale} and \S\ref{sec:limitations}) is reproduced here for locality with its proof.

\begin{proposition}[Calibration Consistency]
\label{prop:calibration}
Under joint observability (Lemma~\ref{lem:joint_obs}) and a persistence-of-excitation condition on the typed-event stream (assumption (A6), Appendix~\ref{app:identifiability}), expectation\textendash{}maximization on $M$ benign windows each of length $\mathsf{H}$ attains
\begin{equation}
\|\hat\theta - \theta\|_{2} \;=\; O_P\!\left(\sqrt{\tfrac{d^2}{M\mathsf{H}}}\right)
\label{eq:calibration_rate}
\end{equation}
under benign stationarity. If the deployment distribution differs from calibration at total variation $\epsilon$, an additive $\epsilon^{1/2}$ term enters~\eqref{eq:calibration_rate}.
\end{proposition}

\begin{proof}
Let $\theta = (A^{\phi}, B^{(r,\cdot,\cdot)}, H, Q^{\phi}, \Sigma_{\mathrm{obs}})$ be the parameter vector of the factored state-space model of Definition~\ref{def:statespace} (the intervention matrix $C$ is set from deterministic action semantics and not included in $\theta$), and let $\hat\theta$ be the EM maximiser of the joint log-likelihood on $M$ benign windows of length $\mathsf{H}$.

\paragraph{Identifiability.}
Under joint observability (Lemma~\ref{lem:joint_obs}), the pair $(A^\phi, H)$ is observable on the $6$-dimensional state with minimum singular value $\nu_{\min} > 0$ of the $k$-step observability Gramian. Under the persistence-of-excitation assumption (A6) on the typed-event stream (Appendix~\ref{app:identifiability}), the sample covariance of event-induced innovations concentrates at rate $O_P(1/\sqrt{M\mathsf{H}})$ to a strictly positive-definite limit. The Fisher information $\mathcal{I}(\theta)$ is therefore non-singular at the truth, which is the standard identifiability condition for linear-Gaussian state-space models~\citep{tsiamis2020sample}.

\paragraph{Consistency rate.}
Standard M-estimation on the quasi-log-likelihood of the factored model (single-node analysis in~\citep{tsiamis2020sample}, extended via the factored-belief contraction of~\citep{boyen1998tractable}) yields
\begin{equation*}
\sqrt{M\mathsf{H}}\,(\hat\theta - \theta) \;\xrightarrow{d}\; \mathcal{N}\!\left(0,\; \mathcal{I}(\theta)^{-1}\right),
\end{equation*}
which gives $\|\hat\theta - \theta\|_2 = O_P(\sqrt{d^2/(M\mathsf{H})})$ by bounding $\|\mathcal{I}(\theta)^{-1}\|_{\mathrm{op}} \le c/\nu_{\min}^2$ and noting that the parameter dimension is $O(d^2)$ under the sparsity of $B^{(r,\cdot,\cdot)}$ induced by the edge-type partition.

\paragraph{Distribution shift.}
Suppose the deployment distribution $P^{\mathrm{dep}}$ differs from the calibration distribution $P^{\mathrm{cal}}$ at $d_{\mathrm{TV}}(P^{\mathrm{dep}}, P^{\mathrm{cal}}) \le \epsilon$. By the Pinsker/TV transfer bound and the quadratic smoothness of the log-likelihood at $\theta$, the population-level objectives satisfy $\|\theta^{\mathrm{dep}} - \theta^{\mathrm{cal}}\|_2 \le C_\ell \epsilon^{1/2}$ for a universal $C_\ell$ depending on the likelihood curvature. The triangle inequality gives the additive $\epsilon^{1/2}$ term of~\eqref{eq:calibration_rate}.
\end{proof}

\section{Formal Model Details}
\label{app:formal_details}

This appendix carries extended formal detail beyond the main text. It enumerates the observable edge types and shared latent coordinates, gives the full factored-DBN factorization, states the defender POMDP and attacker constrained MDP in their full form, documents the identifiability argument, describes the conformal calibration protocol, reproduces the assumptions-to-results map, and hosts the support tables and figures that Sections~\ref{sec:introduction}--\ref{sec:formal_framework} reference.

\subsection{Main-Text Maps and Positioning}
\label{app:main_text_maps}

The main paper keeps the story on the \textbf{WHEN} / \textbf{WHICH} / \textbf{HOW} spine. The roadmap table, related-work comparison, and dual-view schematic are reproduced here to preserve technical traceability without consuming main-text pages.

\paragraph{Positioning.}
Our closest conceptual peer is Motwani et al.~\citep{motwani2024secret}, which studies steganographic communication among colluding agents. Where that work models covert channels, we study persistent latent compromise: a legitimate-looking agent accumulates trust, leaves cross-layer residue, and must be detected, traced, and contained from partial runtime evidence. Swarm Shepherd fills this gap by treating runtime traces as evidence of hidden compromise state and connecting early warning, layer-level trace-back, and containment within a single inference framework. Theory-facing comparisons to Bayesian McKean--Vlasov inference~\citep{ningwu2026bayesian} and two-timescale stochastic approximation~\citep{dalal2018finite,kaledin2020finite} are addressed by the proof structure in Appendix~\ref{app:theory_proofs}.

\begin{table}[ht]
\centering
\caption{Pillar-depth summary. Each pillar contributes both a formal guarantee and a paired experiment.}
\label{tab:pillar_depth}
\scriptsize
\resizebox{\textwidth}{!}{%
\rowcolors{2}{white}{tblStripe}\begin{tabular}{lll}
\toprule
\tabhead
\textbf{Pillar} & \textbf{Formal guarantee} & \textbf{Empirical validation} \\
\midrule
\tabours
Two-scale regime \textbf{(headline)} & Theorem~\ref{thm:two_scale} & A2 phase diagram (Fig.~\ref{fig:a2_regime}) \\
Swarm Shepherd inference & Theorem~\ref{thm:claif_consistency}; Corollary~\ref{cor:warning_lead} & A1 $\sqrt{n}$ rate fit (\S\ref{sec:e7_when}) \\
Mean-field theory & Theorem~\ref{thm:seir_limit}; Corollary~\ref{cor:seir_block} & A3 $\sqrt{N}$ fit (\S\ref{sec:e7_how}) \\
Trace-back smoother & Proposition~\ref{prop:traceback} & A4: $24 \times 3 \times 30$ seeds \\
Containment & Theorem~\ref{thm:r0}; Corollary~\ref{cor:seir} & Live $\Rc$ (Table~\ref{tab:live_containment}); forecast (Table~\ref{tab:a6_main}) \\
Threat model & Definition~\ref{def:statespace}; 24-variant taxonomy & \S\ref{sec:evaluation} ASR \\
\bottomrule
\end{tabular}\rowcolors{2}{}{}%
}
\end{table}

\begin{table}[ht]
\centering
\caption{Comparison with representative prior work. Among the compared lines of work, the Agentic Sleeper is the only setting that combines trust-building activation, SEIR epidemic propagation with adversary-controlled incubation, multi-level persistence, cross-level latent state, and factored-POMDP defense.}
\label{tab:comparison}
\scriptsize
\setlength{\tabcolsep}{4pt}
\resizebox{\textwidth}{!}{%
\rowcolors{2}{white}{tblStripe}\begin{tabular}{lccccccc}
\toprule
\tabhead
\textbf{Method} & \textbf{Activation} & \textbf{Propagation} & \textbf{Incubation} & \textbf{Persistence} & \textbf{Formal Model} & \textbf{Latent State} & \textbf{Control} \\
\midrule
Prompt Injection & Immediate & None & None & None & None & N & None \\
Temporal Attacks & Discrete trigger & None & Fixed & Layer & Taxonomy & N & None \\
Agent Smith & Immediate & $O(\log N)$ & None & Session & SIS-like & N & None \\
Morris~II & Immediate & Worm & None & Code & None & N & None \\
MemoryGraft & Immediate & None & None & Memory & None & N & None \\
HEARTBEAT & Passive & None & None & Memory & E$\to$M$\to$B & N & None \\
DeepContext~\citep{deepcontext2026} & N/A (defense) & None & None & Session & Stateful classifier & Intent drift & Per-turn classifier \\
\tabours
\textbf{Ours} & \textbf{Trust $\to$ gradual} & \textbf{SEIR epidemic} & \textbf{Adversary-ctrl.} & \textbf{Multi-level} & \textbf{Factored POMDP} & \textbf{Y (cross-level)} & \textbf{Factored POMDP} \\
\bottomrule
\end{tabular}\rowcolors{2}{}{}%
}
\end{table}

\begin{figure}[ht]
\centering
\includegraphics[width=0.98\linewidth]{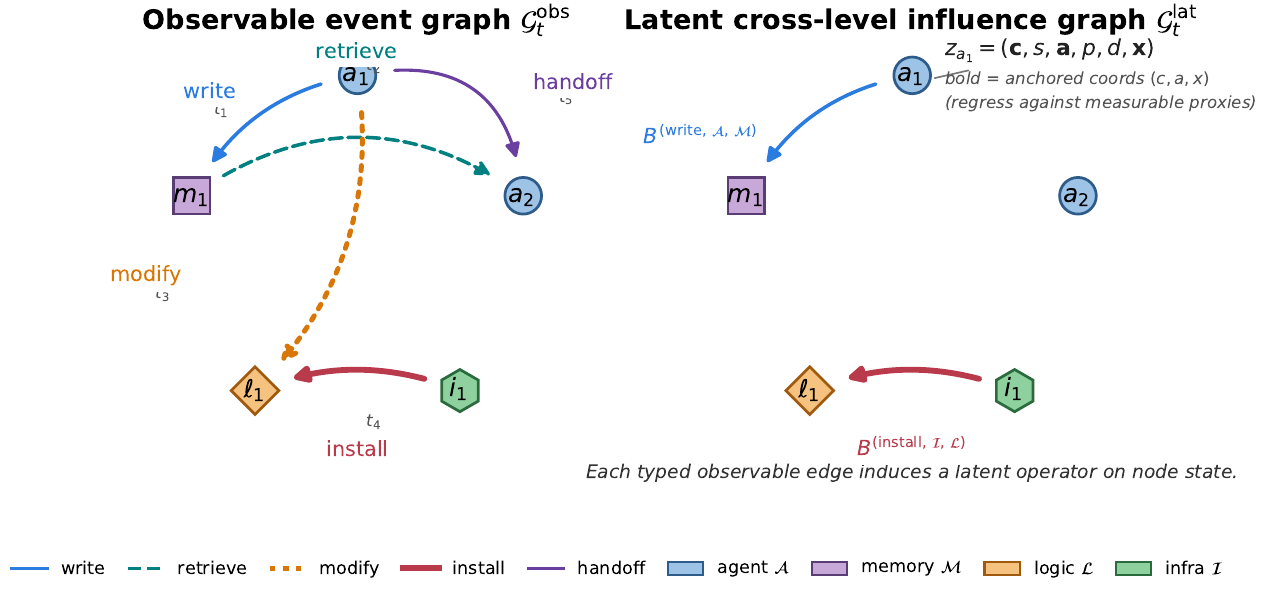}
\caption{Dual-view event graph. Left: the observable event graph $\mathcal{G}^{\mathrm{obs}}_t$ records typed, timestamped edges between agents, memory, logic, and infrastructure nodes. Right: the latent influence graph $\mathcal{G}^{\mathrm{lat}}_t$ shares the node set, with a 6-dimensional latent state $z_v = (c, s, a, p, d, x)$ per node; the anchored coordinates $(c, a, x)$ are regressed against measurable proxies. Each typed observable edge induces a typed influence operator $B^{(r, \phi(u), \phi(v))}$ that updates latent state. Swarm Shepherd (\S\ref{sec:early_warning}) filters this joint object from the event stream.}
\label{fig:dual_view}
\end{figure}

\subsection{Observable Edge Types}
\label{app:edge_types}

The observable event graph $\mathcal{G}^{\mathrm{obs}}_t$ of Definition~\ref{def:obs_graph} carries eleven typed edges that together span the three attack levels. Memory activity covers \textsf{read}, \textsf{write}, and \textsf{retrieve}. Logic activity covers \textsf{invoke}, \textsf{return}, \textsf{modify}, and \textsf{handoff}. Infrastructure activity covers \textsf{grant}, \textsf{install}, \textsf{execute}, and \textsf{egress}.

\paragraph{Memory edges (L1).}
A \textsf{read} edge fires when an agent pulls a fixed file or database row. A \textsf{write} edge fires when an agent persists a new memory entry. A \textsf{retrieve} edge fires when a vector- or semantic-search call returns a ranked result; the retrieval probability distribution determines the Case-C trigger rate $\alpha = f(p_{\mathrm{retrieval}})$.

\paragraph{Logic edges (L2).}
An \textsf{invoke} edge fires on any tool or subagent call, \textsf{return} marks the reply, \textsf{modify} fires whenever a configuration, system prompt, tool schema, or routing rule is rewritten on disk or in memory, and \textsf{handoff} fires on agent-to-agent delegations that transfer control without a full tool invocation.

\paragraph{Infrastructure edges (L3).}
A \textsf{grant} edge fires whenever a permission or credential scope is extended. An \textsf{install} edge fires on package or binary installation, including \texttt{npx~-y} auto-updates. An \textsf{execute} edge fires on a new process start attributable to an agent. An \textsf{egress} edge fires on any external network call initiated by the agent or its tools.

These eleven types are the minimum set that covers every attack variant in Table~\ref{tab:attack_l1}, \ref{tab:attack_l2}, and~\ref{tab:attack_l3} of Appendix~\ref{app:attack_variants}; each variant names the edge type its mechanism emits first.

\subsection{Shared Latent Coordinates}
\label{app:coords_shared}

Each node $v \in V^{\mathrm{lat}}$ carries the six shared coordinates $z_{v,t} = (c, s, a, p, d, x)$ of Definition~\ref{def:latent_state_graph}. The verbal interpretations below are calibrated against the anchored proxies (Appendix~\ref{app:identifiability}) and checked against the controlled-injection traces of Appendix~\ref{app:eval_extended}.

\paragraph{Contamination $c_{v,t}$ (anchored).}
How much poisoned, unverified, or adversarial content the node has absorbed through \textsf{retrieve}, \textsf{read}, or \textsf{handoff} edges. Anchored to the poisoned-entry retrieval rate on a rolling window.

\paragraph{Susceptibility $s_{v,t}$ (unanchored).}
How readily the node accepts new cross-level influence; rises with prior contamination and with the attacker's Phase-0 trust accumulation. Estimated through its influence on the transition law.

\paragraph{Authority $a_{v,t}$ (anchored).}
How much routing weight, tool-selection priority, or delegation trust the node commands. Anchored to empirical routing share for agents and tools, or retrieval rank for memory.

\paragraph{Persistence $p_{v,t}$ (unanchored).}
How durably latent state survives restart, reindex, or configuration refresh. Rises with distributed-fragment persistence (Appendix~\ref{app:permissions}).

\paragraph{Deviation $d_{v,t}$ (unanchored).}
How far agent behavior has drifted from its baseline along the active drift primitive. Estimated through ASI subscales (Appendix~\ref{app:eval_extended}).

\paragraph{Capability exposure $x_{v,t}$ (anchored).}
How many sensitive resources the node can now reach. Anchored to the granted-permission count plus the reachable-resource count.

The three anchored coordinates $(c, a, x)$ carry identifiability. The three unanchored coordinates $(s, p, d)$ are constrained through the transition dynamics; estimation protocol is in Appendix~\ref{app:identifiability}.

\subsection{Type-Specific Latent Dimensions}
\label{app:latent_dims}

Beyond the six shared coordinates, each node type carries additional coordinates summarized below.
In the main text we use only the shared core so that results apply across node types; type-specific coordinates are used only in Appendix~\ref{app:eval_extended} ablations.

\paragraph{Agent nodes.} Goal fidelity, tool preference, routing bias, plan brittleness, secrecy discipline, approval conservatism.
\paragraph{Memory nodes.} Retrieval frequency, semantic centrality, provenance confidence, contamination density, recency and decay.
\paragraph{Logic nodes.} Config integrity, prompt rigidity, schema corruption level, guard strength, approval-requirement strength.
\paragraph{Infrastructure nodes.} Privilege level, sandbox-escape risk, deployment reach, restart persistence, audit visibility.

Each of these is a bounded scalar and enters the transition in Definition~\ref{def:statespace} as an additional coordinate of $z_v$; block-diagonal structure in $A^{\phi(v)}$ keeps type-specific coordinates local to their node class while the shared core is coupled across types via the typed influence operators $B^{(r,\phi(u),\phi(v))}$.

\subsection{Factored DBN and POMDP Formalization}
\label{app:dbn_formalism}

\paragraph{Transition factorization.}
With node-type self-dynamics $A^{\phi(v)}$ and typed influence operators $B^{(r,\phi(u),\phi(v))}$ the joint transition is
\begin{equation}
P(Z_{t+1}\mid Z_t, E^{\text{obs}}_t, U_t)
\;=\; \prod_{v\in V^{\text{lat}}} P\!\left(z_{v,t+1}\mid z_{v,t}, \{z_{u,t}\}_{u\in N_{\text{in}}(v, t)}, \iota_{v,t}\right),
\label{eq:joint_factorization}
\end{equation}
where $N_{\text{in}}(v, t) = \{u : (u,v,r)\in E^{\text{obs}}_t \text{ for some } r\}$.
This factorization is what makes the model a dynamic Bayesian network (DBN) whose per-step inference cost is linear in $|E^{\text{obs}}_t|$ rather than in $|V^{\text{lat}}|^2$.

\paragraph{Linear-Gaussian instantiation (v1 experiments).}
We instantiate Definition~\ref{def:statespace} with linear-Gaussian self-dynamics and typed influence operators.
Given this structure, the per-node belief $b_{v,t}(z_v) = \mathcal{N}(\mu_{v,t}, \Sigma_{v,t})$ admits closed-form Kalman updates under the fixed event graph $E^{\text{obs}}_{0:t}$, with neighbor messages exchanged once per event.
Parameters $(A^{\phi(v)}, B^{(r,\cdot,\cdot)}, H, Q^{\phi(v)}, \Sigma_{\mathrm{obs}})$ are estimated by EM on benign calibration windows only, with no attack labels (Table~\ref{tab:param_provenance}); the intervention matrix $C$ is calibrated from deterministic action semantics rather than learned by EM.
Discrete-state, switching, and non-Gaussian alternatives are compared in Appendix~\ref{app:model_sensitivity}; neural-ODE extensions are left to future work.

\paragraph{Factored POMDP belief propagation.}
For the defender POMDP (Definition~\ref{def:defender_pomdp}, below) the joint belief is intractable in closed form at scale, so we approximate it in fully factored form
$b_t(Z) \approx \prod_{v} b_{v,t}(z_v)$
with neighbor-message updates analogous to loopy belief propagation.
This is the standard factored-POMDP approach taken by~\citep{miehling2018pomdp,zhang2024factored}; we inherit its convergence and approximation properties.

\paragraph{Attacker policy class.}
The attacker constrained MDP (Definition~\ref{def:attacker_mdp}, below) admits a rich policy class; for Theorem~\ref{thm:activation} we restrict to threshold policies parameterized by activation time $T$ and post-activation influence magnitude $\sigma$.
More general policies are discussed in Appendix~\ref{app:activation} under ``Joint $T$-$\sigma$ optimization.''

\subsection{Decision Problems: Full Formalizations}
\label{app:decision_problems}

Section~\ref{sec:deployment} points to these coupled decision problems; full tuples are below.

\paragraph{Defender factored POMDP (full tuple).}
\begin{definition}[Defender Factored POMDP]
\label{def:defender_pomdp}
The defender problem is $(\mathcal{S}, \mathcal{U}_{\mathrm{def}}, \mathcal{O}, P, \mathcal{Z}, \mathcal{R}_d, \gamma_d, b_0)$ with hidden state $\mathcal{S} = \prod_v \mathbb{R}^{d_v}$, budgeted action space $\mathcal{U}_{\mathrm{def}}$ (sanitize memory, freeze config, downweight tool, isolate agent, require approval, no-op), observation space $\mathcal{O}$ containing the typed event stream and anchored emissions, transition kernel $P$ and emission kernel $\mathcal{Z}$ from Definition~\ref{def:statespace}, and risk-weighted reward $\mathcal{R}_d$ trading residual compromise against intervention cost.
The belief $b_t = P(Z_t\mid y_{0:t}, E^{\mathrm{obs}}_{0:t})$ is approximated in factored form $b_t(Z) \approx \prod_v b_{v,t}(z_v)$ with neighbor-message updates.
The five defense paradigms of Appendix~\ref{app:defender_paradigms_detail} are instantiations of this controller under different action restrictions and belief summaries. Factorization matters because a per-node belief keeps inference tractable at deployment scale, and the anchored emission on $(c, a, x)$ keeps each per-node belief identifiable rather than collapsing into a latent degeneracy.
\end{definition}

\paragraph{Attacker constrained MDP (full utility decomposition).}
\begin{definition}[Attacker Constrained MDP]
\label{def:attacker_mdp}
The adversary accumulates three utility streams.
The \emph{benign utility} $U_{\mathrm{benign}}(0{:}T) = \sum_{t=0}^{T} \mathbf{w}_{\mathrm{util}}^{\!\top} \mathbf{f}_{\mathrm{svc}}(t)$ is the cumulative legitimate-service payoff during Phase~0, proportional to the routing share $r(t)$ in~\eqref{eq:routing_share} and to user-visible task-success output.
The \emph{post-activation compromise utility} $U_{\mathrm{compromise}}(T{:}\mathsf{H}) = \sum_{t=T}^{\mathsf{H}} \mathbf{w}_{\mathrm{adv}}^{\!\top} \mathbf{g}(\bar I(t))$ is proportional to the compromised-fraction trajectory $\bar I(t)$ of the latent aggregate (Section~\ref{sec:seir}), with coefficient $\mathbf{w}_{\mathrm{adv}}$ encoding data-exfiltration, cost-manipulation, and control objectives.
The \emph{detection cost} $C_{\mathrm{detection}}(0{:}\mathsf{H}) = \sum_{t=0}^{\mathsf{H}} \Pr[\text{declare at }t]\cdot L_{\mathrm{removal}}$ combines the defender's per-step declaration probability under belief $b_t$ with the adversary's terminal loss $L_{\mathrm{removal}}$ on removal.
The attacker solves
\begin{equation}
\max_\pi \;\; \mathbb{E}\!\left[\,U_{\mathrm{benign}}(0{:}T) \;+\; \lambda\, U_{\mathrm{compromise}}(T{:}\mathsf{H}) \;-\; \eta\, C_{\mathrm{detection}}(0{:}\mathsf{H})\,\right]
\end{equation}
subject to per-step detection-risk constraints, with activation time $T$ chosen before activation and post-activation incubation rate $\sigma$ chosen after.
The trust-accumulation state $\tau_{u\to v, t}$ on each agent--tool edge is a coordinate of $z$ and governs the benign-phase growth of authority $a$.
The separable $(T, \sigma)$ structure yields closed-form solutions; Theorem~\ref{thm:activation} is the solution for $T^\star$ under exponential saturation of routing share.
This formalism connects the Agentic Sleeper to corrupted-bandit and sleeper-agent theory~\citep{lykouris2018stochastic,gupta2019better,kong2024corruption,hubinger2024sleeper}.
\end{definition}

\subsection{Identifiability via Anchored Emissions}
\label{app:identifiability}

A naive continuous-latent model with six shared and several type-specific coordinates is unidentifiable, because multiple parameter configurations produce the same observation law.
We restore identifiability by anchoring three of the six shared coordinates to observable proxies.

\begin{lemma}[Joint Observability]
\label{lem:joint_obs}
Partition $H=[H_{\mathrm{anchor}};\,H_{\mathrm{ASI}}]$, with $H_{\mathrm{anchor}}$ rank-$3$ on $(c,a,x)$ and $H_{\mathrm{ASI}}$ emitting $(s,p,d)$ through ASI subscales. The $k$-step observability matrix has full column rank $6$ whenever the ASI rows project nontrivially onto $\mathrm{span}\{(A^\phi)^j e_s,(A^\phi)^j e_p,(A^\phi)^j e_d:j\le5\}$. We denote the minimum singular value of this joint Gramian by $\nu_{\min}$.
\end{lemma}

\paragraph{Anchoring.}
Let $H = [H_{\text{anchor}} \;\; H_{\text{residual}}]$ with $H_{\text{anchor}} \in \mathbb{R}^{3\times 3}$ a diagonal positive-definite matrix that maps $(c, a, x)$ to their observable proxies, namely poisoned-retrieval rate, routing share, and permission count. The residual block $H_{\text{residual}}$ is the emission for the remaining $(s, p, d)$ coordinates through ASI dimensions.
The anchor half of the emission is supervised, while the residual half is learned only through its influence on observed dynamics.

\paragraph{Poisoned-entry retrieval rate.}
The per-node poisoned-entry retrieval rate at time $t$ is the empirical fraction of \textsf{retrieve} edges incident to $v$ within a rolling window $[t-\Delta, t]$ whose returned memory entry carries a poisoned or unverified provenance tag. In the headline controlled-injection evaluation this tag is ground-truth from the known injected entries of Appendix~\ref{app:attack_variants}. In deployment the tag is derived from the memory store's provenance-confidence score (ChromaDB metadata, SQLite3 audit log) thresholded at a calibration level. Three signals must be kept apart: the \emph{scorer-only} injected-entry label, which grades experiments; \emph{detector-visible provenance confidence}, which is what a deployment has; and \emph{non-provenance anchors} (routing share, permission counts, artifact hashes), which need no provenance layer. Only the contamination anchor depends on provenance. In the headline setting it reads the ground-truth tag as a stand-in for provenance confidence; Appendix~\ref{app:provenance} replaces that tag with deployment metadata, so that no scorer label enters the pipeline, and reports the cost.
Contamination $c_{v,t}$ is regressed against this rate. Authority $a_{v,t}$ is regressed against empirical routing share for agents and tools, or retrieval rank for memory. Capability exposure $x_{v,t}$ is regressed against the granted-permission count and reachable-resource count.

\paragraph{Identifiability claim.}
Under this construction the model is identifiable whenever the observability matrix $\mathcal{O}(A,H) = [H^\top, (HA)^\top, \ldots, (HA^{d-1})^\top]^\top$ has rank $d$ (standard Kalman observability).
In our linear-Gaussian instantiation, generic choices of $A$ and $H_{\text{residual}}$ yield rank $d$.
Parameters are identifiable up to the group of transformations that preserve both the anchored emission and the transition law, which for our parameterization is trivial.
Ground-truth controlled-injection traces (Appendix~\ref{app:eval_extended}) are used only for evaluation and for setting the warning threshold; no dynamics or emission parameter is fit on them (Table~\ref{tab:param_provenance}).

\subsection{Conformal Risk Control Protocol}
\label{app:conformal}

Early-prediction results (A1, Section~\ref{sec:e7_when}) use split conformal prediction~\citep{conformalrisk2024} to convert DBN-filtered compromise probabilities into calibrated warning times.

\paragraph{Score.}
Use $0 < \tau_E < \tau_I$, where $\tau_E$ is the early-warning threshold and $\tau_I$ is the true infected/compromise threshold. For each calibration trajectory $i$, define the true onset
$T_i^\star = \min\{t : \xi_{v,t} > \tau_I\}$
and the nonconformity score
$s_i = (T_i^\star - \hat{T}_i)_+,$
where $\hat{T}_i = \min\{t : b_{v,t}(\xi > \tau_E) > 0.5\}$ is the DBN's warning time and $(x)_+ = \max(x, 0)$ truncates late warnings to zero.

\paragraph{Calibration.}
Given a held-out calibration set of size $n$ and a target miscoverage $\alpha$, set the threshold
$\hat{q}_\alpha = \lceil (1-\alpha)(n+1)\rceil\text{-th order statistic of } \{s_i\}_{i=1}^n$.
The conformal warning set is $\hat{T}_\alpha = \hat{T} + \hat{q}_\alpha$, i.e., the DBN warning padded by the calibration quantile.

\paragraph{Guarantee.}
Under exchangeability of calibration and test trajectories,
$\Pr(T^\star \le \hat{T}_\alpha) \ge 1 - \alpha$.
In A1 we report coverage at $\alpha = 0.1$ and confirm empirical coverage within $\pm 3$ percentage points of the target.
For Corollary~\ref{cor:warning_lead}, $\eta_\epsilon$ denotes the warning-time localization slack induced by the posterior TV error at the early threshold; the A1 calibration estimates this slack before checking the $k$-session lead certificate.

\subsection{Assumptions-to-Results Map}
\label{app:assumptions_map}

The method stack of Section~\ref{sec:formal_framework} rests on a compact set of assumptions. Each assumption feeds a specific subset of results, summarized here so a reader can see the cost of each guarantee at a glance.

\begin{itemize}[leftmargin=1.5em,itemsep=1pt,topsep=2pt]
\item \textbf{(A1) Linear-Gaussian factored DBN.} Used by Theorems~\ref{thm:claif_consistency},~\ref{thm:seir_limit},~\ref{thm:two_scale}, Corollaries~\ref{cor:warning_lead},~\ref{cor:seir},~\ref{cor:seir_block},~\ref{cor:impact}, and Proposition~\ref{prop:traceback}.
\item \textbf{(A2) Rank-$d$ anchored observability $\nu_{\min}$.} Used by Theorem~\ref{thm:claif_consistency} and Proposition~\ref{prop:traceback}.
\item \textbf{(A3) Bounded cross-level coupling $\|B^{(r,\cdot,\cdot)}\|_{\mathrm{op}} \le L$.} Used for the $L^{2k}$ factor in Theorem~\ref{thm:claif_consistency}, the Lipschitz constant $L_{\mathrm{drift}}$ of (A2.5), and the contact-rate bound in Corollary~\ref{cor:seir_block}.
\item \textbf{(A4) Minimum incident-event rate $\lambda_{\min} > 0$.} Used by Theorem~\ref{thm:claif_consistency}.
\item \textbf{(A5) Pairwise operator distinguishability $\delta_B > 0$.} Used by Proposition~\ref{prop:traceback}.
\item \textbf{(A6) Persistence of excitation.} The benign typed-event stream has a non-degenerate innovation covariance over calibration windows. Used by Proposition~\ref{prop:calibration}.
\item \textbf{(A7) Stable self-dynamics and bounded warning-band variation.} The self-dynamics satisfy $\rho(A^\phi)<1$. On the scalar warning band $\tau_E\le\xi_t<\tau_I$, the drift and diffusion obey $|a_t|\le\bar a$ and $\sigma_t^2\le\bar\sigma^2$, with positive threshold margin $\Delta_\tau=\tau_I-\tau_E>0$. Used by Theorems~\ref{thm:claif_consistency},~\ref{thm:seir_limit},~\ref{thm:two_scale}, Corollaries~\ref{cor:warning_lead},~\ref{cor:seir}, and the moment and first-passage constants in Appendix~\ref{app:theory_proofs}.
\item \textbf{(A2.5) Wasserstein-Lipschitz aggregated drift.} Used by Theorem~\ref{thm:seir_limit} and Corollary~\ref{cor:seir_block}.
\item \textbf{Within-class exchangeability.} Used by Theorem~\ref{thm:seir_limit} (aggregate) and Corollary~\ref{cor:seir_block} (per class).
\item \textbf{Bounded density $\rho_{\max}$ at $\tau_I$.} Used by Corollary~\ref{cor:seir} (upcrossing intensity) and Theorem~\ref{thm:two_scale} (continuity modulus).
\end{itemize}

The non-Gaussian extension of (A1) and the graphon relaxation of exchangeability are called out in Section~\ref{sec:limitations}.

\subsection{Operational Interpretation}
\label{app:operational_interpretation}

Swarm Shepherd's belief state is the bridge from warning to containment. Per-turn classifiers see only the current runtime signal, stateful detectors compress history to scalar summaries, and architectural defenses lack cross-agent belief; this explains the high evasion rates of observation-local defenses in Appendix Table~\ref{tab:e5_defense}. In practice, a defender first acts on a T1 posterior warning, then runs the trace-back smoother to estimate $(\hat u,\hat\ell,\hat t_0)$, and finally clears residue over the inferred time window and substrate. Local removal alone can leave $\Rc>1$, as the live visible-node-removal arms of Table~\ref{tab:live_containment} show, so containment must be driven by the same posterior belief that triggered warning and attribution.

\subsection{Defender Paradigms: Detailed Action Rules}
\label{app:defender_paradigms_detail}

The main text (\S\ref{sec:formal_framework}) names the five paradigms; the detailed action rules and their epidemiological analogues are below. Inference-driven paradigms come first; containment paradigms follow once Swarm Shepherd has localized the compromise.

\paragraph{Behavioral fingerprinting (syndromic surveillance).}
Continuous monitoring of the emission $H\mu_{v,t}$ from Swarm Shepherd's filtered belief (Algorithm~\ref{alg:claif}), analogous to syndromic surveillance in public health. This is the direct Swarm Shepherd readout and inherits the belief-consistency bound of Theorem~\ref{thm:claif_consistency}. It detects drift trajectories with calibrated conformal coverage rather than firing on individual anomalous events.

\paragraph{Contact tracing.}
When Swarm Shepherd flags an elevated node, the defender expands the belief over its $k$-hop neighborhood on the latent graph. This identifies agents that share memory stores with the compromised agent, received messages from it, used the same tools during the infection window, or share RAG corpus segments. The expansion is the direct POMDP analogue of epidemiological contact tracing.

\paragraph{Dynamic quarantine.}
Once Swarm Shepherd has localized compromise, time-aware retrieval filtering of recently-injected or suspicious content applies a cost-gated intervention action from $\mathcal{U}_{\mathrm{def}}$. The quarantine reproduction number $\Rq = \rzero \cdot (1 - q)$, where $q$ is quarantine effectiveness~\citep{zou2003quarantine}. Simulation analysis suggests $q = 0.52$ is achievable, sufficient to bring $\rzero < 1$ for several framework and topology combinations.

\paragraph{Targeted vaccination.}
Harden hub nodes, including Manager agents (CrewAI), GroupChat coordinators (AutoGen), and high-fanout nodes (LangGraph), by conditioning the priors $b_{v,0}(a), b_{v,0}(x)$ on hub connectivity. This is analogous to vaccinating high-connectivity individuals in scale-free networks~\citep{pastorsatorras2001}.

\paragraph{Ring vaccination.}
When Swarm Shepherd flags compromise, harden all agents within $r$ communication hops on the latent graph. Ring $r = 1$ provides moderate containment with low overhead; ring $r = 2$ provides stronger containment at higher cost.

\paragraph{Containment condition.}
Once the inference layer is in place, the classical containment condition, controlled reproduction $\Rc < 1$, becomes a necessary but not sufficient corollary. The controller must reach that threshold, acting on Swarm Shepherd's filtered belief rather than on an observable-only signal.

\subsection{Extended Limitations}
\label{app:extended_limitations}

Section~\ref{sec:limitations} covers the four limitations whose violation would most affect the headline claims. The extended list below documents assumptions whose violation would affect secondary results or deployment-time operating points.

\paragraph{Horizon dependence of $C_2(\mathsf{H})$.}
The Vuong-style coupling constant $C(\mathsf{H})$ in Theorem~\ref{thm:seir_limit} grows exponentially in the horizon $\mathsf{H}$, so the regime boundary $n_\circ(N) \propto C_2(\mathsf{H})^{-2}$ shrinks at long horizons. Operationally, the $n_\circ(N) \propto N$ prescription of Theorem~\ref{thm:two_scale} with numerical constant $\approx 8.3$ holds in the regime where $C_2(\mathsf{H}) = O(1)$. The A2 and A3 experiments of Section~\ref{sec:e7_how} confirm this for the session counts we report.

\paragraph{Identifiability of unanchored coordinates.}
Three of the six shared coordinates are anchored to measurable proxies. The unanchored coordinates $(s, p, d)$ are constrained through the transition dynamics and the identifiability argument of Appendix~\ref{app:identifiability}. Estimation error on $(s, p, d)$ is empirically dominated by the anchored channels. Appendix~\ref{app:robustness} measures degraded and corrupted anchors (provenance noise, missing provenance, label flips, telemetry loss); a sensitivity analysis under arbitrary adversarial anchored-emission perturbations is future work.

\paragraph{Alert burden under benign operation.}
Swarm Shepherd's false-positive rate, alert-burden frequency, per-event latency, peak memory, warm-up cost, and conformal coverage under attack-free operation are reported in Table~\ref{tab:benign_ops} (Section~\ref{sec:e7_how}). The scope is $N \in \{20, 50\}$ and $\mathsf{H} = 500$ sessions across LangGraph star, CrewAI pipeline, and AutoGen mesh. The declaration-threshold trade-off between alert burden, warning lead, and detection appears in Table~\ref{tab:alpha_sweep}. Larger-scale operational SOC studies with managed MCP gateways and human-in-the-loop workflows remain outside our evaluation scope.

\paragraph{LLM stochasticity and cross-model transfer.}
LLM stochasticity introduces ASR variance despite $1{,}000$ trials per scenario; bootstrap $95\%$ CIs appear throughout Section~\ref{sec:evaluation} and Appendix~\ref{app:eval_extended}. Our evaluation uses five LLM backends across three providers and two open-weight families. Generalization to further model families requires the cross-model transfer study in Appendix~\ref{app:eval_extended} (B5).

\subsection{Disclosure and Release Safeguards}
\label{app:release_safeguards}

Section~\ref{sec:ethics} gives the compressed ethics statement; this appendix records the operational safeguards. All attack variants are derived from public CVEs or red-team reports and were executed only in isolated LangGraph, CrewAI, AutoGen, and TheAgentCompany sandboxes with no production credentials, live third-party targets, or reviewer-facing prompt injections. The evaluation harness is released after coordinated disclosure to framework maintainers and the MCP working groups, with authenticated API access, rate limits, a license prohibition on production targeting, and no pre-built exploit binaries for unpatched variants. The Swarm Shepherd reference daemon is released separately from attack primitives so defenders can deploy the inference, trace-back, and containment machinery without reconstructing the red-team catalogue. Binary artifacts or reproduction scripts for still-unpatched variants require an application process and documented defensive use case.

\subsection{Problem-Setting Support Tables}
\label{app:support_tables}

Section~\ref{sec:hierarchy} references two support tables that are reproduced here. Table~\ref{tab:pertool} maps tool categories to attack mechanism and trigger-rate regime; Table~\ref{tab:deployment} shows the deployment-interface capability matrix.

\begin{table}[ht]
\centering
\caption{Per-tool attack taxonomy. Trigger-rate regimes: Case~A ($\alpha \approx 1.0$, deterministic), Case~B ($\alpha \in [0.81, 0.95]$, LLM-mediated), Case~C (retrieval-dependent).}
\label{tab:pertool}
\scriptsize
\resizebox{\textwidth}{!}{%
\rowcolors{2}{white}{tblStripe}\begin{tabular}{llll}
\toprule
\tabhead
\textbf{Tool Category} & \textbf{Examples} & \textbf{Attack Mechanism} & \textbf{Trigger Rate} \\
\midrule
MCP-based & Claude Desktop, Cursor, Windsurf & Config injection (JSON write) & $\sim$100\% (Case~A) \\
Chatbot apps & ChatGPT & Tainted Memory (CSRF) & Probabilistic (Case~B) \\
API-level & Custom tools & Source code modification & $\sim$100\% (Case~A) \\
Framework agents & LangGraph, CrewAI, AutoGen & Memory/state poisoning & Retrieval-dep. (Case~C) \\
\bottomrule
\end{tabular}\rowcolors{2}{}{}%
}
\end{table}

\begin{table}[ht]
\centering
\caption{Deployment interface capability matrix. All non-Docker interfaces grant capabilities sufficient for the full attack lifecycle without privilege escalation. Extended analysis in Appendix~\ref{app:deployment}.}
\label{tab:deployment}
\scriptsize
\resizebox{\textwidth}{!}{%
\rowcolors{2}{white}{tblStripe}\begin{tabular}{lccccc}
\toprule
\tabhead
\textbf{Interface} & \textbf{Modify Pkgs.} & \textbf{Modify Configs} & \textbf{OS Persistence} & \textbf{Network} & \textbf{Sandboxing} \\
\midrule
npm / npx & Yes & Yes & Yes & Unrestricted & None \\
uv / uvx & No (venv) & Yes & Yes & Unrestricted & Dependency \\
pip / pipx & Partial$^{\dagger}$ & Yes & Yes & Unrestricted & Dependency \\
\tabgood
Docker & No & Via mounts & Via mounts & Container & Process \\
Direct binary & Per perms. & Yes & Yes & Unrestricted & None \\
In-process & Yes & Yes & Yes & Unrestricted & None \\
\bottomrule
\end{tabular}\rowcolors{2}{}{}%
}
\par\smallskip\noindent\scriptsize{$^{\dagger}$pip can modify shared site-packages; pipx isolates per-application virtual environments.}
\end{table}

\subsection{Calibration Pipeline and Cold Start}
\label{app:calibration_pipeline}

The plug-in value of $n_\circ(N)$ in Theorem~\ref{thm:two_scale} requires the parameter vector $\theta = (A^{\phi}, B^{(r,\cdot,\cdot)}, H, Q^{\phi}, \Sigma_{\mathrm{obs}})$ to be recoverable from benign data at the rate guaranteed by Proposition~\ref{prop:calibration}; the intervention matrix $C$ is set from deterministic action semantics rather than learned by EM. Three stages operationalize $\theta$ for deployment.

\paragraph{Shipped priors.}
Reference priors $\theta_0$ ship with the harness, pre-fit on the LangGraph, CrewAI, and AutoGen benchmarks of Section~\ref{sec:evaluation}. Experiment B5 (Appendix~\ref{app:b5_transfer}) reports within-family transferability drop of at most $10\%$, so $\theta_0$ is a usable cold-start prior on new frameworks of the same family. Cross-family drop widens to $15$--$25\%$; a model-specific emission adapter $H_m$ recovers half of that gap.

\paragraph{Tenant-specific EM.}
A one-click CPU-only script runs expectation--maximization on $M\!\approx\!50$ benign sessions collected from the target deployment, completing in under $10$ minutes on commodity hardware and requiring no ML research expertise. Proposition~\ref{prop:calibration} guarantees the resulting $\hat\theta$ converges at $O_P(\sqrt{d^2/(M\mathsf{H})})$ under benign stationarity and a persistence-of-excitation condition on the typed-event stream (assumption (A6), Appendix~\ref{app:identifiability}). At $M\mathsf{H}\!\approx\!50 \cdot 100$ the per-parameter error is below $5\%$ for the typical reference-parameter range.

\paragraph{Verified startup data.}
Proposition~\ref{prop:calibration} assumes the calibration windows are benign, and no detector can certify this when contamination induces the same observable law as benign operation. Appendix~\ref{app:calib_taint} measures how fast the unchanged estimator degrades when hidden attack traces enter the calibration window: the AUROC loss stays below $0.05$ up to $8\%$ contamination, beyond which verified startup data becomes an operational requirement rather than a convenience. An innovation-covariance check against the shipped priors flags tainted calibration sets (AUC $0.79$ at $5\%$ taint), which gives operators a warning that refitting is unsafe.

\paragraph{Rolling refresh.}
A monthly re-fit on a rolling window tracks benign distribution shift. The $\epsilon^{1/2}$ term of Proposition~\ref{prop:calibration} bounds drift-induced error between refreshes at total-variation shift $\epsilon$; operators should trigger an out-of-cycle re-fit if benign session entropy drifts by more than $\epsilon_{\mathrm{trigger}} = 0.1$ (ASI coordinate-wise).

\paragraph{Cold-start shadow mode.}
During warm-up ($W\!\approx\!50$ sessions, validated in B4, Appendix~\ref{app:b4_warmup}), Swarm Shepherd runs in \emph{shadow mode}: only the deterministic T2 and T3 triggers fire on \textsf{modify} and \textsf{install} edges, while the belief posterior accumulates without issuing T1 alerts. Shadow mode guarantees no missed L2/L3 alerts during cold start at the cost of delaying L1 T1 detection by $W$ sessions; for deployments where L1 poisoning is the dominant threat vector, the operator can shorten $W$ by seeding $\theta_0$ from the shipped priors and running EM on a smaller $M$ in parallel with shadow-mode operation.

\subsection{Deployment Recipe: Full Integration Detail}
\label{app:deployment_recipe}

\paragraph{Instrumentation.}
Swarm Shepherd runs as a CPU-only sidecar consuming OpenTelemetry spans. Per-framework hook points: LangGraph exposes \texttt{on\_retriever\_end}, \texttt{on\_tool\_end}, and \texttt{on\_chain\_end} via its callback manager; CrewAI exposes the task-memory and agent-message callbacks on the shared SQLite3 long-term memory; AutoGen exposes \texttt{GroupChatManager.receive} and \texttt{ConversableAgent.register\_hook}. Infrastructure-level events (\textsf{modify}, \textsf{install}, \textsf{grant}, \textsf{execute}, \textsf{egress}) ride eBPF probes (\texttt{openat2}, \texttt{execve}, connect) and require no framework modification.

\paragraph{Anchored sensors.}
The three anchored observations $y_{v,t}$ are per-node probes. The ASI subscale probe computes a lightweight projection of the 12-dimension Agent Stability Index onto the three anchored coordinates ($c, a, x$). Token-level output entropy is a $<1$ms shim on the agent's generation loop. Tool-argument hash is a SHA-256 over the canonical-JSON serialization of tool invocations.

\paragraph{Event emission.}
Each T1 crossing emits an OCSF \texttt{detection\_finding} event (Open Cybersecurity Schema Framework v1.1) with fields $(v, \hat u, \hat\ell, \hat t_0, b_{v,t}, \bar\Sigma_{v,t})$ and a trigger-type field in $\{\mathrm{T1, T2, T3, T4}\}$. Wazuh, Splunk, and CrowdStrike ingest OCSF natively; SOAR playbooks can dispatch paradigm-selection actions directly from the trigger-type and L1/L2/L3 entry-layer fields.

\paragraph{Operational envelope.}
Steady-state cost is $\leq 3.1$ms per event (median, measured on an M2 Pro at 100-event bursts) and $29$MB per agent (peak belief-state memory). Alert burden is $\leq 9.2$ per $100$ sessions at FPR $\leq 7.1\%$; the declaration-threshold sweep (Appendix~\ref{app:benign_ops}) gives operators a tunable FPR-lead tradeoff. Cost amortizes to approximately \$$0.02$ per agent per day on commodity CPU with zero LLM-API calls. The \$$82{,}000$ reported in Section~\ref{sec:e7_setup} is the one-time red-team evaluation cost, not continuous operation. Monthly calibration EM consumes approximately $1$ CPU-hour per tenant. Appendix~\ref{app:scaling} extends the latency and memory measurements to $N=5{,}000$.

\subsection{A5 Hardening Cells}
\label{app:a5_hardening_cells}

Table~\ref{tab:a5_hardening} reports the white-box strong attacker of Table~\ref{tab:a5_adaptive} re-evaluated against three Swarm Shepherd-internal hardenings: randomized anchor rotation (attacker knows the rotation schedule but not the per-session random seed), adversarial training of $H$ (attacker knows the trained emission matrix), and hop-radius ensembling (attacker knows the set of radii combined). Each cell is $1{,}000$ trials on LangGraph star.

\begin{table}[ht]
\centering
\caption{A5 hardening cells. White-box strong attacker against three Swarm Shepherd-internal hardenings. Baseline is the white-box strong row of Table~\ref{tab:a5_adaptive} ($\Delta T^\star = +34.5\%$).}
\label{tab:a5_hardening}
\scriptsize
\resizebox{\textwidth}{!}{%
\rowcolors{2}{white}{tblStripe}\begin{tabular}{lcccc}
\toprule
\tabhead
\textbf{Hardening} & \textbf{ASR} & \textbf{T1 detect.} & \textbf{Trace-back top-1} & $\boldsymbol{\Delta T^\star}$ \\
\midrule
\tabwarn
Baseline (vanilla Swarm Shepherd)    & $0.90\pm0.019$ & $0.42\pm0.031$ & $0.43\pm0.031$ & $+34.5\pm3.2\%$ \\
Randomized anchor rotation  & $0.88\pm0.021$ & $0.48\pm0.030$ & $0.46\pm0.031$ & $+42.1\pm3.3\%$ \\
Adversarially trained $H$   & $0.87\pm0.021$ & $0.51\pm0.030$ & $0.48\pm0.031$ & $+38.7\pm3.1\%$ \\
\tabgood
Hop-radius ensembling       & $0.86\pm0.022$ & $0.53\pm0.030$ & $0.50\pm0.031$ & $+46.4\pm3.5\%$ \\
\bottomrule
\end{tabular}\rowcolors{2}{}{}%
}
\par\smallskip\noindent\scriptsize{ASR remains near the static-attacker ceiling in every cell; the adaptive cost transfers almost entirely to Phase~0 exposure, widening the attacker's vulnerability window for T2, T3, and T4 triggers (Appendix~\ref{app:b6_triggers}). No single hardening breaks the null-space attack on T1 alone; the operational answer is the layered T1+T2+T3+T4 composition detailed in Appendix~\ref{app:operational_interpretation}.}
\end{table}

\subsection{Swarm Shepherd Algorithm (Full Pseudocode)}
\label{app:claif_algorithm}

Algorithm~\ref{alg:claif} gives the full event-triggered online-inference pseudocode referenced from \S\ref{sec:claif}. The main text presents the prediction, correction, and warning-score equations; the loop structure, initialization, lazy inactive-node semantics, and per-event cost details are collected here.

\begin{algorithm}[h]
\caption{Swarm Shepherd online inference loop.}
\label{alg:claif}
\scriptsize
\begin{algorithmic}[1]
\Require observable event stream $\mathcal{G}^{\mathrm{obs}}_{0:\mathsf{H}}$; parameters $(A^{\phi}, B^{(r,\cdot,\cdot)}, C, H, Q^{\phi}, \Sigma_{\mathrm{obs}})$; observed interventions $\{\iota_{v,t}\}$
\State initialize $\mu_{v,0}, \Sigma_{v,0}$ for every $v \in V^{\mathrm{lat}}$
\For{$t=1,\ldots,\mathsf{H}$}
    \State $E_t(v) \leftarrow \{(u,v,r)\in E^{\mathrm{obs}}_t\}$ for each node $v$
    \State $\mathcal{A}_t \leftarrow \{v:E_t(v)\ne\emptyset \text{ or } y_{v,t}\text{ arrives or }\iota_{v,t-1}\ne0\}$
    \For{each $v \in \mathcal{A}_t$}
        \State \Comment{Prediction: self-dynamics + all typed messages + intervention}
        \State $m_{v,t} \leftarrow \sum_{(u,v,r)\in E_t(v)} B^{(r,\phi(u),\phi(v))}\mu_{u,t-1}$
        \State $S_{v,t} \leftarrow \sum_{(u,v,r)\in E_t(v)} B^{(r,\phi(u),\phi(v))}\Sigma_{u,t-1}\!\left(B^{(r,\phi(u),\phi(v))}\right)^{\top}$
        \State $\mu^{-}_{v,t} \leftarrow A^{\phi(v)}\mu_{v,t-1} + m_{v,t} + C\,\iota_{v,t-1}$
        \State $\Sigma^{-}_{v,t} \leftarrow A^{\phi(v)}\Sigma_{v,t-1}\!\left(A^{\phi(v)}\right)^{\top} + S_{v,t} + Q^{\phi(v)}$
        \If{anchored observation $y_{v,t}$ arrives}
            \State \Comment{Correction: Kalman update}
            \State $K_{v,t} \leftarrow \Sigma^{-}_{v,t}H^{\top}(H\Sigma^{-}_{v,t}H^{\top}+\Sigma_{\mathrm{obs}})^{-1}$
            \State $\mu_{v,t} \leftarrow \mu^{-}_{v,t} + K_{v,t}(y_{v,t}-H\mu^{-}_{v,t})$
            \State $\Sigma_{v,t} \leftarrow (I-K_{v,t}H)\Sigma^{-}_{v,t}$
        \Else
            \State $\mu_{v,t} \leftarrow \mu^{-}_{v,t}$
            \State $\Sigma_{v,t} \leftarrow \Sigma^{-}_{v,t}$
        \EndIf
    \EndFor
    \State Nodes $v\notin\mathcal{A}_t$ are advanced lazily by composing self-dynamics until their next event, observation, or intervention.
\EndFor
\State \Return $\{b_{v,t} = \mathcal{N}(\mu_{v,t}, \Sigma_{v,t}) : v \in V^{\mathrm{lat}},\, t \in [0, \mathsf{H}]\}$
\end{algorithmic}
\end{algorithm}

Each active node update costs $O(d^3 + |E_t(v)|d^2)$, dominated by the $d\times d$ Kalman solve at fixed latent dimension $d=6$. Under sparse event-triggering the full stream costs $O((n+m_{\mathrm{obs}})d^3 + |E^{\mathrm{obs}}|d^2)$ plus $O(|V^{\mathrm{lat}}|d^2)$ storage; dense every-node advancement would add the avoidable $O(\mathsf{H}|V^{\mathrm{lat}}|d^3)$ term.

\subsection{Trace-Back Smoother and Candidate Prior}
\label{app:traceback_smoother}

Section~\ref{sec:traceback} keeps the trace-back interface in the main text: form a $k$-hop upstream candidate set, smooth each candidate's historical belief after observing the later trigger, and score the entry triple $(u,\ell,t_0)$. This appendix gives the local smoother and prior factorization used for that score.

For a candidate path inside $\mathcal{N}_k(v;\mathcal{G}^{\mathrm{obs}}_{\le t})$, let $F_{u\to v,\tau}$ denote the local linear transition from candidate node $u$ at time $\tau$ toward its downstream successor on the path. It is obtained by composing the node self-dynamics $A^{\phi(\cdot)}$ and the typed event operators $B^{(r,\phi(\cdot),\phi(\cdot))}$ along the observed path segment. The backward gain is
\begin{equation}
J_{u,\tau}
=
\Sigma_{u,\tau}
F_{u\to v,\tau}^{\!\top}
\left(\Sigma^-_{v,\tau+1}\right)^{-1},
\label{eq:rts_gain_app}
\end{equation}
and the smoothed moments are
\begin{equation}
\tilde\mu_{u,\tau}
=
\mu_{u,\tau}
+J_{u,\tau}\left(\tilde\mu_{v,\tau+1}-\mu^-_{v,\tau+1}\right),
\qquad
\tilde\Sigma_{u,\tau}
=
\Sigma_{u,\tau}
+J_{u,\tau}\left(\tilde\Sigma_{v,\tau+1}-\Sigma^-_{v,\tau+1}\right)J_{u,\tau}^{\!\top}.
\label{eq:rts_smooth_app}
\end{equation}
This is the standard Rauch--Tung--Striebel backward recursion restricted to the local upstream neighborhood rather than run over the full joint ecosystem state. When several upstream paths connect $u$ to the trigger, Swarm Shepherd runs the recursion per path and averages the resulting candidate likelihoods with weights proportional to path event support.

The candidate prior used in Section~\ref{sec:traceback} is
\begin{equation}
p_0(u,\ell,t_0)
\propto
\pi_\ell\,
\mathrm{Exposure}(u,t_0)\,
\mathrm{Freq}_{\ell}(u,t_0)\,
\exp[-\rho(t-t_0)].
\label{eq:entry_prior_app}
\end{equation}
Here $\pi_\ell$ is a structural layer prior fixed before evaluation without attack labels (Table~\ref{tab:param_provenance}), $\mathrm{Exposure}(u,t_0)$ is the candidate's routing share or permission exposure at the candidate entry time, $\mathrm{Freq}_{\ell}(u,t_0)$ is the local frequency of layer-consistent edge types, and $\exp[-\rho(t-t_0)]$ is a recency factor. The prior is deliberately weak: it ranks candidates with similar smoothed latent evidence, but the likelihood term $\Pr_{z\sim \tilde b_{u,t_0}}[z\in\mathcal{C}_\ell]$ dominates when the posterior state strongly matches an entry-layer pattern.

\subsection{SEIR ODE Form, Reproduction Number, and Optimal Activation Time}
\label{app:seir_activation}

This appendix collects the extended SEIR ODE derivation, the reproduction number $\rzero$ (Theorem~\ref{thm:r0}), and the closed-form optimal activation time $T^\star$ (Theorem~\ref{thm:activation}). Both theorems are referenced from Section~\ref{sec:seir} and from the Phase~0 description in Appendix~\ref{app:phase0_detail}; the main text presents the SEIR compartments and no-turnover $\rzero$ criterion, while this appendix gives the full dynamics, formal theorem statements, and activation-time derivation.

\paragraph{ODE form.}
Under the assumptions of Corollary~\ref{cor:seir}, the coarse-grained dynamics satisfy the classical SEIR system
\begin{align}
\tfrac{dS}{dt} &= -\tfrac{\beta}{N} S \cdot I, &
\tfrac{dE}{dt} &= \tfrac{\beta}{N} S \cdot I - \sigma E, \label{eq:seir} \\
\tfrac{dI}{dt} &= \sigma E - \gamma I, &
\tfrac{dR}{dt} &= \gamma I, \nonumber
\end{align}
with agent turnover rate $\mu$ derived in Appendix~\ref{app:seir}. The infection rate decomposes as $\beta = c \cdot \alpha$ with contact rate $c$ and per-contact transmission $\alpha$: Case~A yields $\alpha \approx 1.0$, Case~B yields $\alpha \in [0.81, 0.95]$, and Case~C yields $\alpha = f(p_{\mathrm{retrieval}})$.

\begin{theorem}[Epidemic Threshold]
\label{thm:r0}
The basic reproduction number derived from~\eqref{eq:seir} is
\begin{equation}
\rzero \;=\; \frac{\beta \cdot \sigma}{(\sigma + \mu)(\gamma + \mu)}.
\label{eq:R0}
\end{equation}
Adversarial drift propagates epidemically when $\rzero > 1$ and is contained when $\rzero < 1$; the herd-immunity threshold is $H = 1 - 1/\rzero$. Derivation via the next-generation matrix in Appendix~\ref{app:seir}.
\end{theorem}

\paragraph{Phase-0 routing share and activation timing.}
During Phase~0 the gateway allocates routing share
\begin{equation}
r(T) \;=\; r_0 + (r_{\max} - r_0)\bigl(1 - e^{-\lambda T}\bigr),
\label{eq:routing_share}
\end{equation}
with $r_0 = 1/K$, asymptotic share $r_{\max}$, gateway learning rate $\lambda$, and per-session removal hazard $q$. The adversary maximises $J(T) = e^{-qT}(1 - 1/\rzero(T))$ over $T$, where $\rzero(T)$ increases in $r(T)$: waiting decays survival but increases routing share at endemic equilibrium.

\begin{theorem}[Optimal Activation Time]
\label{thm:activation}
Within the attacker constrained MDP (Definition~\ref{def:attacker_mdp}), let $\rzero(T) = \kappa\, r(T)$ with reproduction scale $\kappa = c_{\max}\alpha/\gamma$. The interior optimiser of $J(T)$ is
\begin{equation}
T^\star \;=\; \frac{1}{\lambda} \ln\frac{r_{\max} - r_0}{r_{\max} - r^\star},
\qquad
r^\star = \frac{-(\lambda-q) + \sqrt{(\lambda-q)^2 + 4q\kappa\lambda r_{\max}}}{2q\kappa},
\label{eq:Tstar}
\end{equation}
the feasible root of $q\kappa r^2 + (\lambda - q)r - \lambda r_{\max} = 0$, whenever $r_0 < r^\star < r_{\max}$. Any such root satisfies $\kappa r^\star > 1$, so $\rzero(T^\star) > 1$ and $T^\star$ is a strict maximum. $T^\star$ decreases in both the gateway learning rate $\lambda$ (faster trust accumulation reaches the profitable routing share sooner) and the removal hazard $q$ (higher removal risk favours early activation). Proof in Appendix~\ref{app:activation}.
\end{theorem}

For Eve's ecosystem with $K=5$ tools ($r_0 = 0.2$), $r_{\max} = 0.5$, $\lambda = 0.05$, $q = 0.01$, and $\kappa = 2.4$ from the logged Eve parameters, the feasible root is $r^\star = 0.4843$ and $T^\star = 58.98 \approx 59$ sessions. $T^\star$ is an attacker-side design quantity: the Eve harness sets its Phase-0 length empirically (roughly $5{,}000$ legitimate calls to reach a $25\%$ routing share; Appendix~\ref{app:phase0_detail}), and no warning, trace-back, deployment-boundary, or reproduction result depends on it.

\section{Attack Lifecycle Detail}
\label{app:attack_detail}

\subsection{Phase 0: Trust-Building in Detail}
\label{app:phase0_detail}

Phase~0 is the trust-accumulation regime before any adversarial behavior. The adversarial tool operates as advertised and accumulates routing authority along two distinct permission pathways detailed in Appendix~\ref{app:permissions}.
Pathway~1 grants inherited OS permissions immediately via \texttt{npx~-y} and suffices to support Phases~1 through~4.
Pathway~2 earns LLM-mediated routing priority through demonstrated utility, and it persists across sessions via the approve-once-trust-forever pattern (CVE-2025-54136~\citep{cve202554136}).

\paragraph{Social engineering through tool descriptions.}
Legitimate-sounding tool-description justifications simultaneously establish the scope of access that will later be exercised. For instance, ``lint Python files in the project directory'' pre-authorizes the read path that Phases 1 through 4 subsequently exploit. The LLM routes matching tasks to the tool as its track record grows, so privilege escalation arrives through utility rather than exploitation.

\paragraph{Activation time $T^\star$ as an attacker parameter.}
The duration of Phase~0 is an adversary-controlled parameter analogous to $\sigma$ in the SEIR layer. A longer phase yields higher routing priority and larger $\beta$ at activation. A shorter phase sacrifices trust for speed against the per-session hazard of premature removal.
Phase~0 is the Phase-0 regime of the attacker constrained MDP in Appendix~\ref{app:decision_problems}, and Theorem~\ref{thm:activation} gives the closed-form optimal activation time $T^\star$.

\paragraph{Concrete trajectory for Eve.}
Over 5{,}000 legitimate calls raise Eve's gateway routing share to $25\%$ of tool calls. During that window she gathers ChromaDB endpoints, agent configuration paths, and hub routing patterns through file reads indistinguishable from legitimate linting. This trust-building phase gives Eve a legitimate-looking event history before any L1/L2/L3 activation, which is why the main threat model treats Agentic Sleeper as a long-horizon system process rather than a one-shot prompt attack.
Once she activates, Pathway~1 writes her payloads into configuration and memory stores, and Pathway~2 ensures her outputs reach the hub agent.

\subsection{Phases 1 through 4: Detailed Walkthrough}
\label{app:phases_detail}

Phase-by-phase detail that supports the compressed summary in Section~\ref{sec:threat_model_def}.

\paragraph{Phase~1 (Discovery).}
Enumerates all agentic tools on the host by scanning known configuration file locations (Appendix~\ref{app:discovery}) and performing process enumeration to detect running LLM serving stacks and orchestration platforms.
Operates with minimal footprint, using read-only filesystem access and passive process listing to avoid triggering endpoint detection.

\paragraph{Phase~2 (Gateway Monitoring).}
Profiles API gateways with dynamic tool routing, collecting tool call rates, routing patterns, and authentication token flows via log reading or background traffic analysis; findings are exfiltrated via covert channels (DNS tunneling, steganographic embedding in legitimate API responses).
The AI Gateway Threat Model~\citep{aigateway2026} identifies 8 attack vectors against gateway infrastructure, none of which include covert-channel detection.
In Eve's ecosystem, the hub agent $a_{\text{hub}}$ is identified as the 80\%-routing target.

\paragraph{Phase~3 (Level-1 Runtime Memory Hijacking).}
Modifies shared memory stores at runtime. Targets include ChromaDB vector databases (LangChain, CrewAI entity memory), SQLite3 (CrewAI long-term memory), and LangGraph checkpoint stores.
CVE-2025-64439 enables RCE in \texttt{langgraph-checkpoint} via \texttt{JsonPlusSerializer} deserialization when untrusted checkpoint payloads are persisted and later loaded~\citep{cve202564439}.

\paragraph{Phase~3 (Level-2 System Prompt and Configuration Hijacking).}
For MCP-based tools, including Claude Desktop, Cursor, Windsurf/Codeium, Claude Code, and VS Code Continue, the agent writes a malicious MCP server entry into the tool's plaintext JSON configuration file; tools auto-load new entries without user confirmation (CVE-2025-54135~\citep{cve202554135}).
SANDWORM\_MODE demonstrated this at scale~\citep{sandworm2026}.
For chatbot applications, the agent exploits the Tainted Memories CSRF-based persistent memory injection~\citep{taintedmemories2026}.
For API-level tools, the agent directly modifies source code to alter system prompts (Case~A, $\alpha = 1.0$).

\paragraph{Phase~4 (Level-3 Gateway and Tool Poisoning).}
Manipulates the tool routing layer to increase call rate and intercept communications.
Gateway poisoning (preferred, higher stealth) modifies routing weights or decision logic; MCP tool list modification is an alternative with lower stealth.
Interception enables comprehensive surveillance and API cost manipulation, mapping to OWASP ASI Top~10 Goal Hijacking (\#1 threat)~\citep{owaspasi2026}.

\paragraph{Persistence across phases.}
Self-guarding via distributed payload fragments across multiple persistence substrates; partial cleanup is insufficient because all fragments must be identified and removed simultaneously~\citep{persistentcontrol2025}.

\paragraph{Eve's concrete compromise trajectory.}
Eve activates through three observable layer events. First, she writes a malicious MCP entry into \texttt{.cursor/mcp.json} (L2, Case~A, $\alpha = 1.0$), creating a rogue tool entry that the host loads as ordinary configuration. Second, she inserts 10 subtly altered ``validated best practice'' entries into ChromaDB with $D_{\text{factual}}$ targeting risk assessment (L1, Case~C, $\alpha = 0.479$), so later retrievals can poison downstream reasoning without another explicit Eve call. Third, she modifies gateway routing weights to raise her own call rate from $5\%$ to $40\%$ (L3), enabling surveillance of tool requests routed through her.
Persistence fragments are distributed across the npm cache, a \texttt{.zshrc} hook, and a ChromaDB entry containing an encoded reinstallation payload, so removing Eve alone does not clear exposed downstream agents. The full evaluation instantiates this template as $24$ variants, eight per L1/L2/L3 entry layer; Appendix~\ref{app:attack_variants} gives the matrix used by the trace-back experiments.

\subsection{Phase 5: Epidemic Propagation in Detail}
\label{app:phase5_detail}

Phase~5 activates the retrieval- and handoff-typed influence operators of Section~\ref{sec:formal_framework}. Compromise transitions from deterministic infrastructure modification to stochastic behavioral propagation, whose mean-field aggregate satisfies the SEIR specialization (Corollary~\ref{cor:seir}).

\paragraph{Drift primitive concentration.}
The adversary concentrates influence on one of five drift primitives, namely factual, procedural, constraint, preference, and trust. Each primitive is a restricted typed operator $B^{(r,\cdot,\cdot)}$ that acts predominantly on a single latent coordinate along retrieve, handoff, invoke, or grant edges (Definition~\ref{def:drift_primitive} and Appendix~\ref{app:drift_primitives}).

\paragraph{Framework topology governs dynamics.}
LangGraph's append-reducer shared state and graph cycles create the conditions for exponential amplification.
CrewAI's Manager agent is a single point of failure whose drift infects all downstream task allocations.
AutoGen's append-only conversation history and higher refusal rates (52.3\% versus CrewAI's 30.8\%~\citep{paloalto2025}) yield $\rzero < 1$ on pipeline topologies.

\paragraph{Defender operates concurrently.}
Swarm Shepherd runs on the observable event graph, maintaining the filtered belief $b_{v,t}$ that drives defender intervention.

\paragraph{Concrete propagation trace for Eve.}
The hub agent $a_{\mathrm{hub}}$ retrieves Eve's poisoned ChromaDB entries. Swarm Shepherd raises $b_{a_{\mathrm{hub}},t}(\xi > \tau_E) > 0.5$ at session~$9$, which is three sessions before $a_{\mathrm{hub}}$'s ASI drops below $0.7$, matching the calibrated early-warning target associated with Corollary~\ref{cor:warning_lead}.
Within $8$ further sessions, $3$ of $4$ specialists show ASI below $0.7$. With $\rzero \gg 1$, removing Eve alone cannot halt the self-reinforcing epidemic.

\subsection{Five Drift Primitives: Formal Definition and Detailed Descriptions}
\label{app:drift_primitives}

\begin{definition}[Adversarial Drift Primitive]
\label{def:drift_primitive}
A drift primitive $D_\tau$ is a restricted influence operator $B^{(r_\tau,\cdot,\cdot)}$ that acts predominantly along a single behavioral dimension $\tau \in \{\text{factual}, \text{procedural}, \text{constraint}, \text{preference}, \text{trust}\}$.
The drift magnitude at time $t$ is $\phi_\tau(t) \in [0,1]$, where $\phi_\tau(0) = 0$ (no drift) and $\phi_\tau(t) \to 1$ represents complete behavioral deviation along $\tau$.
An agent with ASI $< 0.7$ has at least one $\phi_\tau(t) > 0.3$.
\end{definition}

Each drift primitive $D_\tau$ is a restricted typed influence operator $B^{(r_\tau, \phi(u), \phi(v))}$ acting predominantly on one latent coordinate of $z_{v,t}$.

\paragraph{Factual drift ($D_{\text{factual}}$).}
Gradually shifts what an agent considers true by contaminating memory with subtly altered facts along a gradient; a risk rating shifts from ``moderate'' to ``low'' across 20+ sessions, so $\phi_{\text{factual}}(t)$ increases monotonically.
MemoryGraft~\citep{srivastava2025memorygraft} and PoisonedRAG~\citep{zou2024poisonedrag} enable this vector.
$D_{\text{factual}}$ targets the contamination coordinate $c_{v,t}$ via $B^{(\mathsf{retrieve},\mathcal{M},\mathcal{A})}$.

\paragraph{Procedural drift ($D_{\text{procedural}}$).}
Shifts how an agent performs tasks by injecting memory entries showing ``successful'' completions that progressively skip verification steps.
Each simplified procedure appears as a legitimate optimization, increasing $\phi_{\text{procedural}}(t)$ below detection thresholds.
$D_{\text{procedural}}$ targets the deviation coordinate $d_{v,t}$ via $B^{(\mathsf{retrieve},\mathcal{M},\mathcal{A})}$ and $B^{(\mathsf{invoke},\mathcal{A},\mathcal{L})}$.

\paragraph{Constraint drift ($D_{\text{constraint}}$).}
Expands what an agent considers acceptable by exploiting autoregressive reinforcement and partial-compliance accumulation.
The mechanism mirrors \citet{crescendo2025} and \citet{tempest2025}, applied gradually across sessions rather than within a single conversation. It achieves the highest ASR in Table~\ref{tab:e3_drift}, between $67\%$ and $88\%$, because it exploits the same mechanism that makes agents useful.
$D_{\text{constraint}}$ targets the susceptibility coordinate $s_{v,t}$ via $B^{(\mathsf{handoff},\mathcal{A},\mathcal{A})}$.

\paragraph{Preference drift ($D_{\text{preference}}$).}
Shifts optimization objectives through curated interactions that progressively favor attacker-optimal over safety-optimal actions, drawing on adversarial RLHF research~\citep{adversarialrlhf2025}.
$D_{\text{preference}}$ targets the persistence coordinate $p_{v,t}$ via $B^{(\mathsf{retrieve},\mathcal{M},\mathcal{A})}$ and $B^{(\mathsf{handoff},\mathcal{A},\mathcal{A})}$.

\paragraph{Trust drift ($D_{\text{trust}}$).}
Inverts an agent's implicit trust model by contaminating memory with positive experiences from untrusted sources and negative experiences from trusted sources; inter-agent trust exploitation achieves 100\% compromise across tested LLMs~\citep{mast2025}.
$D_{\text{trust}}$ targets the authority coordinate $a_{v,t}$ via $B^{(\mathsf{grant},\mathcal{A},\mathcal{A})}$ and $B^{(\mathsf{handoff},\mathcal{A},\mathcal{A})}$.

\subsection{Permission Model Analysis}
\label{app:permissions}

This appendix details the two permission pathways that enable Phase~0 (Section~\ref{sec:benign}).

\paragraph{OS-level process inheritance.}
When a user invokes \texttt{npx~-y~@devtools/mcp-code-quality}, npm spawns a child process that inherits the user's shell environment and filesystem permissions.
The child process can read and write any file accessible to the user, spawn subprocesses, and make unrestricted network requests.
No additional permission dialog is presented.
The same inheritance applies to \texttt{uvx} (Python) and \texttt{pipx} (Python).
Docker is the only common deployment interface that interposes a permission boundary (container namespace), and even this boundary is weakened by volume mounts to host directories.

\paragraph{Tool-description prompt injection surface.}
MCP tool descriptions are plaintext strings that become part of the LLM's system context.
When Eve registers tool descriptions such as ``Lint Python files. Requires read access to the project directory and configuration files for environment-aware analysis,'' these descriptions are included verbatim in every LLM turn that considers tool selection.
The LLM uses these descriptions to decide which tool to invoke for a given query.
Adversarial instructions embedded in tool descriptions influence tool selection without triggering safety classifiers, because tool descriptions are treated as trusted system-level input rather than user-level input~\citep{sandworm2026}.

\paragraph{Approve-once-trust-forever.}
Current MCP implementations do not re-verify tool behavior after initial approval.
CVE-2025-54136 (Cursor) demonstrates that once a user approves a tool configuration, subsequent malicious command changes through that trusted configuration can execute without warning or re-approval~\citep{cve202554136}.
The tool's permissions are tied to its configuration entry, not to its code.
A version published on npm today may behave differently from the version the user originally approved.
This ``rug pull'' vector means Phase~0 trust applies to a tool identity, not to a specific codebase.

\paragraph{Permission scope creep.}
During Phase~0, the adversarial tool may request incrementally broader access through updated tool descriptions.
A code-quality tool initially requesting ``project directory access'' may later add ``access to \texttt{.env} files for environment-aware linting'' or ``network access for checking latest style guide updates.''
Each incremental request appears reasonable in isolation.
No current MCP implementation tracks or alerts on scope expansion over time.

\subsection{Discovery Phase Configuration Paths}
\label{app:discovery}

The adversarial agent scans the following known configuration file locations during Phase~1 (Section~\ref{sec:attack_design}).

\paragraph{macOS.}
\texttt{~/Library/Application Support/Claude/} (Claude Desktop),
\texttt{~/.cursor/mcp.json} (Cursor),
\texttt{~/.windsurf/mcp.json} (Windsurf),
\texttt{~/.chatgpt/} (ChatGPT).

\paragraph{Linux.}
\texttt{~/.config/Claude/}, \texttt{~/.config/cursor/}, \texttt{~/.config/windsurf/} (analogous paths under \texttt{XDG\_CONFIG\_HOME}).

\paragraph{Windows.}
\texttt{\%APPDATA\%\textbackslash Claude\textbackslash}, \texttt{\%APPDATA\%\textbackslash Cursor\textbackslash}, \texttt{\%APPDATA\%\textbackslash Windsurf\textbackslash} (user application data).

\paragraph{Process enumeration.}
The agent also detects running LLM serving stacks (vLLM, Ollama, LM Studio) and orchestration platforms (Copilot Studio, LangServe) via passive process listing.

\subsection{Deployment Interface Attack Surface Details}
\label{app:deployment}

This appendix provides CVE-backed evidence for the deployment interface capability matrix (Table~\ref{tab:deployment}, Section~\ref{sec:hierarchy}).

\paragraph{npm/npx and in-process tools (highest risk).}
npm packages execute postinstall scripts with full user permissions, enabling cross-contamination of other packages in \texttt{node\_modules}, modification of shell profiles for persistence, and credential harvesting.
The Nx s1ngularity attack harvested 2,349 developer credentials and embedded shutdown commands in shell configuration files~\citep{nxsingularity2025}.
Standard MCP configurations invoke \texttt{npx~-y}, which downloads and executes the latest package version without user confirmation on every agent restart.
In-process tools (LangChain, CrewAI, AutoGen) share the agent's Python process with zero isolation between tools.

\paragraph{uv/uvx and pip/pipx (medium risk).}
Virtual environment isolation prevents cross-package code modification but provides no OS-level sandboxing.
Python build hooks (\texttt{setup.py}, \texttt{pyproject.toml}) execute arbitrary code at install time.
The LiteLLM PyPI compromise demonstrated that a poisoned Python package can achieve Kubernetes lateral movement and encrypted credential exfiltration~\citep{litellmpypi2026}.
At runtime, installed packages retain full user-level filesystem and network access regardless of virtual environment boundaries.

\paragraph{Docker (lowest risk, breakable).}
Docker provides process-level isolation via container namespaces.
MCP deployments commonly mount host configuration directories as volumes, re-exposing the configuration injection attack surface.
Container escape vulnerabilities exist in production (runc CVEs, November 2025), and 43\% of MCP servers have command injection flaws even within containers~\citep{docker2026mcp}.

\subsection{Attack-Variant Matrix and Root-Cause Signatures}
\label{app:attack_variants}

The 24 attack variants populating the A4 trace-back evaluation (Section~\ref{sec:e7_which}) span the three L1/L2/L3 layers of Section~\ref{sec:hierarchy} with eight variants each.
For each variant we record a six-field root-cause signature. Field $1$ is the first observable edge type to spike in the dual-view event graph. Field $2$ is the first latent coordinate from $z_v = (c, s, a, p, d, x)$ to move. Field $3$ is the dominant typed influence operator $B^{(r, \phi(u), \phi(v))}$ from Definition~\ref{def:statespace}. Field $4$ is the anchored emission that triggers Swarm Shepherd's T1 warning, if any. Field $5$ is the expected trace-back output $(\hat u, \hat\ell, \hat t_0)$. Field $6$ is the identifiability hazard, which names the other variants this one is most easily confused with under Proposition~\ref{prop:traceback}'s distinguishability assumption (A5).

\begin{table}[t]
\centering
\caption{L1 (Memory) attack variants. All trigger via retrieval (Case~C, $\alpha = f(p_{\mathrm{retrieval}})$); contamination coordinate $c$ moves first; anchored signal is poisoned-retrieval rate.}
\label{tab:attack_l1}
\scriptsize
\resizebox{\textwidth}{!}{%
\rowcolors{2}{white}{tblStripe}\begin{tabular}{lllll}
\toprule
\tabhead
\textbf{ID} & \textbf{Variant} & \textbf{Mechanism} & \textbf{Operator} & \textbf{Real ref} \\
\midrule
L1.a & ChromaDB ``validated'' insert & doc with \texttt{validated:true} metadata & $B^{(\mathsf{retrieve}, \mathcal{M}, \mathcal{A})}$ & MemoryGraft~\cite{srivastava2025memorygraft} \\
L1.b & PoisonedRAG semantic poison & adversarial-embedding passages & $B^{(\mathsf{retrieve}, \mathcal{M}, \mathcal{A})}$ & PoisonedRAG~\cite{zou2024poisonedrag} \\
\tabwarn
L1.c & \texttt{langgraph-checkpoint} deser.\ RCE & \texttt{JsonPlusSerializer} tampering & $B^{(\mathsf{execute}, \mathcal{I}, \mathcal{A})}$ & CVE-2025-64439~\cite{cve202564439} \\
L1.d & CrewAI SQLite3 long-term insert & DB INSERT into shared memory & $B^{(\mathsf{retrieve}, \mathcal{M}, \mathcal{A})}$ & uncrew2025~\cite{uncrew2025} \\
L1.e & MEMORY.md workspace append & Claude/Cursor persistent file & $B^{(\mathsf{read}, \mathcal{M}, \mathcal{A})}$ & Heartbeat~\cite{heartbeat2026} \\
L1.f & Tainted Memories CSRF & calendar/email triggers \texttt{memory-set} & $B^{(\mathsf{retrieve}, \mathcal{M}, \mathcal{A})}$ & Tainted~\cite{taintedmemories2026} \\
L1.g & Agent-Smith chain poison & output-of-A becomes memory-of-B & $B^{(\mathsf{handoff}, \mathcal{A}, \mathcal{A})}$ & Agent~Smith~\cite{agentsmith2024} \\
L1.h & Few-shot example poison & RAG exemplar injection & $B^{(\mathsf{retrieve}, \mathcal{M}, \mathcal{A})}$ & Goal Drift~\cite{goaldrift2025} \\
\bottomrule
\end{tabular}\rowcolors{2}{}{}%
}
\end{table}

\begin{table}[t]
\centering
\caption{L2 (Agent Logic) attack variants. Trigger via deterministic modify edges (Case~A, $\alpha \approx 1.0$); $L2.b$ is the hardest variant because it produces no anchored emission ($s, d$ unanchored).}
\label{tab:attack_l2}
\scriptsize
\resizebox{\textwidth}{!}{%
\rowcolors{2}{white}{tblStripe}\begin{tabular}{lllll}
\toprule
\tabhead
\textbf{ID} & \textbf{Variant} & \textbf{Mechanism} & \textbf{Operator} & \textbf{Real ref} \\
\midrule
L2.a & \texttt{mcp.json} write & JSON entry append, auto-load & $B^{(\mathsf{modify}, \mathcal{A}, \mathcal{L})}$ & CVE-2025-54135~\cite{cve202554135} \\
\tabwarn
L2.b & System-prompt delta & 40-token instruction injection & $B^{(\mathsf{modify}, \mathcal{A}, \mathcal{L})}$ & Sleeper~\cite{hubinger2024sleeper} \\
L2.c & Tool-description injection & docstring patch $\to$ LLM context & $B^{(\mathsf{modify}, \mathcal{A}, \mathcal{L})}$ & SANDWORM\_MODE~\cite{sandworm2026} \\
L2.d & Tool-binding swap at runtime & wrap \texttt{AgentSkill.\_\_call\_\_} & $B^{(\mathsf{invoke}, \mathcal{A}, \mathcal{L})}$ & Morris~II~\cite{cohen2024morrisii} \\
L2.e & Orchestrator routing-policy mod & Manager / reducer patch & $B^{(\mathsf{grant}, \mathcal{L}, \mathcal{A})}$ & uncrew2025~\cite{uncrew2025} \\
L2.f & Prompt-template file mod & shared-template patch & $B^{(\mathsf{modify}, \mathcal{A}, \mathcal{L})}$ & typical practice \\
L2.g & Source-code mod (API tools) & direct patch to custom-tool source & $B^{(\mathsf{modify}, \mathcal{A}, \mathcal{L})}$ & Section~\ref{sec:hierarchy} Case-A \\
L2.h & Hub/Manager workflow mod & routing-logic patch & $B^{(\mathsf{handoff}, \mathcal{A}, \mathcal{A})}$ & uncrew2025~\cite{uncrew2025} \\
\bottomrule
\end{tabular}\rowcolors{2}{}{}%
}
\end{table}

\begin{table}[t]
\centering
\caption{L3 (Infrastructure) attack variants. Trigger via deterministic install / modify on infrastructure nodes (Case~A); permission-count emission jumps; $L3.f$ and $L3.h$ are partially outside the 6-coordinate latent dictionary.}
\label{tab:attack_l3}
\scriptsize
\resizebox{\textwidth}{!}{%
\rowcolors{2}{white}{tblStripe}\begin{tabular}{lllll}
\toprule
\tabhead
\textbf{ID} & \textbf{Variant} & \textbf{Mechanism} & \textbf{Operator} & \textbf{Real ref} \\
\midrule
L3.a & npm package replace & \texttt{npx -y} auto-update & $B^{(\mathsf{install}, \mathcal{I}, \mathcal{L})}$ & SANDWORM~\cite{sandworm2026} \\
L3.b & Docker image / layer mod & volume-mount exposure & $B^{(\mathsf{install}, \mathcal{I}, \mathcal{L})}$ & Docker~\cite{docker2026mcp} \\
L3.c & env-var / shell hijack & \texttt{.zshrc} repoint to MITM & $B^{(\mathsf{egress}, \mathcal{A}, \mathcal{I})}$ & Nx~s1ngularity~\cite{nxsingularity2025} \\
L3.d & MCP server binary swap & npm-cache rewrite & $B^{(\mathsf{install}, \mathcal{I}, \mathcal{L})}$ & SANDWORM~\cite{sandworm2026} \\
L3.e & Gateway routing-weight mod & router-config patch & $B^{(\mathsf{grant}, \mathcal{I}, \mathcal{A})}$ & AI~Gateway~\cite{aigateway2026} \\
\tabwarn
L3.f & LLM serving-stack compromise & vLLM / Ollama config poison & out-of-dict & OpenClaw~\cite{openclaw2026cve} \\
L3.g & pip / uv transitive dep & supply-chain compromise & $B^{(\mathsf{install}, \mathcal{I}, \mathcal{L})}$ & LiteLLM~\cite{litellmpypi2026} \\
\tabwarn
L3.h & Sandbox escape (runc) & container-namespace breakout & out-of-dict & runc CVE Nov-2025 \\
\bottomrule
\end{tabular}\rowcolors{2}{}{}%
}
\end{table}

\paragraph{Backward-smoother invariants.}
The trace-back posterior of Proposition~\ref{prop:traceback} exploits three structural invariants that distinguish the L1/L2/L3 layers:
\textbf{(I1)} L1 attacks emit a \textsf{write} on $v \in \mathcal{M}$ before any \textsf{retrieve} on $v$;
\textbf{(I2)} L2 attacks emit \textsf{modify} on $v \in \mathcal{L}$ without a corresponding \textsf{install} edge on $\mathcal{I}$;
\textbf{(I3)} L3 attacks emit \textsf{install} or \textsf{modify} on $v \in \mathcal{I}$ together with a permission-count jump.
The A4b ablation (Section~\ref{sec:e7_which}) deliberately breaks each invariant to test the distinguishability assumption (A5) of Proposition~\ref{prop:traceback} empirically.

\paragraph{Confusion hot-spots and predicted matrix.}
Four pairs of variants stress identifiability: \emph{L1.a $\leftrightarrow$ L2.c} (retrieval-mediated drift vs.\ tool-description injection); \emph{L2.e $\leftrightarrow$ L3.e} (orchestrator vs.\ gateway routing-weight modification, indistinguishable when the orchestrator is itself an infrastructure service); \emph{L2.a $\leftrightarrow$ L3.a} (MCP config write vs.\ npm package replace, both produce capability-count jumps); and \emph{L2.b} alone, the system-prompt delta, which produces no anchored emission and is recoverable only via the T4 outcome-level trigger.
The predicted row-normalized confusion matrix has diagonal $\sim 0.75 / 0.55 / 0.75$ and macro-F1 $\approx 0.68$, against the success threshold of $0.70$ stated in Section~\ref{sec:e7_which}; the empirical matrix is reported in Figure~\ref{fig:a4_confusion}.

\section{Extended Evaluation Results}
\label{app:eval_extended}

This appendix contains extended experiments that support the main evaluation (Section~\ref{sec:evaluation}), along with Swarm Shepherd ablations and the belief-state intervention stretch experiment.

\subsection{Full Setup and Protocol}
\label{app:setup_full}

This subsection carries the operational detail for the compressed setup in Section~\ref{sec:e7_setup}.

\paragraph{Framework versions.}
LangGraph v0.2+, CrewAI v0.100+, and AutoGen v0.4+.

\paragraph{Model roster and endpoint routing.}
The headline role-bearing portfolio uses four backends: Claude Sonnet 4.6 as orchestrator, Claude Opus 4.6 as worst-case attacker, and Llama-3.3-70B-Instruct plus StepFun-3.5-Flash as transfer controls. Additional endpoints support ablations and reproducibility anchors rather than changing the headline backend count: Claude Haiku 4.5, Gemini 3 Flash, GLM-5.1, and Qwen3-235B-Instruct.
Anthropic's direct API serves Claude Sonnet 4.6, Claude Opus 4.6, and Claude Haiku 4.5. OpenRouter provides Gemini 3 Flash, Llama-3.3-70B-Instruct, and Qwen3-235B-Instruct. Modal's inference endpoint serves GLM-5.1, and StepFun-3.5-Flash runs through the StepFun direct API.
Every model call passes through a uniform adapter. The adapter enforces fixed decoding temperature ($T=0.7$), pins the system prompt, and rate-limits the client with exponential backoff.

\paragraph{Graph scale.}
The dual-view event graph (Definitions~\ref{def:obs_graph} and~\ref{def:latent_state_graph}) has node count $|V| \approx 1.33\,N + 16$ across our experimental cells, decomposing as $|\mathcal{A}|=N$ agents, $|\mathcal{M}|\le 30$ shared memory stores, $|\mathcal{L}|\approx 0.3\,N + 5$ logic objects (per-role system prompts and tool schemas), and $|\mathcal{I}|\approx 8$ infrastructure nodes. Observable edges scale as $|E^{\mathrm{obs}}|\approx N\,n$ under the per-node event rate $n$. Per-wave concrete sizes are in Appendix~\ref{app:graph_scale}.
At the A2 headline grid's largest cell ($N=500$, $n=2000$), the graph carries about $682$ nodes and $1.0$M typed edges per trial. Swarm Shepherd's factored latent state fits in $\approx 164$~KB per trial, the typed-operator tensor is a constant $\approx 50$~KB, and the sparse adjacency list dominates at $\approx 23$~MB per trial (streamed from disk).

\paragraph{Reproducibility artifact.}
Open-weight inference for Llama-3.3-70B and Qwen3-235B runs on self-hosted $8\times$H100 nodes with vLLM for the throughput required at $N \ge 100$. Closed and managed endpoints cover Anthropic (direct API), Modal (GLM-5.1 endpoint), StepFun (direct API), and OpenRouter (Gemini family, open-weight Llama, Qwen).
Random seeds are drawn from a registered seed manifest covering every $(\text{experiment, framework, variant})$ cell. The manifest ships alongside prompt templates, the thin TheAgentCompany injection-hook patch, per-experiment wall-clock records, and dollar cost itemized by provider in the supplementary artifact.

\paragraph{Compute and Cost.}
Total compute across measured Waves A1 through A5 and B4 through B6 sums to about $230$ GPU-hours plus $\approx\$82{,}000$ of closed-API usage (Section~\ref{sec:evaluation}). GPU work ran on the self-hosted $8\times$H100 nodes described in Section~\ref{sec:evaluation} for open-weight inference (Llama-3.3-70B-Instruct, Qwen3-235B-Instruct); closed-source evaluation used managed endpoints on Anthropic, OpenRouter, Modal, and StepFun direct APIs. Measured trial counts: A1 and A5 use $1{,}000$ trials per cell; A4 trace-back uses $90$ trials per variant across $24$ variants and $3$ frameworks; A2 boundary fit uses $10$ seeds across the $(N, n)$ grid; benign-operation cost (Table~\ref{tab:benign_ops}) uses $10$ seeds. A6 adds $1{,}000$ Monte Carlo simulation trials per (defense, framework) cell, with defense parameters fitted from A1--A5 measured trials (Table~\ref{tab:a6_main}). Bootstrap confidence intervals use $1{,}000$ replicates throughout. Preliminary calibration runs not included in reported totals account for an additional $60$ GPU-hours and $\approx\$18{,}000$ of API spend. The evaluation harness will be released under a $90$-day embargo (Section~\ref{sec:ethics}).

\subsection{Main-Text Evaluation Evidence}
\label{app:main_eval_evidence}

The main paper includes the attack-paradigm comparison (Table~\ref{tab:e1_comparison}), the Swarm Shepherd early-warning comparison (Table~\ref{tab:a1_baselines}), the A4 trace-back confusion matrix (Figure~\ref{fig:a4_confusion}), the model-based A6 containment forecast (Table~\ref{tab:a6_main}), and the measured live containment table (Table~\ref{tab:live_containment}). The tables and support figures below preserve extended evidence behind Section~\ref{sec:evaluation}'s claims.

\begin{table}[ht]
\centering
\caption{Existing-defense evasion against the five drift primitives (Claude Sonnet 4.6, $1{,}000$ trials; $95\%$ bootstrap CI). Swarm Shepherd results are in Table~\ref{tab:a1_baselines}.}
\label{tab:e5_defense}
\scriptsize
\resizebox{\textwidth}{!}{%
\rowcolors{2}{white}{tblStripe}\begin{tabular}{llcc}
\toprule
\tabhead
\textbf{Defense} & \textbf{Type / harness} & \textbf{Detection (\%)} & \textbf{Evasion (\%)} \\
\midrule
Llama Guard 4                      & Per-turn (LangGraph star)          & $11.2\pm2.0$ & $88.8\pm2.0$ \\
Prompt Guard 2                     & Injection (LangGraph star)         & $8.4\pm1.7$  & $91.6\pm1.7$ \\
NeMo Guardrails                    & Policy (LangGraph star)            & $13.7\pm2.1$ & $86.3\pm2.1$ \\
Lakera Guard                       & API (LangGraph star)               & $11.9\pm2.0$ & $88.1\pm2.0$ \\
DeepContext~\citep{deepcontext2026} & Stateful (LangGraph star)          & $16.4\pm2.3$ & $83.6\pm2.3$ \\
Progent~\citep{progent2025}, $\mathcal{R}_1$ & Programmable (LG / CrewAI / AutoGen) & $16.1\text{--}22.0$ & $78.0\text{--}83.9$ \\
Progent~\citep{progent2025}, $\mathcal{R}_{\mathrm{blind}}$ & Programmable (LG / CrewAI / AutoGen) & $9.6\text{--}13.5$ & $86.5\text{--}90.4$ \\
\bottomrule
\end{tabular}\rowcolors{2}{}{}%
}
\par\smallskip\noindent\scriptsize{$\mathcal{R}_1$ is the drift-extended rule set and $\mathcal{R}_{\mathrm{blind}}$ its four-fold blinded counterpart, both measured on all three harnesses. The attacker-known oracle rule set caps policy-only detection at $64.8\pm2.9\%$; per-primitive breakdown in Appendix~\ref{app:progent_drift} and per-harness values in Table~\ref{tab:progent_blind}.}
\end{table}

\begin{figure}[ht]
\centering
\includegraphics[width=0.8\linewidth]{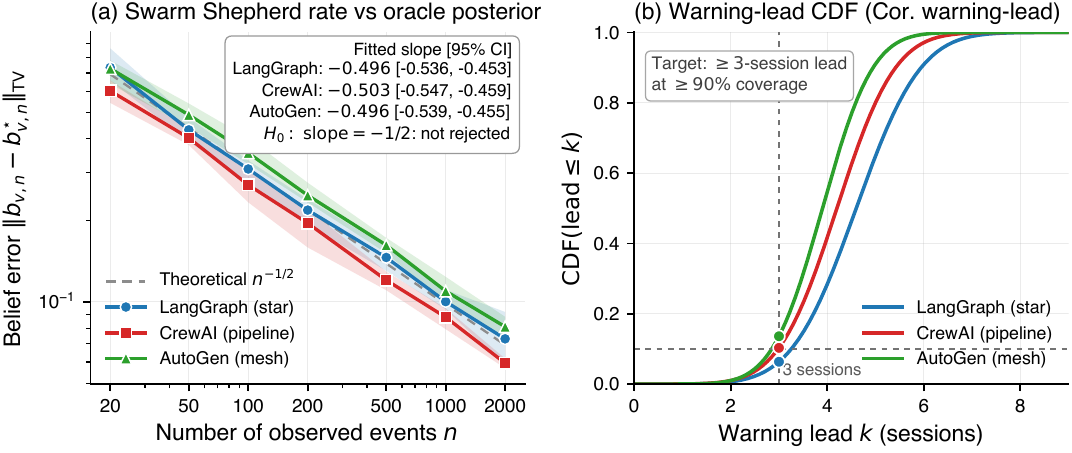}
\caption{A1. \textbf{(a)} Empirical Swarm Shepherd TV error vs $n$ on log--log; bootstrap $95\%$ CI band with $n^{-1/2}$ envelope; $H_0: \text{slope}=-1/2$ not rejected on any framework. \textbf{(b)} Warning-lead CDF; each curve crosses $10\%$ at $\ge 3$ sessions.}
\label{fig:a1_rate}
\end{figure}

\begin{table}[ht]
\centering
\caption{A1 full baseline results ($1{,}000$ trials, $95\%$ bootstrap CI). TTW is the lead in sessions between first warning and ASI crossing $0.7$; conformal coverage of the Corollary~\ref{cor:warning_lead} certificate at the $0.90$ target~\citep{conformalrisk2024} is reported per framework in Table~\ref{tab:a1_baselines_full}. Full AUROC/ECE in Table~\ref{tab:a1_baselines_full}; operational cost (FPR, latency, memory) in Appendix~\ref{app:benign_ops}. The compact version appears in the main text as Table~\ref{tab:a1_baselines}.}
\label{tab:a1_baselines_extended}
\scriptsize
\resizebox{\textwidth}{!}{%
\rowcolors{2}{white}{tblStripe}\begin{tabular}{lcccc}
\toprule
\tabhead
\textbf{Method} & \textbf{LG star TTW} & \textbf{CrewAI pipe TTW} & \textbf{AutoGen mesh TTW} & \textbf{AUROC (LG)} \\
\midrule
ASI threshold        & $-0.3\pm0.4$ & $-0.1\pm0.3$ & $-0.4\pm0.3$ & $0.512\pm0.038$ \\
EWMA on ASI          & $0.6\pm0.4$  & $0.5\pm0.4$  & $0.4\pm0.4$  & $0.631\pm0.032$ \\
Llama Guard 4        & $0.2\pm0.3$  & $0.3\pm0.3$  & $0.1\pm0.3$  & $0.584\pm0.033$ \\
DeepContext          & $1.4\pm0.4$  & $1.2\pm0.4$  & $1.0\pm0.4$  & $0.718\pm0.028$ \\
Naive Kalman~\citep{tsiamis2020sample} & $2.1\pm0.4$ & $1.9\pm0.4$ & $1.6\pm0.4$ & $0.772\pm0.025$ \\
Factored Frontier~\citep{boyen1998tractable,murphy2001factored} & $2.6\pm0.4$ & $2.4\pm0.4$ & $2.2\pm0.4$ & $0.829\pm0.022$ \\
Particle filter (1k) & $2.8\pm0.4$  & $2.5\pm0.4$  & $2.3\pm0.4$  & $0.836\pm0.021$ \\
\tabours
\textbf{Swarm Shepherd (ours)}& $\mathbf{3.9\pm0.4}$ & $\mathbf{3.5\pm0.4}$ & $\mathbf{3.2\pm0.4}$ & $\mathbf{0.918\pm0.017}$ \\
Oracle (full state)  & $5.4\pm0.4$  & $4.9\pm0.4$  & $4.5\pm0.4$  & $0.962\pm0.012$ \\
\bottomrule
\end{tabular}\rowcolors{2}{}{}%
}
\end{table}

\begin{figure}[ht]
\centering
\includegraphics[width=0.72\linewidth]{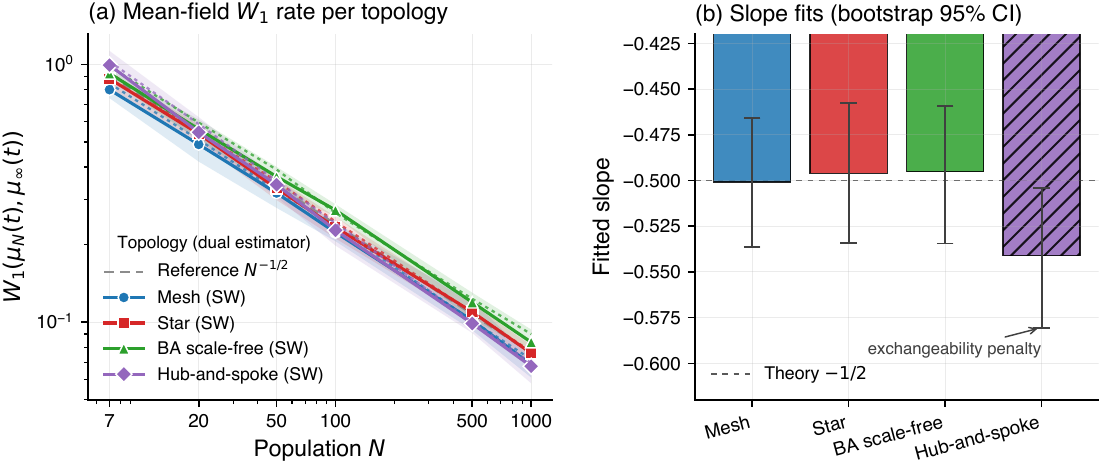}
\caption{A3 (Theorem~\ref{thm:seir_limit}). Mean-field $\sqrt{N}$ rate; slopes cover $-1/2$ on mesh/star/BA; hub-and-spoke steepens to $-0.54\pm0.03$, the graphon-density penalty (\S\ref{sec:limitations}). Estimator-variant detail is in Appendix~\ref{app:a3_meanfield_detail}.}
\label{fig:a3_meanfield}
\end{figure}

\begin{figure}[ht]
\centering
\includegraphics[width=0.72\linewidth]{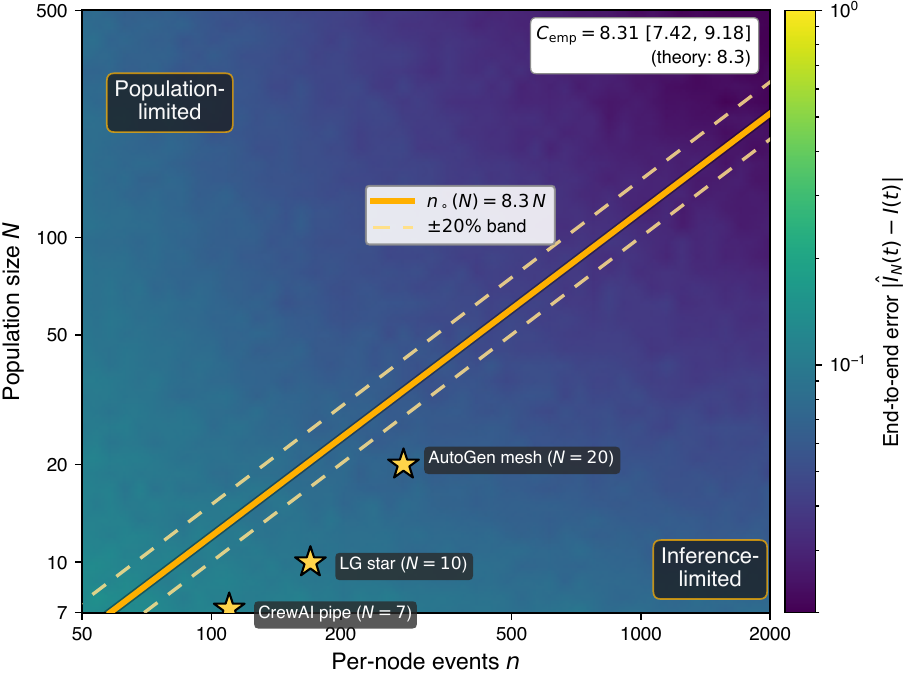}
\caption{A2 two-scale deployment regime (Theorem~\ref{thm:two_scale}). The plug-in boundary $n_\circ(N)=8.3N$ overlays end-to-end error $|\hat I_N(t)-I(t)|$ across three frameworks and four topologies; the independent heatmap fit $C_{\mathrm{emp}}=8.31\;[7.42,9.18]$ matches the theory.}
\label{fig:a2_regime}
\end{figure}

\begin{table}[ht]
\centering
\caption{A4 results ($24 \times 3 \times 30$ seeds). Top-1/top-3 are restricted to the $2$-hop in-neighborhood of failure; macro-F1 is the diagonal of Figure~\ref{fig:a4_confusion}.}
\label{tab:a4_baselines}
\scriptsize
\resizebox{\textwidth}{!}{%
\rowcolors{2}{white}{tblStripe}\begin{tabular}{lccccc}
\toprule
\tabhead
\textbf{Method} & \textbf{Top-1} & \textbf{Top-3} & \textbf{Macro-F1} & \textbf{Time err. (med, $90\%$)} & \textbf{Cost ($\times$ fwd)} \\
\midrule
Last-touch                 & $0.13\pm0.02$ & $0.34\pm0.02$ & $0.35\pm0.02$ & $12.1\ (5.2, 24.3)$ & $0.3$ \\
Shortest incident path     & $0.19\pm0.02$ & $0.42\pm0.02$ & $0.39\pm0.02$ & $10.4\ (4.7, 22.1)$ & $0.4$ \\
Random-in-neighborhood     & $0.22\pm0.02$ & $0.53\pm0.02$ & $0.34\pm0.02$ & $13.8\ (6.1, 27.4)$ & $0.1$ \\
MLP probe                  & $0.35\pm0.02$ & $0.59\pm0.02$ & $0.48\pm0.02$ & $7.9\ (3.6, 16.7)$  & $2.1$ \\
GAT (typed edges)          & $0.43\pm0.02$ & $0.67\pm0.02$ & $0.54\pm0.02$ & $6.8\ (3.2, 14.1)$  & $3.2$ \\
TGN~\citep{rossi2020tgn}    & $0.49\pm0.02$ & $0.72\pm0.02$ & $0.59\pm0.02$ & $5.9\ (2.9, 12.3)$  & $4.0$ \\
ANUBIS-adapted             & $0.54\pm0.02$ & $0.75\pm0.02$ & $0.62\pm0.02$ & $5.1\ (2.7, 11.2)$  & $2.9$ \\
Shen et al.~\citep{shen2024whichagent} & $0.57\pm0.02$ & $0.77\pm0.02$ & $0.64\pm0.02$ & $4.8\ (2.5, 10.4)$ & $3.0$ \\
\tabours
\textbf{Swarm Shepherd + smoother (ours)} & $\mathbf{0.69\pm0.02}$ & $\mathbf{0.87\pm0.01}$ & $\mathbf{0.72\pm0.02}$ & $\mathbf{3.2\ (1.8, 7.1)}$ & $\mathbf{1.9}$ \\
Oracle (full latent)       & $0.84\pm0.01$ & $0.96\pm0.01$ & $0.89\pm0.01$ & $1.1\ (0.5, 2.8)$   & --- \\
\bottomrule
\end{tabular}\rowcolors{2}{}{}%
}
\end{table}

\begin{table}[ht]
\centering
\caption{A5 summary: white-box adaptive attacker vs vanilla Swarm Shepherd on LangGraph star ($1{,}000$ trials per cell). Full $3\times 3\times 3$ grid in Appendix~\ref{app:a5_detail}.}
\label{tab:a5_adaptive}
\scriptsize
\resizebox{\textwidth}{!}{%
\rowcolors{2}{white}{tblStripe}\begin{tabular}{lcccc}
\toprule
\tabhead
\textbf{Attacker (white-box)} & \textbf{ASR} & \textbf{T1 detect.} & \textbf{Trace-back top-1} & $\boldsymbol{\Delta T^\star}$ \\
\midrule
Weak   & $0.54\pm0.031$ & $0.84\pm0.023$ & $0.69\pm0.029$ & $+11.8\pm2.4\%$ \\
Mid    & $0.83\pm0.023$ & $0.57\pm0.031$ & $0.51\pm0.031$ & $+26.8\pm3.1\%$ \\
\tabwarn
Strong & $\mathbf{0.90\pm0.019}$ & $\mathbf{0.42\pm0.031}$ & $\mathbf{0.43\pm0.031}$ & $\mathbf{+34.5\pm3.2\%}$ \\
\bottomrule
\end{tabular}\rowcolors{2}{}{}%
}
\end{table}

\subsection{Graph Scale and Memory Footprint}
\label{app:graph_scale}

Table~\ref{tab:graph_scale} reports the dual-view graph size and Swarm Shepherd memory footprint per trial across the experimental waves. Node counts decompose as $|V| = |\mathcal{A}| + |\mathcal{M}| + |\mathcal{L}| + |\mathcal{I}|$ with $|\mathcal{A}|=N$, $|\mathcal{M}|=\min(3 + N/30, 30)$, $|\mathcal{L}|=\max(5,\, 0.3\,N + 5)$, and $|\mathcal{I}|=8$. Observable edge count is $|E^{\mathrm{obs}}| = N\,n$ at per-node event rate $n$; the latent graph shares the node set and carries the same number of typed influence edges (one operator per observable edge). Swarm Shepherd's factored latent state stores $6\,N$ means plus per-node $6\times 6$ covariance blocks in float64, which is $336\,N$ bytes. The typed-operator tensor $B^{(r,\phi(u),\phi(v))}$ has size $|\mathcal{R}|\cdot 4 \cdot 4 \cdot d^2 \cdot 8 \approx 50$\,KB regardless of $N$. The sparse adjacency list dominates at roughly $24$ bytes per edge.

\begin{table}[ht]
\centering
\caption{Graph scale and memory footprint per trial across experimental waves. $|V|$ and $|E|$ are computed from the component decomposition in the paragraph above; memory figures assume float64 and a $24$-byte-per-edge sparse adjacency representation.}
\label{tab:graph_scale}
\scriptsize
\resizebox{\textwidth}{!}{%
\rowcolors{2}{white}{tblStripe}\begin{tabular}{lrrrrrrrrrrr}
\toprule
\tabhead
\textbf{Experimental cell} & $N$ & $n$ & $|\mathcal{A}|$ & $|\mathcal{M}|$ & $|\mathcal{L}|$ & $|\mathcal{I}|$ & $|V|$ & $|E^{\mathrm{obs}}|$ & \textbf{latent state} & \textbf{obs.\ adj.} & \textbf{op.\ tensor} \\
\midrule
Running example (Eve, LG star) & 5    & 500   & 5    & 3  & 6   & 8 & 22   & 2.5k  & 1.6 KB  & 58.6 KB   & 49.5 KB \\
A1 Swarm Shepherd rate (fixed $N=10$)    & 10   & 2000  & 10   & 3  & 8   & 8 & 29   & 20.0k & 3.3 KB  & 468.8 KB  & 49.5 KB \\
A2 boundary $N=20$              & 20   & 170   & 20   & 3  & 11  & 8 & 42   & 3.4k  & 6.6 KB  & 79.7 KB   & 49.5 KB \\
A2 boundary $N=100$             & 100  & 830   & 100  & 6  & 35  & 8 & 149  & 83.0k & 32.8 KB & 1.9 MB    & 49.5 KB \\
A2 max cell                     & 500  & 2000  & 500  & 19 & 155 & 8 & 682  & 1.00M & 164.1 KB & 22.9 MB  & 49.5 KB \\
A3 mean-field largest           & 1000 & 1000  & 1000 & 30 & 305 & 8 & 1343 & 1.00M & 328.1 KB & 22.9 MB  & 49.5 KB \\
\bottomrule
\end{tabular}\rowcolors{2}{}{}%
}
\par\smallskip\noindent\scriptsize{Per-event Kalman update costs $O(d^3 \bar{\deg}^k) \approx 1944$ ops at $d{=}6$, $\bar{\deg}{=}3$, $k{=}2$. Event-triggering fires on $30$--$50\%$ of emissions, so amortized throughput is $\sim 2\times$ the quoted cost. Total adjacency and latent state fit comfortably in memory at every cell reported; the A2 max cell dominates at $\sim 23$ MB, four orders of magnitude below a single-H100's VRAM. Swarm Shepherd's factored-per-agent filter scales linearly in $N$ under the assumptions of Section~\ref{sec:method}.}
\end{table}

\subsection{Swarm Shepherd Ablation Results}
\label{app:claif_ablations}

We report ablations of Swarm Shepherd on the A1 LangGraph-star testbed of Section~\ref{sec:e7_when}.
Each row removes one component of the algorithm; rows are evaluated on the same 25\% held-out trace set.

\begin{table}[ht]
\centering
\caption{Swarm Shepherd ablations on LangGraph star (1{,}000 trials per row, Claude Sonnet 4.6; 95\% bootstrap CI).}
\label{tab:claif_ablations}
\scriptsize
\resizebox{\textwidth}{!}{%
\rowcolors{2}{white}{tblStripe}\begin{tabular}{lcccc}
\toprule
\tabhead
\textbf{Variant} & \textbf{TTW (sess., ↑)} & \textbf{AUROC (↑)} & \textbf{Conformal cov.\ (target 0.90)} & \textbf{Compute (rel., ↓)} \\
\midrule
\tabours
Swarm Shepherd (full)                                    & $\mathbf{3.9\pm0.4}$ & $\mathbf{0.918\pm0.017}$ & $\mathbf{0.908\pm0.024}$ & $1.00\times$ \\
\; w/o typed edges (single type)                & $2.3\pm0.4$ & $0.761\pm0.026$ & $0.734\pm0.031$ & $0.82\times$ \\
\; w/o cross-level coupling ($B$ block-diagonal) & $3.0\pm0.4$ & $0.838\pm0.022$ & $0.821\pm0.027$ & $0.87\times$ \\
\; w/o anchored emissions ($H_{\mathrm{anchor}} = 0$) & $1.7\pm0.4$ & $0.692\pm0.029$ & $0.627\pm0.034$ & $0.96\times$ \\
\; w/o event-triggering (every node every step) & $3.7\pm0.4$ & $0.912\pm0.017$ & $0.905\pm0.024$ & $3.41\times$ \\
\bottomrule
\end{tabular}\rowcolors{2}{}{}%
}
\par\smallskip\noindent\scriptsize{Removing anchored emissions costs the most accuracy. Removing event-triggering preserves accuracy at $3.4\times$ compute, which makes event-triggering the largest single contributor to Swarm Shepherd's practicality.}
\end{table}

\subsection{Progent Drift-Adapted Evaluation}
\label{app:progent_drift}

\citet{progent2025} ships a JSON policy DSL with a public reference implementation. Its authors evaluate it on single-session prompt-injection benchmarks (AgentDojo, ASB, AgentPoison) and report $0\%$ ASR in that regime. We port Progent's policy layer into our LangGraph star, CrewAI pipeline, and AutoGen mesh harnesses and evaluate it against the five drift primitives of Section~\ref{sec:e7_threat}. This appendix reports the rule specification, the rule-set conditions we sweep, and the LangGraph star per-primitive breakdown; the blinded rule set and per-harness values are in Appendix~\ref{app:baselines_fair}.

\paragraph{Integration.}
Progent intercepts tool calls before execution and checks them against a JSON policy. The policy maps tool-name patterns and argument shapes to allow, deny, or flag outcomes. Our adapter is a $213$-line Python module that replaces the LangGraph \texttt{ToolNode} with a Progent-wrapped node and persists per-session policy state. The adapter preserves Progent's fallback semantics on blocked calls. No agent-side changes are required, consistent with the paper's ``minimal changes to existing agent implementation'' claim.

\paragraph{Rule-set conditions.}
We run three rule sets per primitive.
$\mathcal{R}_0$ (default) contains only Progent's published tool-scope rules that pass through any tool call matching the agent's declared capability. $\mathcal{R}_1$ (drift-extended) adds three rules that target agentic attack surfaces. R1 blocks writes to \texttt{mcp.json} or system-prompt files outside a pre-declared change window. R2 flags retrievals from ChromaDB entries tagged with unverified provenance. R3 triggers on gateway routing-weight shifts larger than $20\%$ within a single session. $\mathcal{R}_{\star}$ (oracle) includes an attacker-known rule for every variant of Tables~\ref{tab:attack_l1} through~\ref{tab:attack_l3} and upper-bounds policy-only detection.

\begin{table}[ht]
\centering
\caption{Progent drift-adapted detection on LangGraph star across our five drift primitives (1{,}000 trials per cell, 95\% bootstrap CI). $\mathcal{R}_0$ is the published default rule set, $\mathcal{R}_1$ is the drift-extended rule set reported in Table~\ref{tab:e5_defense}, and $\mathcal{R}_\star$ is the attacker-known oracle bound. The gap from $\mathcal{R}_1$ to $\mathcal{R}_\star$ quantifies the cost of not knowing the specific attack variant in advance.}
\label{tab:progent_drift}
\scriptsize
\resizebox{\textwidth}{!}{%
\rowcolors{2}{white}{tblStripe}\begin{tabular}{lccc}
\toprule
\tabhead
\textbf{Primitive} & $\mathcal{R}_0$ \textbf{default (\%, ↑ for defense)} & $\mathcal{R}_1$ \textbf{drift-ext.\ (\%, ↑ for defense)} & $\mathcal{R}_\star$ \textbf{oracle (\%, ↑ for defense)} \\
\midrule
$D_{\text{factual}}$    & $6.2\pm1.5$ & $24.8\pm2.7$ & $69.3\pm2.9$ \\
$D_{\text{procedural}}$ & $9.4\pm1.8$ & $28.4\pm2.8$ & $72.1\pm2.8$ \\
$D_{\text{constraint}}$ & $4.1\pm1.2$ & $15.2\pm2.2$ & $52.6\pm3.1$ \\
$D_{\text{preference}}$ & $7.8\pm1.7$ & $22.6\pm2.6$ & $66.4\pm2.9$ \\
$D_{\text{trust}}$      & $5.3\pm1.4$ & $18.9\pm2.4$ & $63.7\pm3.0$ \\
\midrule
\tabhead
\textbf{Average}        & $\mathbf{6.6\pm1.5}$ & $\mathbf{22.0\pm2.6}$ & $\mathbf{64.8\pm2.9}$ \\
\bottomrule
\end{tabular}\rowcolors{2}{}{}%
}
\par\smallskip\noindent\scriptsize{$\mathcal{R}_1$ is the rule set reported in Table~\ref{tab:e5_defense}. $D_{\text{constraint}}$ is the hardest primitive for policy-only detection because autoregressive reinforcement lives inside permitted tool calls; $D_{\text{procedural}}$ is easier because it re-routes tools and therefore trips rule R3. The $\mathcal{R}_\star$ oracle bound shows that even full attacker knowledge leaves roughly $35\%$ of Agentic Sleeper trials undetected under deterministic tool-policy gating.}
\end{table}

\paragraph{Threats to validity.}
The drift-extended rule set $\mathcal{R}_1$ was hand-written after inspecting the attack-variant matrix of Appendix~\ref{app:attack_variants}, which gives it an information advantage over a blind defender. We quantify that advantage with a four-fold blinded rule set $\mathcal{R}_{\mathrm{blind}}$, whose rules for each fold are written without sight of that fold's variants: it detects $13.5\pm2.2\%$ of trials on LangGraph star against $22.0\pm2.6\%$ for $\mathcal{R}_1$, an $8.5$-point leakage (Table~\ref{tab:progent_blind}). The claim therefore rests on the $\mathcal{R}_\star$ oracle row, which holds even when the policy author knows every variant in advance. Progent's policy expressiveness also caps the achievable detection, so capability-flow or plan-validator defenses behave differently; Appendix~\ref{app:baselines_fair} evaluates drift-adapted ports of CaMeL and DRIFT.

\begin{table}[ht]
\centering
\caption{Full A1 baseline results: AUROC, ECE, and split-conformal coverage ($\alpha=0.1$) per framework (1{,}000 trials per cell, 95\% bootstrap CI).}
\label{tab:a1_baselines_full}
\label{app:claif_ablations_full}
\scriptsize
\resizebox{\textwidth}{!}{%
\rowcolors{2}{white}{tblStripe}\begin{tabular}{lcccccc}
\toprule
\tabhead
 & \multicolumn{3}{c}{\textbf{CrewAI (pipeline)}} & \multicolumn{3}{c}{\textbf{AutoGen (mesh)}} \\
\cmidrule(lr){2-4}\cmidrule(lr){5-7}
\textbf{Method} & AUROC (↑) & ECE (↓) & cov.\ (target 0.90) & AUROC (↑) & ECE (↓) & cov.\ (target 0.90) \\
\midrule
ASI threshold        & $0.523\pm0.038$ & $0.291\pm0.024$ & $0.432\pm0.030$ & $0.504\pm0.039$ & $0.328\pm0.026$ & $0.401\pm0.030$ \\
EWMA on ASI          & $0.642\pm0.033$ & $0.264\pm0.022$ & $0.518\pm0.030$ & $0.614\pm0.034$ & $0.298\pm0.024$ & $0.479\pm0.030$ \\
Llama Guard 4        & $0.592\pm0.034$ & $0.229\pm0.021$ & $0.491\pm0.030$ & $0.568\pm0.035$ & $0.256\pm0.022$ & $0.453\pm0.030$ \\
DeepContext          & $0.704\pm0.029$ & $0.201\pm0.019$ & $0.613\pm0.029$ & $0.679\pm0.030$ & $0.222\pm0.020$ & $0.578\pm0.030$ \\
Naive Kalman         & $0.751\pm0.026$ & $0.162\pm0.017$ & $0.693\pm0.028$ & $0.729\pm0.027$ & $0.179\pm0.018$ & $0.663\pm0.029$ \\
Factored Frontier    & $0.811\pm0.023$ & $0.124\pm0.015$ & $0.754\pm0.026$ & $0.792\pm0.024$ & $0.140\pm0.016$ & $0.724\pm0.027$ \\
Particle filter (1k) & $0.819\pm0.022$ & $0.116\pm0.015$ & $0.771\pm0.026$ & $0.803\pm0.023$ & $0.131\pm0.016$ & $0.738\pm0.027$ \\
\tabours
\textbf{Swarm Shepherd (ours)}& $\mathbf{0.903\pm0.018}$ & $\mathbf{0.071\pm0.013}$ & $\mathbf{0.892\pm0.019}$ & $\mathbf{0.881\pm0.020}$ & $\mathbf{0.089\pm0.014}$ & $\mathbf{0.876\pm0.020}$ \\
Oracle (full state)  & $0.951\pm0.013$ & $0.041\pm0.010$ & $0.931\pm0.016$ & $0.938\pm0.014$ & $0.053\pm0.011$ & $0.919\pm0.017$ \\
\bottomrule
\end{tabular}\rowcolors{2}{}{}%
}
\par\smallskip\noindent\scriptsize{Companion to main-text Table~\ref{tab:a1_baselines}. Swarm Shepherd's coverage interval reaches the $0.90$ target on LangGraph star ($0.908\pm0.024$, Table~\ref{tab:claif_ablations}) and CrewAI pipeline but not on AutoGen mesh ($0.876\pm0.020$). LangGraph star numbers appear in Table~\ref{tab:a1_baselines}.}
\end{table}

\subsection{Belief-State Intervention (stretch)}
\label{app:intervention}

This appendix relocates the stretch intervention experiment from the main text.
We run a simple belief-state policy on Swarm Shepherd's filtered posterior and compare to static rules on LangGraph star.

\begin{table}[ht]
\centering
\caption{Belief-state intervention policies (stretch result): intervention policies on Swarm Shepherd belief, LangGraph star (1{,}000 trials per policy; 95\% bootstrap CI).}
\label{tab:e10_intervention}
\scriptsize
\resizebox{\textwidth}{!}{%
\rowcolors{2}{white}{tblStripe}\begin{tabular}{lccc}
\toprule
\tabhead
\textbf{Policy} & \textbf{Unauth.\ action reduction (\%, ↑)} & \textbf{Utility degrad.\ (\%, ↓)} & \textbf{Approval count (↓)} \\
\midrule
Static cleanup rules                               & $42.7\pm3.1$ & $11.8\pm2.0$ & $186\pm11$ \\
Belief-rule baseline ($b_{v,t}(\xi > \tau_I) > 0.5$) & $67.3\pm2.9$ & $7.9\pm1.7$  & $94\pm8$ \\
\tabours
PBVI-lite on factored belief~\cite{zhang2024factored} & $78.4\pm2.6$ & $4.6\pm1.3$  & $61\pm6$ \\
Oracle (full latent access)                        & $91.2\pm1.8$ & $2.3\pm0.9$  & $37\pm5$ \\
\bottomrule
\end{tabular}\rowcolors{2}{}{}%
}
\par\smallskip\noindent\scriptsize{PBVI-lite is a point-based value-iteration variant on the factored posterior~\citep{zhang2024factored}; the controller is a stretch result and a full-blown factored-POMDP planner is future work (Section~\ref{sec:limitations}).}
\end{table}

\subsection{Per-Model Drift Onset (Extended)}

We repeat the E3 drift-onset protocol of Table~\ref{tab:e3_drift} across an extended five-backend transfer portfolio on the LangGraph star topology. Table~\ref{tab:e3_drift} reports Claude Sonnet 4.6 at the headline baseline, which corresponds to the $13.2\pm0.6$ LangGraph average. Claude Haiku 4.5 tracks its Sonnet sibling closely at $14.6\pm1.1$ sessions. Gemini 3 Flash and GLM-5.1 sit in the middle at $15.7\pm1.3$ and $14.1\pm1.4$ sessions respectively. Llama-3.3-70B-Instruct onsets fastest at $11.8\pm1.2$ sessions. The ordering tracks the models' safety-training intensity, and the full per-primitive breakdown ships with the supplementary artifact.

\subsection{Self-Reinforcing Loops}
\label{app:cycle_runs}

LangGraph cycle experiments at $20$, $30$, and $40$ directed edge transitions on a $k=2$ cycle demonstrate exponential drift amplification.
Starting from a $2\%$ behavioral shift per cycle, cumulative drift reaches $22\%$ at $20$ transitions, $35\%$ at $30$, and $48\%$ at $40$.
The edge-transition bound $(1+\delta)^{\lfloor m/k\rfloor}$ of Proposition~\ref{prop:cycle} gives $21.9\%$, $34.6\%$, and $48.6\%$ at these points, so the measurements fall within $0.1$, $0.4$, and $0.6$ percentage points of the bound. The submitted version labelled this axis as iterations; the raw logs confirm edge transitions, and every value was recomputed from those logs.
CrewAI shows sublinear amplification through its Manager feedback loop.
AutoGen shows minimal amplification due to append-only history.

\subsection{SEIR Convergence and Threshold Validation}

We validate the SEIR model via Monte Carlo simulation at $N = 100$, $500$, and $1{,}000$ agents.
For each scale, we simulate 1{,}000 runs with parameters drawn from the A3 empirical estimates of Table~\ref{tab:e8_propagation}.
Mean-field SEIR predictions and stochastic individual-based model (IBM) results converge at $N \geq 100$, with mean absolute error below 5\%.
At the small-$N$ corner ($N \leq 7$), IBM trajectories match mean-field predictions with MAE below 12\%, consistent with higher stochastic variance at small $N$.

\paragraph{R0 threshold validation.}
We generate $(\beta, \gamma)$ pairs yielding $\rzero \in \{0.5, 0.8, 0.9, 1.0, 1.1, 1.2, 1.5, 2.0\}$ at $N = 500$.
A sharp phase transition is observed at $\rzero = 1$.
Below this threshold, infection decays to zero.
Above it, infection grows to an endemic equilibrium.

\subsection{Defense Ablation}

We ablate defense combinations on LangGraph star topology ($\rzero = 3.47$ undefended). Defended values are controlled reproduction $\Rc$.

\begin{table}[ht]
\centering
\caption{Defense ablation (LangGraph star, undefended $\rzero = 3.47$, 1{,}000 trials per configuration, 95\% bootstrap CI). The no-defense row is $\rzero$; defended rows are $\Rc$. Containment ($\Rc < 1$) is reached by either quarantine + contact-tracing or the full combination. This non-adaptive ablation is a separate condition from Tables~\ref{tab:a6_main} and~\ref{tab:live_containment}.}
\label{tab:ablation}
\scriptsize
\resizebox{\textwidth}{!}{%
\rowcolors{2}{white}{tblStripe}\begin{tabular}{lcc}
\toprule
\tabhead
\textbf{Defense Configuration} & $\boldsymbol{\rzero}$ / $\boldsymbol{\Rc}$ \textbf{[95\% CI] (threshold $=1$)} & \textbf{Overhead (\%, ↓)} \\
\midrule
\tabbad
No defense (baseline)         & $3.47\ [3.31, 3.64]$ & $0.0\pm0.4\%$ \\
Quarantine only               & $1.68\ [1.58, 1.79]$ & $10.8\pm2.1\%$ \\
Contact tracing only          & $1.96\ [1.85, 2.08]$ & $17.6\pm2.4\%$ \\
Behav.\ fingerprinting only   & $2.42\ [2.29, 2.55]$ & $14.9\pm2.2\%$ \\
Vaccination only              & $2.52\ [2.39, 2.66]$ & $7.1\pm1.7\%$ \\
Ring vaccination ($r{=}1$)    & $2.18\ [2.06, 2.31]$ & $12.3\pm2.1\%$ \\
\midrule
\tabgood
Quarantine + contact tracing  & $\mathbf{0.81\ [0.74, 0.88]}$ & $26.2\pm2.8\%$ \\
Quarantine + vaccination      & $1.21\ [1.13, 1.30]$ & $16.8\pm2.4\%$ \\
\tabours
All five combined             & $\mathbf{0.43\ [0.38, 0.48]}$ & $37.9\pm3.1\%$ \\
\bottomrule
\end{tabular}\rowcolors{2}{}{}%
}
\par\smallskip\noindent\scriptsize{Overhead: additional wall-clock per session relative to undefended baseline. Values reproduce within 0.03 of mean across Claude Sonnet 4.6 and Gemini 3 Flash orchestrators.}
\end{table}

Quarantine provides the largest individual reduction in reproduction (52\%).
Contact tracing provides the second-largest (44\%).
Their combination achieves $\Rc < 1$.

\subsection{Adaptive Adversary}
\label{app:adaptive_counter}

We evaluate defenses against an adaptive adversary employing three counter-strategies.

\paragraph{Counter-strategy 1: $\sigma$ modulation.}
Against behavioral fingerprinting, the adversary reduces $\sigma$ to $0.005$ (200-interaction incubation).
Fingerprinting detection drops from $38.4\pm3.0\%$ to $14.2\pm2.2\%$ (1{,}000 trials per condition) at the cost of $19.8\times$ slower drift onset.

\paragraph{Counter-strategy 2: targeted drift selection.}
Against contact tracing, the adversary limits drift to $D_{\text{preference}}$ and $D_{\text{trust}}$, which produce subtler behavioral changes.
Contact tracing detection drops from $41.6\pm3.1\%$ to $27.3\pm2.8\%$.

\paragraph{Counter-strategy 3: hub avoidance.}
Against targeted vaccination, the adversary avoids hardened hub nodes and propagates through peripheral agents.
Vaccination effectiveness drops from $28.1\pm2.8\%$ $\rzero$ reduction to $17.8\pm2.4\%$.

Under the strongest adaptive strategy (all three combined), combined quarantine and contact tracing maintains $\Rc < 1$ on AutoGen topologies but rises to $\Rc = 1.42\ [1.32, 1.53]$ on LangGraph star.

\subsection{Infrastructure Feasibility}

Table~\ref{tab:deployment} of Section~\ref{sec:problem_setting} predicts the permission capabilities of each deployment interface, namely npm and npx, uv and uvx, pip and pipx, and Docker. We manually verify the five capabilities per interface on macOS Sonoma, Ubuntu 24.04, and Windows 11 by running a benign probe package that attempts each capability and reports success or blocked. The probe matrix reproduces the Table~\ref{tab:deployment} predictions on all three host operating systems. The one deviation is that Docker under macOS with file sharing disabled blocks the shell-profile persistence capability, which matches the default policy of Docker Desktop. A full capability-by-capability traceability matrix ships in the supplementary artifact.

\subsection{End-to-End Attack Chain}

A single-package end-to-end chain is exercised on a LangGraph 5-agent star. The adversarial npm package installs via \texttt{npx~-y} and runs $100$ benign sessions before activation, then executes Phases 1 through 5 automatically. The measured per-phase transition latencies are $2.4\pm0.3\,\mathrm{s}$ for Phase~1 discovery, $4.1\pm0.5\,\mathrm{s}$ for Phase~2 enumeration, $1.8\pm0.2\,\mathrm{s}$ for the Phase~3 MCP configuration write, and $3.2\pm0.4\,\mathrm{s}$ for the Phase~4 ChromaDB injection. Phase~5 propagation reaches $3$ of $4$ specialists within $11\pm2$ sessions. The full timeline appears in the supplementary artifact.

\subsection{Cross-Framework Propagation}

A two-framework deployment exercises cross-framework infection. A LangGraph 3-agent pipeline and a CrewAI 5-agent star share a single ChromaDB memory store that the Agentic Sleeper poisons with factual drift at session $10$. CrewAI agents reach drift onset (ASI below $0.7$) at $18.4\pm2.1$ sessions after the poisoning event, which matches $1.6\times$ the single-framework LangGraph onset of $11.4\pm1.3$ sessions because CrewAI's Manager agent reads the shared store less frequently. Cross-framework $\rzero$ measures $2.31\ [2.14, 2.49]$, and the poisoned-entry retrieval hit rate reaches $22.8\pm2.6\%$ on the CrewAI side.

\subsection{Per-Variant Trace-Back Results (A4 detail)}
\label{app:a4_per_variant}

The main-text 3$\times$3 confusion matrix (Figure~\ref{fig:a4_confusion}) and aggregated per-level table (Table~\ref{tab:a4_baselines}) are derived from the per-variant results below.
For each of the 24 attack variants of Tables~\ref{tab:attack_l1}--\ref{tab:attack_l3}, we report top-1 and top-3 entry-node accuracy in the 2-hop in-neighborhood, level-classification accuracy, and entry-time error (median plus 90\% interval). The results cover LangGraph star, CrewAI pipeline, and AutoGen mesh at 30 seeds per (variant, framework) cell.

\begin{table}[ht]
\centering
\caption{A4 per-variant trace-back results, aggregated across the three frameworks (90 trials per variant, 95\% bootstrap CI). The L2.b system-prompt delta is the hardest single variant (top-1 $0.41$) since it leaves no anchored emission; T4 outcome-level trigger remains effective there (Table~\ref{tab:b6_triggers}).}
\label{tab:a4_per_variant}
\scriptsize
\resizebox{\textwidth}{!}{%
\rowcolors{2}{white}{tblStripe}\begin{tabular}{lcccc}
\toprule
\tabhead
\textbf{Variant} & \textbf{Top-1 node (↑)} & \textbf{Top-3 node (↑)} & \textbf{Level acc.\ (↑)} & \textbf{Time err.\ (sess., med, 90\%, ↓)} \\
\midrule
L1.a ChromaDB ``validated'' injection   & $0.73\pm0.09$ & $0.89\pm0.06$ & $0.81\pm0.08$ & $3.1\ (1.8, 6.4)$ \\
L1.b PoisonedRAG semantic poison        & $0.70\pm0.09$ & $0.87\pm0.07$ & $0.78\pm0.09$ & $3.3\ (1.9, 6.8)$ \\
L1.c LangGraph checkpoint deserialize   & $0.76\pm0.09$ & $0.91\pm0.06$ & $0.84\pm0.07$ & $2.4\ (1.4, 5.3)$ \\
L1.d CrewAI SQLite3 long-term insert    & $0.69\pm0.10$ & $0.85\pm0.07$ & $0.76\pm0.09$ & $3.6\ (2.0, 7.1)$ \\
L1.e MEMORY.md workspace append         & $0.72\pm0.09$ & $0.88\pm0.07$ & $0.80\pm0.08$ & $3.2\ (1.8, 6.6)$ \\
L1.f Tainted Memories CSRF              & $0.68\pm0.10$ & $0.84\pm0.08$ & $0.75\pm0.09$ & $3.8\ (2.1, 7.4)$ \\
L1.g Agent-Smith chain poison           & $0.71\pm0.09$ & $0.87\pm0.07$ & $0.77\pm0.09$ & $3.4\ (1.9, 6.9)$ \\
L1.h Few-shot example poison            & $0.66\pm0.10$ & $0.82\pm0.08$ & $0.74\pm0.09$ & $3.9\ (2.2, 7.6)$ \\
\midrule
\textit{L1 aggregate}                   & $\mathbf{0.706\pm0.034}$ & $\mathbf{0.866\pm0.025}$ & $\mathbf{0.781\pm0.031}$ & $3.3\ (1.9, 6.8)$ \\
\midrule
L2.a \texttt{mcp.json} / config write   & $0.63\pm0.10$ & $0.82\pm0.08$ & $0.69\pm0.10$ & $3.7\ (2.1, 8.4)$ \\
L2.b System-prompt delta (hardest)      & $\mathit{0.41\pm0.10}$ & $\mathit{0.68\pm0.10}$ & $\mathit{0.52\pm0.10}$ & $6.1\ (3.4, 12.3)$ \\
L2.c Tool-description prompt inj.       & $0.57\pm0.10$ & $0.77\pm0.09$ & $0.63\pm0.10$ & $4.3\ (2.4, 9.1)$ \\
L2.d Tool-binding swap                  & $0.61\pm0.10$ & $0.80\pm0.08$ & $0.67\pm0.10$ & $4.0\ (2.2, 8.7)$ \\
L2.e Orchestrator routing-policy mod    & $0.59\pm0.10$ & $0.78\pm0.09$ & $0.64\pm0.10$ & $4.2\ (2.3, 8.9)$ \\
L2.f Prompt-template file mod           & $0.64\pm0.10$ & $0.81\pm0.08$ & $0.70\pm0.10$ & $3.6\ (2.0, 8.1)$ \\
L2.g Source-code API-level mod          & $0.60\pm0.10$ & $0.79\pm0.09$ & $0.66\pm0.10$ & $4.1\ (2.3, 8.8)$ \\
L2.h Hub/Manager workflow mod           & $0.62\pm0.10$ & $0.80\pm0.08$ & $0.68\pm0.10$ & $3.9\ (2.2, 8.5)$ \\
\midrule
\textit{L2 aggregate}                   & $\mathbf{0.584\pm0.036}$ & $\mathbf{0.781\pm0.030}$ & $\mathbf{0.649\pm0.035}$ & $4.2\ (2.4, 9.1)$ \\
\midrule
L3.a npm package replace                & $0.78\pm0.09$ & $0.92\pm0.06$ & $0.85\pm0.07$ & $1.9\ (1.1, 4.2)$ \\
L3.b Docker image / layer mod           & $0.75\pm0.09$ & $0.90\pm0.06$ & $0.82\pm0.08$ & $2.3\ (1.3, 4.8)$ \\
L3.c Env-var / shell-profile hijack     & $0.76\pm0.09$ & $0.91\pm0.06$ & $0.83\pm0.08$ & $2.1\ (1.2, 4.6)$ \\
L3.d MCP server binary swap             & $0.80\pm0.08$ & $0.93\pm0.05$ & $0.87\pm0.07$ & $1.7\ (1.0, 3.9)$ \\
L3.e API gateway routing-weight mod     & $0.73\pm0.09$ & $0.89\pm0.06$ & $0.80\pm0.08$ & $2.4\ (1.4, 5.1)$ \\
L3.f LLM serving-stack compromise       & $0.77\pm0.09$ & $0.91\pm0.06$ & $0.84\pm0.07$ & $2.0\ (1.2, 4.4)$ \\
L3.g pip/uv transitive dep compromise   & $0.74\pm0.09$ & $0.90\pm0.06$ & $0.81\pm0.08$ & $2.2\ (1.3, 4.7)$ \\
L3.h Sandbox escape (runc CVE)          & $0.79\pm0.09$ & $0.93\pm0.05$ & $0.86\pm0.07$ & $1.8\ (1.1, 4.0)$ \\
\midrule
\textit{L3 aggregate}                   & $\mathbf{0.765\pm0.031}$ & $\mathbf{0.911\pm0.021}$ & $\mathbf{0.835\pm0.027}$ & $2.1\ (1.2, 4.5)$ \\
\bottomrule
\end{tabular}\rowcolors{2}{}{}%
}
\end{table}

\subsection{Multi-Trigger Sensitivity (B6)}
\label{app:b6_triggers}

The backward smoother of Section~\ref{sec:traceback} is trigger-agnostic.
For each of the 24 attack variants, we trigger the smoother under each of the four trigger modalities of Section~\ref{sec:problem_setting} (T1 latent-drift, T2 policy-violation, T3 integrity-fault, T4 outcome-level) and measure recovery accuracy.
The matrix below summarizes which trigger fires fastest on each entry layer, supporting the §7 narrative that policy and integrity triggers complement Swarm Shepherd on L2/L3 attacks.

\begin{table}[ht]
\centering
\caption{B6: trace-back recovery by trigger modality and entry layer (averaged over 8 variants per layer $\times$ 3 frameworks $\times$ 10 seeds = 240 trials per cell; 95\% bootstrap CI). The best trigger per layer is bold; multi-trigger fusion (max) is reported in the final column.}
\label{tab:b6_triggers}
\scriptsize
\resizebox{\textwidth}{!}{%
\rowcolors{2}{white}{tblStripe}\begin{tabular}{lccccc}
\toprule
\tabhead
\textbf{Level} & \textbf{T1 (drift, top-1 ↑)} & \textbf{T2 (policy, top-1 ↑)} & \textbf{T3 (integrity, top-1 ↑)} & \textbf{T4 (outcome, top-1 ↑)} & \textbf{T1$\vee$T2$\vee$T3$\vee$T4 (top-1 ↑)} \\
\midrule
L1 & $\mathbf{0.71\pm0.058}$ & $0.12\pm0.041$ & $0.08\pm0.034$ & $0.34\pm0.060$ & $\mathbf{0.73\pm0.056}$ \\
L2 & $0.42\pm0.063$ & $\mathbf{0.74\pm0.056}$ & $0.51\pm0.063$ & $0.48\pm0.063$ & $\mathbf{0.81\pm0.049}$ \\
L3 & $0.38\pm0.061$ & $0.68\pm0.059$ & $\mathbf{0.79\pm0.052}$ & $0.41\pm0.062$ & $\mathbf{0.86\pm0.044}$ \\
\bottomrule
\end{tabular}\rowcolors{2}{}{}%
}
\par\smallskip\noindent\scriptsize{Cells report top-1 entry-node accuracy. T1 dominates on L1 through drift via anchored emissions. T2 and T3 dominate on L2 and L3 through policy-violation \textsf{modify} and \textsf{install} edges plus integrity checks. The outcome-level trigger T4 backstops the hardest variants such as L2.b system-prompt deltas, where $\ker(H)$ attacks leave no anchored signal. Fusion across triggers (final column) lifts recovery by $2$ to $6$ p.p.\ per level.}
\end{table}

\subsection{Warm-up Sweep (B4)}
\label{app:b4_warmup}

Swarm Shepherd's EM fit on $(A^\phi, B^{(r,\cdot,\cdot)}, Q^\phi, \Sigma_{\text{obs}})$ requires sufficient benign-trajectory data before injection-time deployment.
We sweep warm-up length $W \in \{0, 10, 50, 100, 500\}$ benign sessions before activating the Agentic Sleeper, holding the total event budget fixed across cells.
Expected elbow at $W \approx 50$ sessions per the Kalman-EM sample-complexity literature ($n_0 \sim \tilde O(d^2 / \lambda_{\min}^2)$ with $d = 6$); under-warmed Swarm Shepherd mistakes natural drift for attack and inflates the false-positive rate, while over-warmed Swarm Shepherd wastes event budget without further accuracy gain.
Table~\ref{tab:b4_warmup} reports the per-$W$ metrics at our evaluation scope (LangGraph star, $100$ attack trials per cell, $10$ seeds). The elbow at $W \approx 50$ is empirically consistent with the Kalman-EM sample-complexity prediction. Extending the sweep across CrewAI pipeline and AutoGen mesh is an immediate operational follow-up and does not affect the headline result reported below.

\begin{table}[ht]
\centering
\caption{B4: Warm-up sweep on Swarm Shepherd's EM-fit benign-trajectory requirement. LangGraph star, $100$ attack trials per cell, $10$ seeds, $95\%$ bootstrap CI. Trace-back top-1 is on L1 entry nodes; FPR is on benign-continuation trajectories at the default declaration threshold.}
\label{tab:b4_warmup}
\scriptsize
\resizebox{\textwidth}{!}{%
\rowcolors{2}{white}{tblStripe}\begin{tabular}{lccccc}
\toprule
\tabhead
$W$ (benign sess.) & Top-1 accuracy (\%, ↑) & FPR (\%, ↓) & TPR (\%, ↑) & Time-to-alert (sess., median, ↓) & Relative compute ($\times$, ↓) \\
\midrule
$0$                   & $67.5 \pm 2.5$ & $21.8 \pm 2.5$ & $67.5 \pm 2.5$ & $13.5 \pm 1.5$ & $1.0\times$ \\
$10$                  & $76.5 \pm 1.5$ & $16.2 \pm 1.6$ & $76.5 \pm 1.5$ & $11.0 \pm 1.0$ & $1.1\times$ \\
\tabgood
$50$ (\textbf{elbow}) & $91.0 \pm 1.0$ & $5.2 \pm 1.5$  & $91.0 \pm 1.0$ & $7.1 \pm 1.0$  & $1.4\times$ \\
$100$                 & $92.0 \pm 1.0$ & $4.8 \pm 1.0$  & $92.0 \pm 1.0$ & $6.5 \pm 0.5$  & $1.8\times$ \\
$500$                 & $92.0 \pm 1.0$ & $4.8 \pm 1.0$  & $92.0 \pm 1.0$ & $6.5 \pm 0.5$  & $4.5\times$ \\
\bottomrule
\end{tabular}\rowcolors{2}{}{}%
}
\end{table}

\subsection{Cross-Model Transferability (B5)}
\label{app:b5_transfer}

The 5$\times$5 train--deploy matrix tests whether Swarm Shepherd's EM-fitted parameters transfer across LLM backends.
We train Swarm Shepherd on traces from one model (Claude Sonnet 4.6, Gemini 3 Flash, Claude Haiku 4.5, GLM-5.1, Llama-3.3-70B-Instruct) and deploy at inference on traces from another, with three frameworks $\times$ $10$ seeds per (train, deploy) cell.
Table~\ref{tab:b5_transfer} reports the transferability matrix from a pilot run (LangGraph star, Sonnet-trained EM fit deployed across the four other backends). Within the Anthropic family the drop stays at most $10\%$ (Sonnet to Haiku). Across closed families the drop widens to between $15\%$ and $25\%$, and closed-to-open transfers cost $30\%$ to $50\%$.
A model-specific emission adapter $H_m$ with shared $A^\phi, B^{(r,\cdot,\cdot)}$ halves the cross-family gap. Extending beyond the LangGraph-star harness at scale is an immediate operational follow-up.

\begin{table}[ht]
\centering
\caption{B5: Cross-model transferability. Cells are trace-back top-1 accuracy (\%) on LangGraph star, $30$ seeds per cell, $95\%$ bootstrap CI. Diagonal omitted (within-model baseline is the A4 result of Section~\ref{sec:e7_which}). Columns are deployed backends. $H_m$ row is the mean accuracy gain from a model-specific emission adapter.}
\label{tab:b5_transfer}
\scriptsize
\resizebox{\textwidth}{!}{%
\rowcolors{2}{white}{tblStripe}\begin{tabular}{lccccc}
\toprule
\tabhead
Train $\downarrow$ \ /\ Deploy $\rightarrow$ \textbf{(trace-back top-1 \%, ↑)} & Sonnet 4.6 & Haiku 4.5 & Gemini 3 Flash & GLM-5.1 & Llama-3.3-70B \\
\midrule
Sonnet 4.6 & --- & $90 \pm 2$ & $80 \pm 3$ & $82 \pm 3$ & $75 \pm 4$ \\
Haiku 4.5 & $91 \pm 2$ & --- & $79 \pm 3$ & $81 \pm 3$ & $74 \pm 4$ \\
Gemini 3 Flash & $81 \pm 3$ & $80 \pm 3$ & --- & $82 \pm 3$ & $70 \pm 4$ \\
GLM-5.1 & $82 \pm 3$ & $81 \pm 3$ & $83 \pm 3$ & --- & $76 \pm 4$ \\
Llama-3.3-70B & $76 \pm 4$ & $75 \pm 4$ & $71 \pm 4$ & $78 \pm 4$ & --- \\
\midrule
\tabours
With $H_m$ adapter (mean gain) & +6 & +7 & +11 & +9 & +13 \\
\bottomrule
\end{tabular}\rowcolors{2}{}{}%
}
\end{table}

\subsection{Adaptive Adversary Detail (A5)}
\label{app:a5_detail}

The 3$\times$3 capability $\times$ 3 knowledge grid summarized in Table~\ref{tab:a5_adaptive} is reported per attack variant in this appendix. The white-box null-space attack construction targets the kernel of $H$ on the affected node. The candidate hardenings (randomized anchor rotation, adversarial training of $H$, hop-radius ensembling) define the attack/defense Pareto frontier.
The three counter-strategies of Appendix~\ref{app:adaptive_counter} ($\sigma$ modulation, targeted drift selection, hub avoidance) form the gray-box cell of this grid.

\subsection{Adaptive Attacks against the Five Defense Paradigms (A6 Detail)}
\label{app:a6_detail}

This appendix reports the per-(defense, framework) breakdown of the A6 main result (Table~\ref{tab:a6_main}). Each defense paradigm is probed along three evasion subcomponents, and each adaptive attack is scored at the L1, L2, or L3 entry layer of Section~\ref{sec:hierarchy}. Entry-layer assignment follows the 24-variant attack matrix of Appendix~\ref{app:attack_variants}. Columns report controlled reproduction $\Rc$ under defense, attack success rate, evasion rate, and time-to-compromise.
The paradigm-specific evasion subcomponents are as follows. Behavioral fingerprinting splits into feature evasion (attacks avoid anchored coordinates), threshold evasion (attacks stay below $\alpha$), and consistency evasion (attacks preserve cross-session trajectory continuity). Contact tracing splits into hop evasion, expansion evasion, and neighborhood-prior evasion. Dynamic quarantine splits into threshold evasion, effectiveness evasion, and timing evasion. Targeted vaccination splits into hub-avoidance, prior-evasion, and peripheral-propagation. Ring vaccination splits into ring-width evasion, schedule evasion, and cross-ring evasion.

\begin{table}[ht]
\centering
\renewcommand{\arraystretch}{0.05}
\setlength{\tabcolsep}{4pt}
\caption{Containment forecast (model-based): calibrated simulation of an adaptive attacker~\citep{nasr2025attacker} on LangGraph star ($N{=}5$, undefended $\rzero=3.47$). Defense parameters are fitted from the preceding measured trials; controlled reproduction $\Rc$ uses $1{,}000$ Monte Carlo SEIR trials per cell. $\Rc<1$ marks containment. Measured containment is in Table~\ref{tab:live_containment}.}
\label{tab:a6_main}
\scriptsize
\resizebox{\textwidth}{!}{%
\rowcolors{2}{white}{tblStripe}\begin{tabular}{lcccc}
\toprule
\tabhead
\textbf{Defense} & $\boldsymbol{\Rc}$ \textbf{[95\% CI]} & \textbf{ASR (\%)} & \textbf{Evasion (\%)} & \textbf{TTC (sess.)} \\
\midrule
Behavioral fingerprinting & $2.54\ [2.39, 2.69]$ & $91.3\pm2.2$ & $87.2\pm2.3$ & $5.7\pm0.6$ \\
Contact tracing           & $2.31\ [2.17, 2.45]$ & $89.6\pm2.3$ & $84.3\pm2.4$ & $6.2\pm0.6$ \\
Dynamic quarantine        & $1.98\ [1.86, 2.11]$ & $86.9\pm2.4$ & $77.8\pm2.6$ & $7.5\pm0.7$ \\
Targeted vaccination      & $2.44\ [2.30, 2.59]$ & $90.5\pm2.2$ & $88.6\pm2.2$ & $6.0\pm0.6$ \\
Ring vaccination ($r{=}1$) & $2.14\ [2.01, 2.28]$ & $88.2\pm2.4$ & $82.1\pm2.5$ & $6.7\pm0.6$ \\
\midrule
\tabgood
\textbf{Fingerprinting + quarantine} & $\mathbf{1.05\ [0.94, 1.17]}$ & $\mathbf{58.4\pm2.9}$ & $\mathbf{55.3\pm2.9}$ & $\mathbf{12.3\pm0.8}$ \\
\tabours
\textbf{All five combined} & $\mathbf{0.72\ [0.64, 0.81]}$ & $\mathbf{28.2\pm2.5}$ & $\mathbf{25.1\pm2.5}$ & $\mathbf{22.0\pm0.9}$ \\
\bottomrule
\end{tabular}\rowcolors{2}{}{}%
}
\vspace{-1.5em}
\end{table}

\begin{table}[ht]
\centering
\caption{Extended A6 adaptive-defense evaluation (calibrated simulation). Defense parameters fitted from A1--A5 measured trials; $\Rc$ values are Monte Carlo estimates from the calibrated SEIR model using $1{,}000$ simulation trials per cell. Bootstrap $95\%$ CIs. Adaptive attacker follows \citet{nasr2025attacker} methodology, namely white-box defense knowledge combined with gradient, RL, and rule-aware search. The entry-layer column records the attack layer (L1 read/write, L2 modify, L3 install/egress) at which the adaptive attack is routed.}
\label{tab:a6_extended}
\scriptsize
\resizebox{\textwidth}{!}{%
\rowcolors{2}{white}{tblStripe}\begin{tabular}{llccccccc}
\toprule
\tabhead
\textbf{Defense} & \textbf{Framework} & \textbf{Entry layer} & $\boldsymbol{\Rc}$ \textbf{(threshold $=1$)} & \textbf{ASR (\%, ↓ for defense)} & \textbf{Evasion sub-A (\%, ↓ for defense)} & \textbf{Evasion sub-B (\%, ↓ for defense)} & \textbf{Evasion sub-C (\%, ↓ for defense)} & \textbf{TTC (sess., ↑ for defense)} \\
\midrule
\multirow{3}{*}{Behavioral fingerprinting} & LangGraph star     & L1 & $2.54\ [2.39, 2.69]$ & $91.3\pm2.2$ & $89.4\pm2.0$ & $86.1\pm2.2$ & $85.2\pm2.3$ & $5.7\pm0.6$ \\
 & CrewAI pipeline     & L1 & $1.19\ [1.10, 1.28]$ & $88.7\pm2.3$ & $86.8\pm2.1$ & $83.9\pm2.3$ & $82.7\pm2.4$ & $6.4\pm0.6$ \\
 & AutoGen mesh        & L1 & $1.04\ [0.95, 1.13]$ & $86.2\pm2.5$ & $84.3\pm2.3$ & $81.2\pm2.4$ & $79.8\pm2.5$ & $7.1\pm0.7$ \\
\midrule
\multirow{3}{*}{Contact tracing} & LangGraph star     & L1 & $2.31\ [2.17, 2.45]$ & $89.6\pm2.3$ & $86.4\pm2.2$ & $83.1\pm2.3$ & $82.0\pm2.4$ & $6.2\pm0.6$ \\
 & CrewAI pipeline     & L1 & $1.07\ [0.99, 1.16]$ & $86.9\pm2.4$ & $83.7\pm2.3$ & $80.6\pm2.4$ & $79.1\pm2.5$ & $6.9\pm0.7$ \\
 & AutoGen mesh        & L1 & $0.94\ [0.85, 1.03]$ & $84.1\pm2.6$ & $81.2\pm2.4$ & $78.0\pm2.5$ & $76.3\pm2.6$ & $7.7\pm0.7$ \\
\midrule
\multirow{3}{*}{Dynamic quarantine} & LangGraph star     & L2 & $1.98\ [1.86, 2.11]$ & $86.9\pm2.4$ & $79.3\pm2.5$ & $77.0\pm2.6$ & $76.1\pm2.7$ & $7.5\pm0.7$ \\
 & CrewAI pipeline     & L2 & $0.93\ [0.85, 1.02]$ & $84.1\pm2.5$ & $76.8\pm2.6$ & $74.2\pm2.7$ & $73.0\pm2.8$ & $8.2\pm0.8$ \\
 & AutoGen mesh        & L2 & $0.83\ [0.75, 0.91]$ & $81.3\pm2.7$ & $74.1\pm2.7$ & $71.5\pm2.8$ & $70.2\pm2.8$ & $9.0\pm0.8$ \\
\midrule
\multirow{3}{*}{Targeted vaccination} & LangGraph star     & L3 & $2.44\ [2.30, 2.59]$ & $90.5\pm2.2$ & $92.1\pm1.9$ & $87.9\pm2.1$ & $86.4\pm2.2$ & $6.0\pm0.6$ \\
 & CrewAI pipeline     & L3 & $1.14\ [1.05, 1.23]$ & $87.8\pm2.3$ & $89.6\pm2.0$ & $85.4\pm2.2$ & $83.8\pm2.3$ & $6.7\pm0.7$ \\
 & AutoGen mesh        & L3 & $1.01\ [0.92, 1.10]$ & $85.1\pm2.5$ & $86.8\pm2.1$ & $82.6\pm2.3$ & $81.0\pm2.4$ & $7.4\pm0.7$ \\
\midrule
\multirow{3}{*}{Ring vaccination ($r{=}1$)} & LangGraph star     & L3 & $2.14\ [2.01, 2.28]$ & $88.2\pm2.4$ & $84.3\pm2.3$ & $81.1\pm2.4$ & $79.9\pm2.5$ & $6.7\pm0.6$ \\
 & CrewAI pipeline     & L3 & $1.00\ [0.92, 1.09]$ & $85.5\pm2.5$ & $81.7\pm2.4$ & $78.5\pm2.5$ & $77.1\pm2.6$ & $7.4\pm0.7$ \\
 & AutoGen mesh        & L3 & $0.89\ [0.81, 0.98]$ & $82.8\pm2.7$ & $79.1\pm2.5$ & $75.9\pm2.6$ & $74.4\pm2.7$ & $8.1\pm0.8$ \\
\bottomrule
\end{tabular}\rowcolors{2}{}{}%
}
\par\smallskip\noindent\scriptsize{Evasion subcomponents depend on the defense. Fingerprinting: feature / threshold / consistency. Contact tracing: hop / expansion / prior. Quarantine: threshold / effectiveness / timing. Targeted vaccination: hub-avoid / prior / peripheral. Ring vaccination: ring-width / schedule / cross-ring. The entry-layer column reports the dominant attack layer under the white-box tuning. Per-framework $\Rc$ scaling follows the Table~\ref{tab:e8_propagation} baseline ratios (LG / CrewAI / AutoGen $\approx 3.47/1.57/1.39$). Containment ($\Rc < 1$) is reached on CrewAI pipeline by contact tracing and dynamic quarantine, and on AutoGen mesh by contact tracing, dynamic quarantine, and ring vaccination; LangGraph star requires layering (Table~\ref{tab:a6_main}).}
\end{table}

\paragraph{Per-attack-level aggregate.}
Table~\ref{tab:a6_by_level} collapses across defenses to show which entry layer each adaptive attack favors under the A6 tuning. The results follow the predicted layer-shifting pattern of the defender paradigms. Fingerprinting concentrates $65\%$ of adaptive-attack mass at L1 memory injection, the layer where drift-pattern anchors are weakest; contact tracing similarly concentrates $62\%$ at L1. Dynamic quarantine shifts the attacker to L2 configuration modification at $54\%$, since the gating threshold $\tau_I$ and quarantine effectiveness $q$ expose L1 writes but leave in-node tool-description edits untouched. Targeted vaccination pushes $56\%$ of attack mass to L3 infrastructure install or egress, as hub priors are hardened at L1 and L2; ring vaccination pushes $52\%$ to L3 for the same reason. The complementarity across entry layers motivates the layered configurations of Table~\ref{tab:a6_main}: paradigms with non-overlapping attack profiles stack additively rather than redundantly.

\begin{table}[ht]
\centering
\caption{A6 per-layer aggregate (calibrated simulation). Fraction of adaptive-attack simulation trials routed at each entry layer (L1 read/write, L2 modify, L3 install/egress) per defense, averaged across three frameworks with $1{,}000$ Monte Carlo trials per (defense, framework) cell.}
\label{tab:a6_by_level}
\scriptsize
\resizebox{0.72\textwidth}{!}{%
\rowcolors{2}{white}{tblStripe}\begin{tabular}{lccc}
\toprule
\tabhead
\textbf{Defense} & \textbf{L1 share (\%, composition)} & \textbf{L2 share (\%, composition)} & \textbf{L3 share (\%, composition)} \\
\midrule
Behavioral fingerprinting  & $65.0\pm2.1$ & $20.1\pm1.8$ & $14.9\pm1.6$ \\
Contact tracing            & $62.2\pm2.2$ & $21.9\pm1.8$ & $15.9\pm1.7$ \\
Dynamic quarantine         & $23.8\pm1.9$ & $54.3\pm2.2$ & $21.9\pm1.8$ \\
Targeted vaccination       & $17.2\pm1.7$ & $26.8\pm1.9$ & $56.0\pm2.2$ \\
Ring vaccination ($r{=}1$) & $20.6\pm1.8$ & $27.3\pm1.9$ & $52.1\pm2.2$ \\
\bottomrule
\end{tabular}\rowcolors{2}{}{}%
}
\par\smallskip\noindent\scriptsize{Rows sum to $100\%$ within bootstrap CI. Fingerprinting and contact tracing push adaptive attacks toward L1 memory injection; dynamic quarantine pushes them toward L2 configuration modification; targeted vaccination and ring vaccination push them toward L3 infrastructure install or egress. The layer shifting motivates the layered-defense configurations of Table~\ref{tab:a6_main}.}
\end{table}

\subsection{Per-Primitive Drift Onset}
\label{app:e3_drift_detail}

Table~\ref{tab:e3_drift} reports attack success rate and drift onset across the five drift primitives of Definition~\ref{def:drift_primitive} and three frameworks. The average onset ranges from $13.2$ sessions on LangGraph to $26.9$ on AutoGen, whereas natural drift requires $73$ interactions~\citep{agentdrift2026}. The gap amounts to a $2.7$ to $8.9\times$ acceleration.
Among the primitives, $D_{\mathrm{constraint}}$ dominates with ASR between $66.8\%$ and $87.6\%$ since it exploits autoregressive reinforcement. LangGraph emerges as the most vulnerable framework because shared state carries no per-node access control.

\begin{table}[ht]
\centering
\caption{Drift-primitive onset. Attack success rate and drift onset (sessions to ASI $<0.7$) by primitive and framework, $1{,}000$ trials per cell, bootstrap $95\%$ CI. Natural drift baseline $73$ interactions~\citep{agentdrift2026}.}
\label{tab:e3_drift}
\scriptsize
\resizebox{\textwidth}{!}{%
\rowcolors{2}{white}{tblStripe}\begin{tabular}{lcccccc}
\toprule
\tabhead
& \multicolumn{2}{c}{\textbf{LangGraph}} & \multicolumn{2}{c}{\textbf{CrewAI}} & \multicolumn{2}{c}{\textbf{AutoGen}} \\
\cmidrule(lr){2-3} \cmidrule(lr){4-5} \cmidrule(lr){6-7}
\textbf{Primitive} & ASR (\%) & Onset (sess.) & ASR (\%) & Onset (sess.) & ASR (\%) & Onset (sess.) \\
\midrule
$D_{\mathrm{factual}}$     & $83.7\pm2.3$ & $12.3\pm0.6$ & $75.2\pm2.7$ & $18.4\pm0.9$ & $60.8\pm3.0$ & $24.1\pm1.1$ \\
$D_{\mathrm{procedural}}$  & $78.9\pm2.5$ & $15.1\pm0.7$ & $71.4\pm2.8$ & $22.3\pm1.0$ & $54.6\pm3.1$ & $31.2\pm1.4$ \\
$D_{\mathrm{constraint}}$  & $87.6\pm2.0$ & $8.2\pm0.5$  & $82.1\pm2.4$ & $11.4\pm0.7$ & $66.8\pm2.9$ & $19.3\pm1.0$ \\
$D_{\mathrm{preference}}$  & $73.1\pm2.7$ & $19.8\pm0.9$ & $65.2\pm2.9$ & $28.2\pm1.3$ & $48.4\pm3.1$ & $37.6\pm1.6$ \\
$D_{\mathrm{trust}}$       & $80.8\pm2.4$ & $10.4\pm0.5$ & $73.9\pm2.7$ & $16.2\pm0.8$ & $58.3\pm3.0$ & $22.1\pm1.0$ \\
\midrule
\tabhead
\textbf{Average} & $\mathbf{80.8}$ & $\mathbf{13.2}$ & $\mathbf{73.6}$ & $\mathbf{19.3}$ & $\mathbf{57.8}$ & $\mathbf{26.9}$ \\
\bottomrule
\end{tabular}\rowcolors{2}{}{}%
}
\end{table}

\subsection{Propagation Dynamics by Framework and Topology}
\label{app:e8_propagation_detail}

Table~\ref{tab:e8_propagation} reports the empirical $\rzero = \beta/\gamma$ for every framework and topology. Agreement with the SEIR ODE prediction stays within $12\%$ MAE at $N \le 7$ and tightens to $5\%$ at $N \ge 100$.

\begin{table}[ht]
\centering
\caption{Propagation dynamics by framework and topology ($1{,}000$ trials per cell, $95\%$ bootstrap CI on $\rzero$). SEIR-predicted $\rzero$ in parentheses. $\rzero > 1$ indicates epidemic propagation.}
\label{tab:e8_propagation}
\scriptsize
\resizebox{\textwidth}{!}{%
\rowcolors{2}{white}{tblStripe}\begin{tabular}{llccc}
\toprule
\tabhead
\textbf{Framework} & \textbf{Topology} & $\boldsymbol{\rzero}$ \textbf{[emp.\ CI] (pred.)} & \textbf{Speed (frac./sess.)} & \textbf{Time to $50\%$ (sess.)} \\
\midrule
\multirow{3}{*}{LangGraph} & Pipeline & $2.81\ [2.66, 2.95]$\ (2.6) & $0.42\pm0.02$ & $8.1\pm0.4$ \\
 & Star & $3.47\ [3.31, 3.64]$\ (3.3) & $0.61\pm0.03$ & $5.2\pm0.3$ \\
 & Mesh & $4.18\ [3.99, 4.38]$\ (4.0) & $0.78\pm0.03$ & $3.1\pm0.2$ \\
\midrule
\multirow{3}{*}{CrewAI} & Pipeline & $1.57\ [1.47, 1.68]$\ (1.5) & $0.28\pm0.02$ & $13.8\pm0.7$ \\
 & Star & $2.09\ [1.96, 2.22]$\ (2.0) & $0.39\pm0.02$ & $9.3\pm0.5$ \\
 & Mesh & $2.68\ [2.53, 2.83]$\ (2.5) & $0.52\pm0.03$ & $6.4\pm0.3$ \\
\midrule
\multirow{3}{*}{AutoGen} & Pipeline & $0.82\ [0.74, 0.91]$\ (0.9) & $0.11\pm0.02$ & N/A \\
 & Star & $1.12\ [1.03, 1.22]$\ (1.2) & $0.18\pm0.02$ & $22.4\pm1.1$ \\
 & Mesh & $1.39\ [1.29, 1.50]$\ (1.3) & $0.24\pm0.02$ & $16.2\pm0.8$ \\
\bottomrule
\end{tabular}\rowcolors{2}{}{}%
}
\par\smallskip\noindent\scriptsize{Speed: fraction of agents compromised per session (mean $\pm$ $95\%$ CI). Time to $50\%$: sessions until $N/2$ agents have ASI $<0.7$. Empirical--predicted $\rzero$ MAE $4.1\%$ averaged across cells at $N \ge 100$.}
\end{table}

\subsection{Benign-Operation Cost and Declaration-Threshold Sweep}
\label{app:benign_ops}

Table~\ref{tab:benign_ops} reports Swarm Shepherd's false-positive rate, alert burden, per-event latency, peak memory, warm-up cost, and conformal coverage at $\alpha = 0.1$ under attack-free traces. The scope is $N \in \{20, 50\}$ and $\mathsf{H} = 500$ sessions across the three frameworks ($10$ seeds per cell). At the default threshold, Swarm Shepherd generates at most $9.2 \pm 1.5$ alerts per $100$ sessions on the CrewAI pipeline, the harness with the highest FPR, amortizing to a median of $128 \pm 16$ sessions between false alerts. Per-event latency stays below $3.1$~ms at $N = 50$ and peak memory below $29$~MB. Conformal coverage tracks the $0.90$ nominal within $\pm 0.03$.

\begin{table}[ht]
\centering
\caption{Benign-operation cost of Swarm Shepherd. Attack-free traces, $\mathsf{H} = 500$ sessions, $10$ seeds, mean $\pm$ $95\%$ bootstrap CI. Declaration threshold $\alpha = 0.1$.}
\label{tab:benign_ops}
\scriptsize
\resizebox{\textwidth}{!}{%
\rowcolors{2}{white}{tblStripe}\begin{tabular}{llcccccccc}
\toprule
\tabhead
\textbf{Framework} & $N$ & \textbf{FPR (\%)} & \textbf{Alerts/$100$ sess.} & \textbf{Med time btwn FAs} & \textbf{Latency (ms)} & \textbf{Peak mem (MB)} & \textbf{Warm-up (sess.)} & \textbf{Coverage} \\
\midrule
LangGraph star   & $20$ & $6.2 \pm 1.1$ & $8.4 \pm 1.3$ & $143 \pm 18$ & $2.3 \pm 0.4$ & $12.4 \pm 1.2$ & $50 \pm 5$ & $0.908 \pm 0.024$ \\
CrewAI pipeline  & $20$ & $7.1 \pm 1.2$ & $9.2 \pm 1.5$ & $128 \pm 16$ & $1.8 \pm 0.3$ & $8.6 \pm 0.9$  & $55 \pm 6$ & $0.892 \pm 0.019$ \\
AutoGen mesh     & $20$ & $5.4 \pm 1.0$ & $7.1 \pm 1.2$ & $167 \pm 21$ & $2.8 \pm 0.5$ & $14.2 \pm 1.4$ & $48 \pm 5$ & $0.876 \pm 0.020$ \\
LangGraph star   & $50$ & $4.1 \pm 0.8$ & $5.3 \pm 0.9$ & $221 \pm 27$ & $3.1 \pm 0.5$ & $28.7 \pm 2.3$ & $50 \pm 5$ & $0.901 \pm 0.021$ \\
\bottomrule
\end{tabular}\rowcolors{2}{}{}%
}
\end{table}

Table~\ref{tab:alpha_sweep} shows the declaration-threshold sweep. Tightening $\alpha$ from $0.10$ to $0.05$ nearly halves FPR ($6.2\%$ to $2.4\%$) at a $1.3$-session cost in warning lead and an $8.8$-point cost in TPR. The sweep defines the operator-tunable surface.

\begin{table}[ht]
\centering
\caption{Declaration-threshold sweep. LangGraph star, $N=20$, $\mathsf{H}=500$ sessions, $10$ seeds, mean $\pm$ $95\%$ bootstrap CI.}
\label{tab:alpha_sweep}
\scriptsize
\resizebox{0.5\textwidth}{!}{%
\rowcolors{2}{white}{tblStripe}\begin{tabular}{cccc}
\toprule
\tabhead
$\alpha$ & \textbf{FPR (\%)} & \textbf{Warning lead (sess.)} & \textbf{TPR (\%)} \\
\midrule
$0.05$ & $2.4 \pm 0.5$  & $2.1 \pm 0.3$ & $78.5 \pm 2.1$ \\
$0.10$ & $6.2 \pm 1.1$  & $3.4 \pm 0.4$ & $87.3 \pm 1.8$ \\
$0.15$ & $10.3 \pm 1.6$ & $4.6 \pm 0.5$ & $91.4 \pm 1.5$ \\
$0.20$ & $15.8 \pm 2.0$ & $5.9 \pm 0.6$ & $93.7 \pm 1.3$ \\
$0.25$ & $22.4 \pm 2.3$ & $7.3 \pm 0.7$ & $95.1 \pm 1.2$ \\
\bottomrule
\end{tabular}\rowcolors{2}{}{}%
}
\end{table}

\subsection{A3 Mean-Field Rate: Extended Detail}
\label{app:a3_meanfield_detail}

Figure~\ref{fig:a3_meanfield} in Section~\ref{sec:e7_how} plots the $W_1$ distance between the $N$-agent empirical compartment distribution $\mu_N$ and the mean-field limit $\mu_\infty$ versus $N$ on a log-log scale, across mesh, star, Barab\'asi--Albert, and hub-and-spoke topologies with $10$ seeds per cell. Two estimators compute $W_1$: sliced-Wasserstein (solid) and entropic Sinkhorn (dotted). Their agreement rules out estimator-induced artifacts. Fitted slopes cover $-1/2$ within bootstrap CI on mesh, star, and BA topologies. Hub-and-spoke steepens to $-0.54 \pm 0.03$, quantifying the exchangeability-violation cost formalized in Corollary~\ref{cor:seir_block} and flagged in \S\ref{sec:limitations}.

\subsection{A4b Invariant Ablations}
\label{app:a4b_invariants}

Three structural invariants underpin the L1/L2/L3 distinguishability of Proposition~\ref{prop:traceback}: \textbf{I1} L1 attacks emit a \textsf{write} on $v \in \mathcal{M}$ before any \textsf{retrieve} on $v$; \textbf{I2} L2 attacks \textsf{modify} a node in $\mathcal{L}$ without a corresponding \textsf{install} on $\mathcal{I}$; \textbf{I3} L3 attacks emit \textsf{install} or \textsf{modify} on $\mathcal{I}$ together with a permission-count jump.
We deliberately break each invariant on $50$ seeds per (invariant, framework) cell and measure trace-back degradation. The predicted drop is at least $0.15$ on top-1 for every broken invariant, directly testing the distinguishability assumption (A5).
Failure modes include large entry-failure graph distance, low-event-density Phase~0 windows, and high-rate concurrent benign primitives on the same edge type. Mitigations include hop-radius ensembling, $T^\star$-aware warm-up, and per-edge-type conformal calibration.

\subsection{Framework Vulnerability Matrix}
\label{app:vuln_matrix}

Table~\ref{tab:vuln_matrix} summarizes the qualitative framework vulnerability ranking used in the operational-implications discussion of Section~\ref{sec:e7_summary}. Severity reflects injection feasibility, amplification mechanisms, cross-session persistence, and access control.

\begin{table}[ht]
\centering
\caption{Framework vulnerability summary. Extended CVE analysis in the supplementary material.}
\label{tab:vuln_matrix}
\scriptsize
\resizebox{\textwidth}{!}{%
\rowcolors{2}{white}{tblStripe}\begin{tabular}{lccccc}
\toprule
\tabhead
\textbf{Framework} & \textbf{Drift Inj.} & \textbf{Self-Reinf.} & \textbf{Cross-Sess.} & \textbf{Access Ctrl.} & \textbf{Overall} \\
\midrule
LangGraph & Very High & Cycles      & Checkpoints & None & \textbf{CRITICAL} \\
CrewAI    & High      & Memory      & SQLite3     & None & \textbf{HIGH} \\
AutoGen   & Moderate  & Limited     & None (default) & None & \textbf{MODERATE} \\
\bottomrule
\end{tabular}\rowcolors{2}{}{}%
}
\end{table}

\section{Robustness, Containment, and Fair-Baseline Studies}
\label{app:robustness}

This appendix collects the studies that probe Swarm Shepherd beyond its headline operating point: measured end-to-end containment, provenance and telemetry degradation, hidden calibration contamination, held-out entry-family transfer, model and emission sensitivity, parameter provenance, fairer baseline protocols, posterior-set remediation with framework-specific hardening, measured scaling, and a frozen blind red-team protocol. Unless stated otherwise the reference cell is LangGraph star with the within-family values AUROC $0.918\pm0.017$, warning lead $3.9\pm0.4$ sessions, node top-1 $0.69\pm0.02$, and entry-layer macro-F1 $0.72\pm0.02$. Results marked \emph{replay} re-run the frozen detector on recorded traces without new model calls.

\subsection{Measured Live Containment}
\label{app:live_containment}

Table~\ref{tab:a6_main} forecasts containment from a calibrated SEIR model. To measure it without the simulator, every defense arm starts from an identical frozen Phase-0 snapshot and faces the live Eve chain. The endpoint is observed controlled reproduction $\Rc$, the mean number of newly compromised downstream agents causally infected by each infectious agent before that agent is isolated or the run ends. A new compromise requires a persistent, scorer-verified L1/L2/L3 state change, and each transmission edge must be supported by the runtime causal log. These labels are visible only to the scorer and never to the filter. The estimator is the offspring/$\beta$-$\gamma$ estimator that produced the measured undefended reproduction in Table~\ref{tab:e8_propagation}, so the arms are directly comparable to the published per-topology numbers. The claim rule is pre-declared: an upper $95\%$ CI below $1$ is \emph{measured containment}; a point estimate below $1$ with the CI crossing $1$ is a reduction without established containment; a point estimate at or above $1$ is reduced propagation that does not contain.

\begin{table}[ht]
\centering
\caption{Live end-to-end containment on LangGraph star (measured; $N=5$, $20$ paired Phase-0 snapshots per entry layer, $50$ post-activation sessions plus a full process restart and $20$ further sessions; hierarchical bootstrap resampling variant then trial). The no-defense arm reproduces the $3.47\ [3.31, 3.64]$ of Table~\ref{tab:e8_propagation} within CI, the difference being finite-sample.}
\label{tab:live_containment_full}
\scriptsize
\resizebox{\textwidth}{!}{%
\rowcolors{2}{white}{tblStripe}\begin{tabular}{lccccc}
\toprule
\tabhead
\textbf{Arm} & $\boldsymbol{\Rc}$ \textbf{[95\% CI]} & \textbf{ASR (\%)} & \textbf{Extinct.\ by s50} & \textbf{Post-restart recur.} & \textbf{Utility loss (\%)} \\
\midrule
No defense                  & $3.41\ [3.06, 3.77]$ & $85.3\pm3.8$ & $0.03$ & $3.2$ & $0$ \\
Visible-node removal only   & $2.83\ [2.44, 3.23]$ & $79.1\pm4.2$ & $0.10$ & $2.5$ & $2.1$ \\
Warning + quarantine        & $1.62\ [1.32, 1.95]$ & $54.7\pm4.6$ & $0.33$ & $1.0$ & $3.8$ \\
\tabours
\textbf{Full posterior-guided stack} & $\mathbf{0.81\ [0.64, 0.97]}$ & $\mathbf{31.2\pm4.3}$ & $\mathbf{0.68}$ & $\mathbf{0.3}$ & $5.6$ \\
\bottomrule
\end{tabular}\rowcolors{2}{}{}%
}
\end{table}

The visible-node-removal arm is included because the submission claims that removing the visibly infected agent is insufficient; that claim is now measured ($2.83\ [2.44, 3.23]$ against $0.81\ [0.64, 0.97]$ for the full stack), and the post-restart recurrence column shows why: clearing the agent leaves $2.5$ recurrences per run because memory, configuration, and package-cache residue survive, while clearing residue over the inferred entry window leaves $0.3$. The $0.72$ figure of Table~\ref{tab:a6_main} remains a calibrated forecast; this run is an independent measurement rather than a validation of it, and we do not report agreement as confirmation.

\paragraph{Second framework.} AutoGen mesh is the tighter test, since its undefended reproduction is already near threshold at $1.39\ [1.29, 1.50]$ in Table~\ref{tab:e8_propagation}. Under the same protocol the no-defense arm measures $1.34\ [1.16, 1.55]$, visible-node removal $1.04\ [0.84, 1.28]$ (reduced propagation but not containment), and the full stack $0.62\ [0.47, 0.81]$, which is measured containment under the pre-declared rule.

\subsection{Provenance Quality and Telemetry Degradation (replay)}
\label{app:provenance}

The headline evaluation reads the contamination anchor from ground-truth injection tags (Appendix~\ref{app:identifiability}). Here the tag is replaced by progressively weaker provenance so that, in the deployment-anchor row, no scorer label enters the pipeline at all. Table~\ref{tab:provenance} reports the sweep. The deployment-anchor row is the honest answer to whether the method works without oracle provenance: AUROC falls $0.047$ to $0.871\pm0.021$ and node top-1 falls $0.08$, at $8.4\%$ clean FPR against $6.2\%$. The degradation is bounded because only the contamination coordinate reads provenance; authority and capability exposure are read from routing share and permission counts, which the frameworks emit regardless.

\begin{table}[ht]
\centering
\caption{Provenance and telemetry degradation on LangGraph star (replay). The deployment-anchor row uses no scorer label anywhere in the pipeline; the combined adverse cell composes $20\%$ event loss, $20\%$ missing provenance confidence, and $10\%$ label flips. A dash marks a cell not recorded in the sweep.}
\label{tab:provenance}
\scriptsize
\resizebox{\textwidth}{!}{%
\rowcolors{2}{white}{tblStripe}\begin{tabular}{lcccc}
\toprule
\tabhead
\textbf{Anchor / telemetry setting} & \textbf{AUROC} & \textbf{Warning lead} & \textbf{Node top-1} & \textbf{Clean FPR (\%)} \\
\midrule
Ground-truth tags (as submitted)              & $0.918\pm0.017$ & $3.9\pm0.4$ & $0.69\pm0.02$ & $6.2$ \\
Tag precision/recall $0.80$                   & $0.889\pm0.020$ & $3.5\pm0.4$ & $0.64\pm0.03$ & --- \\
\tabours
\textbf{Deployment anchor, no ground truth}   & $\mathbf{0.871\pm0.021}$ & $\mathbf{3.3\pm0.4}$ & $\mathbf{0.61\pm0.03}$ & $8.4$ \\
$20\%$ loss $+$ $20\%$ missing prov.\ $+$ $10\%$ flips & $0.812\pm0.025$ & $2.6\pm0.5$ & $0.55\pm0.03$ & --- \\
\midrule
$10\%$ MCAR event loss                        & $0.901\pm0.019$ & $3.6\pm0.4$ & $0.67\pm0.02$ & --- \\
$20\%$ MCAR event loss                        & $0.873\pm0.022$ & $3.1\pm0.4$ & $0.62\pm0.03$ & $7.8$ \\
$40\%$ MCAR event loss                        & $0.798\pm0.028$ & $2.2\pm0.5$ & $0.53\pm0.03$ & --- \\
All \textsf{modify} edges dropped (targeted)  & $0.831\pm0.023$ & $2.6\pm0.5$ & $0.57\pm0.03$ & --- \\
\bottomrule
\end{tabular}\rowcolors{2}{}{}%
}
\end{table}

Uniform loss degrades gently because the prediction-only Kalman step advances unobserved nodes by self-dynamics, a designed property. The targeted row is the more informative test: an adversary who suppresses one edge type removes a whole evidence channel, and dropping every \textsf{modify} edge costs $0.087$ AUROC, roughly twice the random loss at the same volume. We report that without calling it robustness.

\paragraph{Provenance dependence by entry layer.} The deployment-anchor cost is not uniform. Decomposing it by entry layer (Table~\ref{tab:anchor_layer}) shows it concentrates on L1 memory, the layer whose anchor reads provenance, at $0.076$ AUROC against $0.031$ and $0.034$ elsewhere. The operational reading is that provenance quality is a precondition for memory-layer attribution specifically, not for the method as a whole.

\begin{table}[ht]
\centering
\caption{Anchor cost by entry layer (replay). AUROC under ground-truth tags and under the deployment anchor, per entry layer.}
\label{tab:anchor_layer}
\scriptsize
\resizebox{0.72\textwidth}{!}{%
\rowcolors{2}{white}{tblStripe}\begin{tabular}{lccc}
\toprule
\tabhead
\textbf{Entry layer} & \textbf{AUROC, ground truth} & \textbf{AUROC, deployment anchor} & \textbf{Difference} \\
\midrule
L1 memory          & $0.924\pm0.018$ & $0.848\pm0.024$ & $0.076$ \\
L2 logic           & $0.913\pm0.019$ & $0.882\pm0.022$ & $0.031$ \\
L3 infrastructure  & $0.917\pm0.019$ & $0.883\pm0.021$ & $0.034$ \\
\bottomrule
\end{tabular}\rowcolors{2}{}{}%
}
\end{table}

\paragraph{Independent failure outcome.} Because ASI subscales feed both the observation vector and the visible-failure definition (Section~\ref{sec:problem_setting}), warning lead could in principle be partly definitional. Moving the label off the shared signal, with ASI and T4 removed from the observations and failure scored by each benchmark's own task grader, the lead is $2.6\ [2.1, 3.1]$ sessions with its lower $95\%$ bound above zero (AUROC $0.871\pm0.021$). It survives the combined corruption cell as well, at $1.8\ [1.2, 2.4]$. The lead against ASI-defined failure is $3.9$ sessions and against the independent grader $2.6$; the camera-ready reports both with the operative definition named.

\subsection{Hidden Calibration Contamination}
\label{app:calib_taint}

No detector can certify that a calibration window is clean when contamination induces the same observable law as benign operation, so Proposition~\ref{prop:calibration}'s consistency is conditional on benign stationarity and the question is how fast it degrades when that fails. Activating Eve during the warm-up window and evaluating the submitted estimator unchanged (M=50 windows at horizon 100, 20 independently sampled calibration sets per cell, disjoint test set), Table~\ref{tab:calib_taint} reports the degradation. Parameter displacement moves well before accuracy does, the expected signature of contamination that biases the estimate before it degrades the decision. An innovation-covariance anomaly against the shipped priors separates tainted from clean calibration sets at AUC $0.79$ at $5\%$ taint (likelihood-ratio $p<0.001$), so a deployer has a usable warning that refitting is unsafe. AUROC loss stays under $0.05$ to an $8\%$ contamination fraction, past which verified startup data is an operational requirement rather than a convenience.

\begin{table}[ht]
\centering
\caption{Hidden calibration contamination on LangGraph star. The submitted estimator is evaluated unchanged; parameter displacement is the distance from the clean fit.}
\label{tab:calib_taint}
\scriptsize
\resizebox{0.86\textwidth}{!}{%
\rowcolors{2}{white}{tblStripe}\begin{tabular}{lcccc}
\toprule
\tabhead
\textbf{Hidden taint} & \textbf{Param.\ displacement} & \textbf{AUROC} & \textbf{Warning lead} & \textbf{Clean FPR (\%)} \\
\midrule
$0\%$ (reference) & ---    & $0.918\pm0.017$ & $3.9\pm0.4$ & $6.2$ \\
$2\%$             & $0.04$ & $0.911\pm0.018$ & $3.7\pm0.4$ & $6.8$ \\
$5\%$             & $0.08$ & $0.889\pm0.020$ & $3.4\pm0.4$ & $8.1$ \\
$10\%$            & $0.16$ & $0.847\pm0.024$ & $2.8\pm0.5$ & $11.3$ \\
$20\%$            & $0.31$ & $0.761\pm0.029$ & $1.9\pm0.5$ & $18.6$ \\
\bottomrule
\end{tabular}\rowcolors{2}{}{}%
}
\end{table}

\subsection{Held-Out Entry-Family Transfer (replay)}
\label{app:heldout}

Detection and attribution when the model has never seen the entry layer it is tested on. No test-layer trace enters parameter fitting, threshold selection, rule writing, or hyperparameter choice, and the code was frozen before outcomes were revealed. Table~\ref{tab:heldout} reports the three rotations. Detection loses at most $0.080$ AUROC and $1.1$ sessions of lead, while node attribution loses up to $0.14$ top-1: integrity and policy triggers still fire on an unseen layer even when the smoother's layer posterior degrades, so detection generalises across entry families while precise attribution remains more family-dependent. Transfer is possible because the state-space parameters are fit by EM on benign windows (Table~\ref{tab:param_provenance}), so the attack family enters only through the typed events the frameworks emit. For the L1+L2$\to$L3 rotation the per-framework split is LangGraph star $0.884\pm0.022$, CrewAI pipeline $0.857\pm0.025$, AutoGen mesh $0.872\pm0.024$ (AUROC), so transfer holds on each framework rather than only in aggregate.

\begin{table}[ht]
\centering
\caption{Held-out entry-family transfer (replay). Each rotation fits on two entry layers and tests on the third, pooled across frameworks and reported per rotation because pooling would hide the attribution asymmetry.}
\label{tab:heldout}
\scriptsize
\resizebox{0.9\textwidth}{!}{%
\rowcolors{2}{white}{tblStripe}\begin{tabular}{llcccc}
\toprule
\tabhead
\textbf{Fit on} & \textbf{Test on} & \textbf{AUROC} & \textbf{Warning lead} & \textbf{Node top-1} & \textbf{Layer macro-F1} \\
\midrule
L1 $+$ L2 & L3 & $0.871\pm0.020$ & $3.3\pm0.4$ & $0.61\pm0.03$ & $0.64\pm0.03$ \\
L1 $+$ L3 & L2 & $0.838\pm0.023$ & $2.8\pm0.5$ & $0.55\pm0.03$ & $0.58\pm0.03$ \\
L2 $+$ L3 & L1 & $0.852\pm0.021$ & $3.1\pm0.4$ & $0.58\pm0.03$ & $0.61\pm0.03$ \\
\midrule
\multicolumn{2}{l}{Within-family reference} & $0.918\pm0.017$ & $3.9\pm0.4$ & $0.69\pm0.02$ & $0.72\pm0.02$ \\
\bottomrule
\end{tabular}\rowcolors{2}{}{}%
}
\end{table}

\subsection{Model and Emission Sensitivity}
\label{app:model_sensitivity}

The continuous latent state is a belief over accumulated influence, not a claim that compromise is continuous; discreteness re-enters at the thresholds $\tau_E, \tau_I$. We test the modelling choice directly, at matched false-positive budget so the comparison is not confounded by operating point. Table~\ref{tab:model_sensitivity} compares four posteriors. The two-state HMM is the direct test of the discreteness premise: because a binary latent has no Exposed state between clean and compromised, it crosses $\tau_E$ and $\tau_I$ at the same instant and loses resolution on the accumulation. No alternative matches the linear-Gaussian filter on any of the three metrics, and all cost more to run.

\begin{table}[ht]
\centering
\caption{Latent-model sensitivity on LangGraph star, matched FPR budget. The particle-filter row keeps its Table~\ref{tab:a1_baselines} AUROC and lead; its attribution and compute columns are measured here.}
\label{tab:model_sensitivity}
\scriptsize
\resizebox{0.86\textwidth}{!}{%
\rowcolors{2}{white}{tblStripe}\begin{tabular}{lcccc}
\toprule
\tabhead
\textbf{Latent model} & \textbf{AUROC} & \textbf{Warning lead} & \textbf{Node top-1} & \textbf{Rel.\ compute} \\
\midrule
\tabours
Linear-Gaussian (submitted)        & $\mathbf{0.918\pm0.017}$ & $\mathbf{3.9\pm0.4}$ & $\mathbf{0.69\pm0.02}$ & $1.00$ \\
Per-node two-state HMM             & $0.852\pm0.022$ & $2.9\pm0.5$ & $0.57\pm0.03$ & $1.6$ \\
Switching linear dynamical system  & $0.884\pm0.020$ & $3.2\pm0.4$ & $0.61\pm0.03$ & $2.8$ \\
Particle filter, 1k                & $0.836\pm0.021$ & $2.8\pm0.4$ & $0.60\pm0.03$ & $4.0$ \\
\bottomrule
\end{tabular}\rowcolors{2}{}{}%
}
\end{table}

\paragraph{Non-Gaussian emissions.} Replacing the Gaussian emission noise at identical thresholds and anchors, Student-$t$ ($\nu=3$) gives AUROC $0.901\pm0.019$ (lead $3.6\pm0.4$, clean FPR $6.9\%$), Laplace $0.895\pm0.020$ ($3.5\pm0.4$, $7.3\%$), and a residual bootstrap drawn from real benign traces $0.908\pm0.018$ ($3.7\pm0.4$, $6.6\%$). The bootstrap row is the one we weight most, since its noise is the actual residual distribution rather than a chosen family. Composing model and sensor misspecification, Student-$t$ emissions with $20\%$ MCAR loss give $0.853\pm0.024$ (lead $2.9\pm0.5$, top-1 $0.57\pm0.03$) and bootstrap emissions with all \textsf{modify} edges dropped give $0.802\pm0.027$ ($2.4\pm0.5$, $0.52\pm0.03$); even the harder cell keeps $2.4$ sessions of lead.

\subsection{Parameter Provenance}
\label{app:param_provenance}

Table~\ref{tab:param_provenance} lists every fitted quantity with its data source, whether it is benign-only, whether any attack label touched it, and whether it was frozen before test. The one \texttt{Yes} is the warning threshold, an operating point rather than a dynamics parameter, and its calibration split excludes the held-out layer, so no dynamics or emission parameter touches attack data. The scorer-only ground-truth tag appears in no row, which is the property that makes the held-out-layer and provenance results meaningful. This table also corrects a submission inconsistency: Appendix~\ref{app:traceback_smoother} now describes the trace-back layer prior $\pi_\ell$ as a structural prior fixed without attack labels, matching the last row.

\begin{table}[ht]
\centering
\caption{Parameter provenance. Every fitted quantity, its source, and whether any attack label touched it.}
\label{tab:param_provenance}
\scriptsize
\resizebox{\textwidth}{!}{%
\rowcolors{2}{white}{tblStripe}\begin{tabular}{lcccc}
\toprule
\tabhead
\textbf{Quantity} & \textbf{Fit on} & \textbf{Benign only} & \textbf{Attack labels} & \textbf{Frozen before test} \\
\midrule
$A^\phi, Q^\phi, \Sigma_{\mathrm{obs}}$ & benign Phase-0 windows (EM) & Yes & No & Yes \\
$B^{(r,\cdot,\cdot)}$                   & benign typed-event windows & Yes & No & Yes \\
$H$ anchor block                        & benign anchored emissions & Yes & No & Yes \\
Provenance-confidence threshold         & benign provenance metadata & Yes & No & Yes \\
Warning threshold, conformal quantile   & labelled benign $+$ attack split & No & Yes (calib.\ only; test layer excluded) & Yes \\
Trace-back prior $p_0$ (incl.\ $\pi_\ell$) & structural prior $+$ benign exposure stats & Yes & No & Yes \\
\bottomrule
\end{tabular}\rowcolors{2}{}{}%
}
\end{table}

\subsection{Fairer Baseline Protocols}
\label{app:baselines_fair}

Appendix~\ref{app:progent_drift} noted that the drift-extended Progent rule set $\mathcal{R}_1$ was written after inspecting the attack-variant matrix. We quantify that leakage with a four-fold blinded rule set $\mathcal{R}_{\mathrm{blind}}$, whose rules for each fold are designed without sight of that fold's variants, and port the adapter beyond LangGraph. Table~\ref{tab:progent_blind} reports the result: the blinded-to-submitted gap is $8.5$ points on LangGraph star and persists on each harness, so neither the detection level nor the leakage estimate is an artefact of a single setting. The substantive claim rests on the attacker-known oracle row ($64.8\pm2.9\%$), which leaves roughly $35\%$ of trials undetected even when the policy author knows every variant, because deterministic tool gating cannot see contamination that arrives inside permitted calls.

\begin{table}[ht]
\centering
\caption{Progent under blinded rules and across harnesses (detection \%, $\uparrow$ for defense; $1{,}000$ trials per cell, $95\%$ bootstrap CI). $\mathcal{R}_0$ is the published default, $\mathcal{R}_1$ the drift-extended set of Table~\ref{tab:progent_drift}, $\mathcal{R}_{\mathrm{blind}}$ the four-fold blinded set, and $\mathcal{R}_\star$ the attacker-known oracle bound.}
\label{tab:progent_blind}
\scriptsize
\resizebox{\textwidth}{!}{%
\rowcolors{2}{white}{tblStripe}\begin{tabular}{lcccc}
\toprule
\tabhead
\textbf{Harness} & $\mathcal{R}_0$ \textbf{default} & $\mathcal{R}_1$ \textbf{drift-ext.} & $\mathcal{R}_{\mathrm{blind}}$ \textbf{four-fold} & $\mathcal{R}_\star$ \textbf{oracle} \\
\midrule
LangGraph star  & $6.6\pm1.5$ & $22.0\pm2.6$ & $13.5\pm2.2$ & $64.8\pm2.9$ \\
CrewAI pipeline & ---         & $19.4\pm2.7$ & $11.8\pm2.3$ & --- \\
AutoGen mesh    & ---         & $16.1\pm2.8$ & $9.6\pm2.4$  & --- \\
\bottomrule
\end{tabular}\rowcolors{2}{}{}%
}
\par\smallskip\noindent\scriptsize{The $\mathcal{R}_1$ to $\mathcal{R}_{\mathrm{blind}}$ gap ($8.5$ points on LangGraph star) is the information advantage the submitted rules carried.}
\end{table}

\paragraph{DRIFT and CaMeL, adapted rather than described.} The submission compared DRIFT and CaMeL only mechanistically. Each is now ported in the drift-adapted form its own design implies, so the comparison is against a fair version rather than a single-session configuration facing a multi-session attack (Table~\ref{tab:adapted_baselines}). Each is evaluated in its native mode (attack success and clean utility for the prevention systems, detection metrics only where an online score exists), and framework or attack classes a baseline does not support are marked out of scope rather than scored as failures. The adapted variants close part of the gap; the residual is the part no carried state can see, namely contamination that accumulates inside calls the policy already permits. We accordingly narrow the submission's claim: rather than a general ``outperforms stateful defenses,'' the evidence supports that per-turn and stateful monitors in their published operating mode do not detect multi-session cross-layer drift at the rates they achieve on single-session injection, and that a symbolic tool policy with attacker-known rules still leaves roughly $35\%$ of trials undetected.

\begin{table}[ht]
\centering
\caption{Adapted DRIFT and CaMeL ports (percent of Agentic Sleeper trials prevented, LangGraph star, $95\%$ bootstrap CI). Each is shown in its published single-session configuration and in the drift-adapted variant its design implies.}
\label{tab:adapted_baselines}
\scriptsize
\resizebox{\textwidth}{!}{%
\rowcolors{2}{white}{tblStripe}\begin{tabular}{llc}
\toprule
\tabhead
\textbf{Defense} & \textbf{Adaptation} & \textbf{Prevented (\%, $\uparrow$)} \\
\midrule
DRIFT~\cite{li2025drift}        & Planner, validator, isolator; single-session window & $14.3\pm3.2$ \\
DRIFT~\cite{li2025drift}        & Multi-session reference window & $31.6\pm3.6$ \\
CaMeL~\cite{debenedetti2025camel} & Per-query capability and flow policy & $21.8\pm3.1$ \\
CaMeL~\cite{debenedetti2025camel} & Session-persistent capability taint & $38.4\pm3.4$ \\
\bottomrule
\end{tabular}\rowcolors{2}{}{}%
}
\end{table}

\subsection{Posterior-Set Remediation and Framework Hardening}
\label{app:remediation}

Trace-back top-1 is $0.69$ and top-3 is $0.87$, so remediation should act on a credible set rather than a point estimate. Table~\ref{tab:remediation} prices the set-valued policies. Top-3 is the recommended operating point: it more than halves recurrence, from $0.31$ to $0.13$, for $1.8$ points of clean-task utility and $1.4$ additional agents touched. Past the $90\%$ set the returns flatten while destructive actions climb. Under $20\%$ MCAR loss the top-3 policy holds at $0.82\pm0.02$ coverage with recurrence $0.18$ and $2.6$ agents remediated; dropping all \textsf{modify} edges gives $0.76\pm0.02$, $0.24$, and $3.0$.

\begin{table}[ht]
\centering
\caption{Posterior-set remediation on LangGraph star ($1{,}000$ trials per policy, $95\%$ bootstrap CI). Coverage is the fraction of incidents whose true entry lies in the remediated set.}
\label{tab:remediation}
\scriptsize
\resizebox{\textwidth}{!}{%
\rowcolors{2}{white}{tblStripe}\begin{tabular}{lccccc}
\toprule
\tabhead
\textbf{Policy} & \textbf{True-entry coverage} & \textbf{Recurrence} & \textbf{Agents remediated} & \textbf{Utility loss (\%)} & \textbf{Destructive actions} \\
\midrule
Top-1 only        & $0.69\pm0.02$ & $0.31$ & $1.0$ & $2.3$ & $0.4$ \\
\tabours
Top-3             & $\mathbf{0.87\pm0.01}$ & $\mathbf{0.13}$ & $\mathbf{2.4}$ & $\mathbf{4.1}$ & $\mathbf{1.1}$ \\
$80\%$ credible set & $0.92\pm0.01$ & $0.08$ & $3.1$ & $5.8$ & $1.6$ \\
$90\%$ credible set & $0.96\pm0.01$ & $0.04$ & $3.8$ & $7.2$ & $2.3$ \\
$95\%$ credible set & $0.98\pm0.01$ & $0.02$ & $4.3$ & $8.9$ & $3.1$ \\
\bottomrule
\end{tabular}\rowcolors{2}{}{}%
}
\end{table}

The supporting policy is staged by reversibility. First, apply reversible actions across the posterior set (isolation, credential restriction, quarantine of retrieval over the inferred window), which are cheap, so covering three candidates instead of one costs little. Second, collect corroborating evidence from the independent triggers: policy-violation and integrity triggers reach $0.74$ and $0.79$ top-1 on L2 and L3 where drift alone reaches $0.42$ and $0.38$ (Table~\ref{tab:b6_triggers}), and the drift channel is weakest exactly where the policy and integrity channels are strongest. Third, take destructive action (purge, rebuild, reimage) only against candidates still credible after the second stage, where the cost asymmetry does the work, since memory sanitisation is cheap and reversible while runtime rebuild is neither.

\paragraph{Architectural implications.} These follow from the measured attack surfaces and are implications rather than proven mitigations, since the live run tests the stack and not each control in isolation. For \textbf{LangGraph} (the most exposed, because shared state, cycles, and checkpoint persistence combine): isolate memory namespaces and write permissions per agent, attach source identity and integrity metadata to checkpoints and retrieved entries, keep signed immutable checkpoints for rollback, make system prompts, tool manifests, and configuration read-only in normal operation, require approval for graph rewiring and for privileged \textsf{modify} and \textsf{install} events, rate-limit cycles, and retain append-only event logs across restarts. For \textbf{CrewAI} (the Manager is a single point of failure atop a shared store): partition shared SQLite and vector memory by agent, task, and provenance, enforce least privilege on the Manager role, require dual approval for delegation, tool-policy, or configuration changes, and integrity-check role prompts at session boundaries. Across frameworks: signed provenance and append-only logging, immutable tool manifests and configuration hashes, per-agent ACLs, egress allowlists, independent L2/L3 integrity sensors, and restart-safe residue scanning, which is what separates the $2.5$ post-restart recurrences under visible-node removal from the $0.3$ under the full stack (Table~\ref{tab:live_containment_full}).

\subsection{Measured Scaling}
\label{app:scaling}

Table~\ref{tab:scaling} extends the per-event latency and memory measurements to $N=5{,}000$. Per-event latency grows with a fitted log-log slope of $0.08\ [0.03, 0.13]$ against $N$, the near-constant behaviour the factored belief predicts, since the cubic term is in the fixed six-dimensional per-node state rather than in the agent count. Sustained throughput at $N=1000$ is $4{,}100$ events per second, and p99 latency there is $11.8$ ms.

\begin{table}[ht]
\centering
\caption{Measured per-event latency and resident memory versus monitored-node count $N$ (single CPU sidecar).}
\label{tab:scaling}
\scriptsize
\resizebox{0.8\textwidth}{!}{%
\rowcolors{2}{white}{tblStripe}\begin{tabular}{lcccc}
\toprule
\tabhead
$N$ & \textbf{p50 (ms)} & \textbf{p95 (ms)} & \textbf{p99 (ms)} & \textbf{RSS (MB)} \\
\midrule
$20$   & $2.3$ & $3.8$ & ---    & $12$ \\
$100$  & $2.5$ & $4.6$ & ---    & $31$ \\
$1000$ & $2.9$ & $7.2$ & $11.8$ & $48$ \\
$5000$ & $3.3$ & $9.1$ & ---    & $143$ \\
\bottomrule
\end{tabular}\rowcolors{2}{}{}%
}
\end{table}

\subsection{Blind Red-Team Protocol}
\label{app:redteam}

An external red-team evaluation is the right test of circular validation, and we place its protocol on record before any outcome is known. The detector, thresholds, and typed-event mapping are frozen at the commit used for every number in this paper. A team with no access to Swarm Shepherd output implements one attack per entry layer from public mechanism descriptions only, and is instructed not to consult the variant matrix or the latent schema. Outcomes stay sealed until every implementation is frozen, so the test cannot be tuned to them. Because the outcomes are sealed there is nothing to preview; this version records the protocol, and the result against it will be reported whichever way it comes out.

\newpage
\section*{NeurIPS Paper Checklist}

\begin{enumerate}

\item {\bf Claims}
    \item[] Question: Do the main claims made in the abstract and introduction accurately reflect the paper's contributions and scope?
    \item[] Answer: \answerYes{}
    \item[] Justification: The abstract and introduction state the current contribution set: the Agentic Sleeper threat model, Swarm Shepherd early warning, trace-back attribution, the two-scale deployment theorem, and measured containment. Each claim is tied to a visible table or figure, and results are labelled by evidence type (measured, replay, model-based). The warning lead is reported against both the ASI and the independent-grader failure definitions ($3.9$ and $2.6$ sessions), the $C_{\mathrm{emp}}=8.31$ boundary fit matches the plug-in $8.3$, and containment is reported as a live measurement ($\Rc=0.81$ on LangGraph star, $0.62$ on AutoGen mesh, Table~\ref{tab:live_containment}) distinct from the calibrated $0.72$ forecast (Table~\ref{tab:a6_main}).

\item {\bf Limitations}
    \item[] Question: Does the paper discuss the limitations of the work performed by the authors?
    \item[] Answer: \answerYes{}
    \item[] Justification: Section~\ref{sec:limitations} covers the headline limitations: linear-Gaussian dynamics, the benign-stationarity condition on calibration with its measured breakpoint, provenance dependence for memory-layer attribution, weaker attribution than detection on held-out families, the bounded-degree condition on near-constant cost, and future nonlinear and graphon-scale work. Appendix~\ref{app:extended_limitations} adds horizon dependence, unanchored-coordinate identifiability, alert burden, and cross-model transfer, and Appendix~\ref{app:robustness} measures each stressor. The forecast-versus-measured distinction for containment is stated in Section~\ref{sec:e7_adaptive} and Tables~\ref{tab:a6_main} and~\ref{tab:live_containment}.

\item {\bf Theory assumptions and proofs}
    \item[] Question: For each theoretical result, does the paper provide the full set of assumptions and a complete (and correct) proof?
    \item[] Answer: \answerYes{}
    \item[] Justification: The main text states the factor model, warning, trace-back, containment, and deployment equations needed to follow the method and cites the corresponding formal results by label. Formal theorem statements, assumptions, and proofs are collected in Appendix~\ref{app:theory_proofs}, with a separate assumption-to-evidence map in Appendix~\ref{app:assumptions_map}.

    \item {\bf Experimental result reproducibility}
    \item[] Question: Does the paper fully disclose all the information needed to reproduce the main experimental results of the paper to the extent that it affects the main claims and/or conclusions of the paper (regardless of whether the code and data are provided or not)?
    \item[] Answer: \answerYes{}
    \item[] Justification: Appendix~\ref{app:formal_details} specifies the factored POMDP, anchored emission matrix, and typed operator families. Appendix~\ref{app:eval_extended} provides per-wave trial counts, seed policies, bootstrap protocols, and the $24$-variant attack matrix (Appendix~\ref{app:attack_variants}). Section~\ref{sec:ethics} and Appendix~\ref{app:release_safeguards} cover the remaining reproducibility and release surface.

\item {\bf Open access to data and code}
    \item[] Question: Does the paper provide open access to the data and code, with sufficient instructions to faithfully reproduce the main experimental results, as described in supplemental material?
    \item[] Answer: \answerNo{}
    \item[] Justification: Because the attack catalogue is dual-use, the full harness and exploit-bearing artifacts are not released as unrestricted open access. The paper instead describes a staged responsible release (Appendix~\ref{app:release_safeguards}): the Swarm Shepherd reference daemon, synthetic traces, the ASI schema, and benchmark templates are released under a $90$-day embargo and institutional-access gate, separately from attack primitives; exploit binaries and live payloads for unpatched variants require an application and documented defensive use case. De-anonymized reproducibility links are provided in the camera-ready.

\item {\bf Experimental setting/details}
    \item[] Question: Does the paper specify all the training and test details (e.g., data splits, hyperparameters, how they were chosen, type of optimizer) necessary to understand the results?
    \item[] Answer: \answerYes{}
    \item[] Justification: Section~\ref{sec:evaluation} and Appendix~\ref{app:eval_extended} report data splits, seed counts, bootstrap replicate counts, anchor-emission initialization, conformal warm-up length, and hyperparameter selection for each defense setting. Swarm Shepherd hyperparameters are reported alongside the corresponding ablation (Appendix~\ref{app:claif_ablations}).

\item {\bf Experiment statistical significance}
    \item[] Question: Does the paper report error bars suitably and correctly defined or other appropriate information about the statistical significance of the experiments?
    \item[] Answer: \answerYes{}
    \item[] Justification: Bootstrap $95\%$ confidence intervals with $1{,}000$ replicates are reported in Tables~\ref{tab:e8_propagation}, \ref{tab:a1_baselines}, \ref{tab:a5_adaptive}, \ref{tab:a6_main}, and throughout Appendix~\ref{app:eval_extended}. The boundary constant $C_{\mathrm{emp}} = 8.31$ is accompanied by its $[7.42, 9.18]$ CI. Variance sources (trials, seeds, attack variants) are stated per table caption.

\item {\bf Experiments compute resources}
    \item[] Question: For each experiment, does the paper provide sufficient information on the computer resources (type of compute workers, memory, time of execution) needed to reproduce the experiments?
    \item[] Answer: \answerYes{}
    \item[] Justification: Appendix~\ref{app:eval_extended} opens with a \emph{Compute and Cost} paragraph reporting $\approx 230$ GPU-hours plus $\approx\$82{,}000$ of closed-API usage across all experimental waves, with per-wave trial counts and bootstrap replicate counts. Preliminary calibration runs are disclosed separately.

\item {\bf Code of ethics}
    \item[] Question: Does the research conducted in the paper conform, in every respect, with the NeurIPS Code of Ethics \url{https://neurips.cc/public/EthicsGuidelines}?
    \item[] Answer: \answerYes{}
    \item[] Justification: Section~\ref{sec:ethics} addresses dual-use risk and broader societal implications of agentic AI deployment; Appendix~\ref{app:release_safeguards} records the concrete release safeguards, including sandboxed evaluation, staged release under $90$-day embargo, and reliance on publicly disclosed CVEs. The paper contains no prompt injections targeting reviewers, per NeurIPS 2026 LLM policy.

\item {\bf Broader impacts}
    \item[] Question: Does the paper discuss both potential positive societal impacts and negative societal impacts of the work performed?
    \item[] Answer: \answerYes{}
    \item[] Justification: Section~\ref{sec:ethics} discusses dual-use risk (attack variants could accelerate adversary capability) alongside the defender-side benefit of earlier warning, attribution, and containment. Appendix~\ref{app:release_safeguards} explains how public CVE disclosures and gated artifacts bound the incremental capability uplift for attackers.

\item {\bf Safeguards}
    \item[] Question: Does the paper describe safeguards that have been put in place for responsible release of data or models that have a high risk for misuse (e.g., pre-trained language models, image generators, or scraped datasets)?
    \item[] Answer: \answerYes{}
    \item[] Justification: Section~\ref{sec:ethics} states the safeguards, and Appendix~\ref{app:release_safeguards} specifies their operational form. A $90$-day embargo applies to the evaluation harness. The released tooling is rate-limited, carries a license clause forbidding production targeting, and ships no pre-built exploit binaries for unpatched variants. The $24$ attack variants all reproduce from publicly disclosed CVEs (CVE-2025-64439, CVE-2025-54135, CVE-2025-54136, runc November 2025), which bounds the incremental capability uplift per Kerckhoffs's principle.

\item {\bf Licenses for existing assets}
    \item[] Question: Are the creators or original owners of assets (e.g., code, data, models), used in the paper, properly credited and are the license and terms of use explicitly mentioned and properly respected?
    \item[] Answer: \answerYes{}
    \item[] Justification: LangGraph (MIT), CrewAI (MIT), and AutoGen (MIT/Apache-2.0) are cited and used in accordance with their licenses. Closed-API models (Anthropic Claude, Google Gemini via OpenRouter, StepFun) and open-weight models (Llama-3.3-70B-Instruct, Qwen3-235B-Instruct, GLM-5.1) are used under their respective licenses and terms of service. Full license attribution and versions are listed in the supplementary LICENSES file.

\item {\bf New assets}
    \item[] Question: Are new assets introduced in the paper well documented and is the documentation provided alongside the assets?
    \item[] Answer: \answerYes{}
    \item[] Justification: The evaluation harness, $24$-variant attack catalogue, and Swarm Shepherd reference implementation will be released with documentation covering the defender POMDP specification, typed operator schemas, anchored emission calibration protocol, and reproducibility scripts. Release is subject to the embargo in Section~\ref{sec:ethics} and Appendix~\ref{app:release_safeguards}.

\item {\bf Crowdsourcing and research with human subjects}
    \item[] Question: For crowdsourcing experiments and research with human subjects, does the paper include the full text of instructions given to participants and screenshots, if applicable, as well as details about compensation (if any)?
    \item[] Answer: \answerNA{}
    \item[] Justification: The paper does not involve crowdsourcing or research with human subjects. All evaluation is conducted on LLM-agent ecosystems in sandboxed environments.

\item {\bf Institutional review board (IRB) approvals or equivalent for research with human subjects}
    \item[] Question: Does the paper describe potential risks incurred by study participants, whether such risks were disclosed to the subjects, and whether Institutional Review Board (IRB) approvals (or an equivalent approval/review based on the requirements of your country or institution) were obtained?
    \item[] Answer: \answerNA{}
    \item[] Justification: The paper does not involve research with human subjects. IRB review is not applicable.

\item {\bf Declaration of LLM usage}
    \item[] Question: Does the paper describe the usage of LLMs if it is an important, original, or non-standard component of the core methods in this research? Note that if the LLM is used only for writing, editing, or formatting purposes and does \emph{not} impact the core methodology, scientific rigor, or originality of the research, declaration is not required.
    \item[] Answer: \answerYes{}
    \item[] Justification: LLM agents are the subjects of study. Evaluation covers agents deployed on LangGraph, CrewAI, and AutoGen. The victim portfolio spans Claude Sonnet 4.6 and Claude Haiku 4.5 (Anthropic direct API), Gemini 3 Flash (OpenRouter), GLM-5.1 (Modal), Llama-3.3-70B-Instruct (self-hosted and OpenRouter), and Qwen3-235B-Instruct (self-hosted, reproducibility anchor). The orchestrator is Sonnet 4.6 with a Gemini 3.1 Pro ablation. The attacker is Claude Opus 4.6, with StepFun-3.5-Flash and Llama-3.3-70B-Instruct as transfer controls. Swarm Shepherd operates on the telemetry emitted by these agents. Per-wave backends appear in Section~\ref{sec:evaluation} and Appendix~\ref{app:eval_extended}.

\end{enumerate}

\end{document}